

%
%
\def\unredoffs{} \def\redoffs{\voffset=-.31truein\hoffset=-.48truein}
\def\speclscape{}
%
%
%
%
%
\newbox\leftpage \newdimen\fullhsize \newdimen\hstitle \newdimen\hsbody
\tolerance=1000\hfuzz=2pt
\catcode`\@=11 
\ifx\hyperdef\UNd@FiNeD\def\hyperdef#1#2#3#4{#4}\def\hyperref#1#2#3#4{#4}\fi
\def\bigans{b }
\def\answ{b }
%
\ifx\answ\bigans\message{(This will come out unreduced.}
\magnification=1200\unredoffs\baselineskip=16pt plus 2pt minus 1pt
\hsbody=\hsize \hstitle=\hsize 
\else\message{(This will be reduced.} \let\l@r=L
\magnification=1000\baselineskip=16pt plus 2pt minus 1pt \vsize=7truein
\redoffs \hstitle=8truein\hsbody=4.75truein\fullhsize=10truein\hsize=\hsbody
\output={\ifnum\pageno=0 
  \shipout\vbox{\speclscape{\hsize\fullhsize\makeheadline}
    \hbox to \fullhsize{\hfill\pagebody\hfill}}\advancepageno
  \else
  \almostshipout{\leftline{\vbox{\pagebody\makefootline}}}\advancepageno
  \fi}
\def\almostshipout#1{\if L\l@r \count1=1 \message{[\the\count0.\the\count1]}
      \global\setbox\leftpage=#1 \global\let\l@r=R
 \else \count1=2
  \shipout\vbox{\speclscape{\hsize\fullhsize\makeheadline}
      \hbox to\fullhsize{\box\leftpage\hfil#1}}  \global\let\l@r=L\fi}
\fi
%
\newcount\yearltd\yearltd=\year\advance\yearltd by -2000

\def\Title#1#2{\nopagenumbers\abstractfont\hsize=\hstitle\rightline{#1}%
\vskip 1in\centerline{\titlefont #2}\abstractfont\vskip .5in\pageno=0}
\def\Date#1{\vfill\leftline{#1}\tenpoint\supereject\global\hsize=\hsbody%
\footline={\hss\tenrm\hyperdef\hypernoname{page}\folio\folio\hss}}%
%

\def\draftmode{\message{ DRAFTMODE }\def\draftdate{{\rm preliminary draft:
\number\month/\number\day/\number\yearltd\ \ \hourmin}}%
\headline={\hfil\draftdate}\writelabels\baselineskip=20pt plus 2pt minus 2pt
 {\count255=\time\divide\count255 by 60 \xdef\hourmin{\number\count255}
  \multiply\count255 by-60\advance\count255 by\time
  \xdef\hourmin{\hourmin:\ifnum\count255<10 0\fi\the\count255}}}
\def\nolabels{\def\wrlabeL##1{}\def\eqlabeL##1{}\def\reflabeL##1{}}
\def\writelabels{\def\wrlabeL##1{\leavevmode\vadjust{\rlap{\smash%
{\line{{\escapechar=` \hfill\rlap{\sevenrm\hskip.03in\string##1}}}}}}}%
\def\eqlabeL##1{{\escapechar-1\rlap{\sevenrm\hskip.05in\string##1}}}%
\def\reflabeL##1{\noexpand\llap{\noexpand\sevenrm\string\string\string##1}}}
\nolabels
%
\global\newcount\secno \global\secno=0
\global\newcount\meqno \global\meqno=1
\def\s@csym{}
\def\newsec#1{\global\advance\secno by1%
{\toks0{#1}\message{(\the\secno. \the\toks0)}}%
\global\subsecno=0\eqnres@t\let\s@csym\secsym\xdef\secn@m{\the\secno}\noindent
{\bf\hyperdef\hypernoname{section}{\the\secno}{\the\secno.} #1}%
\writetoca{{\string\hyperref{}{section}{\the\secno}{\the\secno.}} {#1}}%
\par\nobreak\medskip\nobreak}
\def\eqnres@t{\xdef\secsym{\the\secno.}\global\meqno=1\bigbreak\bigskip}
\def\sequentialequations{\def\eqnres@t{\bigbreak}}\xdef\secsym{}
\global\newcount\subsecno \global\subsecno=0
\def\subsec#1{\global\advance\subsecno by1%
{\toks0{#1}\message{(\s@csym\the\subsecno. \the\toks0)}}%
\ifnum\lastpenalty>9000\else\bigbreak\fi
\noindent{\it\hyperdef\hypernoname{subsection}{\secn@m.\the\subsecno}%
{\secn@m.\the\subsecno.} #1}\writetoca{\string\quad
{\string\hyperref{}{subsection}{\secn@m.\the\subsecno}{\secn@m.\the\subsecno.}}
{#1}}\par\nobreak\medskip\nobreak}
\def\appendix#1#2{\global\meqno=1\global\subsecno=0\xdef\secsym{\hbox{#1.}}%
\bigbreak\bigskip\noindent{\bf Appendix \hyperdef\hypernoname{appendix}{#1}%
{#1.} #2}{\toks0{(#1. #2)}\message{\the\toks0}}%
\xdef\s@csym{#1.}\xdef\secn@m{#1}%
\writetoca{\string\hyperref{}{appendix}{#1}{Appendix {#1.}} {#2}}%
\par\nobreak\medskip\nobreak}
%
%
\def\checkm@de#1#2{\ifmmode{\def\f@rst##1{##1}\hyperdef\hypernoname{equation}%
{#1}{#2}}\else\hyperref{}{equation}{#1}{#2}\fi}
\def\eqnn#1{\DefWarn#1\xdef #1{(\noexpand\relax\noexpand\checkm@de%
{\s@csym\the\meqno}{\secsym\the\meqno})}%
\wrlabeL#1\writedef{#1\leftbracket#1}\global\advance\meqno by1}
\def\f@rst#1{\c@t#1a\em@ark}\def\c@t#1#2\em@ark{#1}
\def\eqna#1{\DefWarn#1\wrlabeL{#1$\{\}$}%
\xdef #1##1{(\noexpand\relax\noexpand\checkm@de%
{\s@csym\the\meqno\noexpand\f@rst{##1}}{\hbox{$\secsym\the\meqno##1$}})}
\writedef{#1\numbersign1\leftbracket#1{\numbersign1}}\global\advance\meqno by1}
\def\eqn#1#2{\DefWarn#1%
\xdef #1{(\noexpand\hyperref{}{equation}{\s@csym\the\meqno}%
{\secsym\the\meqno})}$$#2\eqno(\hyperdef\hypernoname{equation}%
{\s@csym\the\meqno}{\secsym\the\meqno})\eqlabeL#1$$%
\writedef{#1\leftbracket#1}\global\advance\meqno by1}
\def\xeqn{\expandafter\xe@n}\def\xe@n(#1){#1}
\def\xeqna#1{\expandafter\xe@n#1}
\def\eqns#1{(\e@ns #1{\hbox{}})}
\def\e@ns#1{\ifx\UNd@FiNeD#1\message{eqnlabel \string#1 is undefined.}%
\xdef#1{(?.?)}\fi{\let\hyperref=\relax\xdef\next{#1}}%
\ifx\next\em@rk\def\next{}\else%
\ifx\next#1\xeqn#1\else\def\n@xt{#1}\ifx\n@xt\next#1\else\xeqna#1\fi
\fi\let\next=\e@ns\fi\next}

\def\DefWarn#1{\ifx\UNd@FiNeD#1\else
\immediate\write16{*** WARNING: the label \string#1 is already defined ***}\fi}
%
\newskip\footskip\footskip14pt plus 1pt minus 1pt 
\def\footnotefont{\ninepoint}\def\f@t#1{\footnotefont #1\@foot}
\def\f@@t{\baselineskip\footskip\bgroup\footnotefont\aftergroup\@foot\let\next}
\setbox\strutbox=\hbox{\vrule height9.5pt depth4.5pt width0pt}
\global\newcount\ftno \global\ftno=0
\def\foot{\global\advance\ftno by1\def\foot@rg{\hyperref{}{footnote}%
{\the\ftno}{\the\ftno}\xdef\foot@rg{\noexpand\hyperdef\noexpand\hypernoname%
{footnote}{\the\ftno}{\the\ftno}}}\footnote{$^{\foot@rg}$}}
%
\newwrite\ftfile
\def\footend{\def\foot{\global\advance\ftno by1\chardef\wfile=\ftfile
\hyperref{}{footnote}{\the\ftno}{$^{\the\ftno}$}%
\ifnum\ftno=1\immediate\openout\ftfile=\jobname.fts\fi%
\immediate\write\ftfile{\noexpand\smallskip%
\noexpand\item{\noexpand\hyperdef\noexpand\hypernoname{footnote}
{\the\ftno}{f\the\ftno}:\ }\pctsign}\findarg}%
\def\footatend{\vfill\eject\immediate\closeout\ftfile{\parindent=20pt
\centerline{\bf Footnotes}\nobreak\bigskip\input \jobname.fts }}}
\def\footatend{}
%
%
\global\newcount\refno \global\refno=1
\newwrite\rfile
\def\ref{[\hyperref{}{reference}{\the\refno}{\the\refno}]\nref}
\def\nref#1{\DefWarn#1%
\xdef#1{[\noexpand\hyperref{}{reference}{\the\refno}{\the\refno}]}%
\writedef{#1\leftbracket#1}%
\ifnum\refno=1\immediate\openout\rfile=\jobname.refs\fi
\chardef\wfile=\rfile\immediate\write\rfile{\noexpand\item{[\noexpand\hyperdef%
\noexpand\hypernoname{reference}{\the\refno}{\the\refno}]\ }%
\reflabeL{#1\hskip.31in}\pctsign}\global\advance\refno by1\findarg}
\def\findarg#1#{\begingroup\obeylines\newlinechar=`\^^M\pass@rg}
{\obeylines\gdef\pass@rg#1{\writ@line\relax #1^^M\hbox{}^^M}%
\gdef\writ@line#1^^M{\expandafter\toks0\expandafter{\striprel@x #1}%
\edef\next{\the\toks0}\ifx\next\em@rk\let\next=\endgroup\else\ifx\next\empty%
\else\immediate\write\wfile{\the\toks0}\fi\let\next=\writ@line\fi\next\relax}}
\def\striprel@x#1{} \def\em@rk{\hbox{}}
\def\lref{\begingroup\obeylines\lr@f}
\def\lr@f#1#2{\DefWarn#1\gdef#1{\let#1=\UNd@FiNeD\ref#1{#2}}\endgroup\unskip}

\def\addref#1{\immediate\write\rfile{\noexpand\item{}#1}} 
\def\listrefs{\footatend\vfill\supereject\immediate\closeout\rfile\writestoppt
\baselineskip=\footskip\centerline{{\bf References}}\bigskip{\parindent=20pt%
\frenchspacing\escapechar=` \input \jobname.refs\vfill\eject}\nonfrenchspacing}
\def\startrefs#1{\immediate\openout\rfile=\jobname.refs\refno=#1}
\def\xref{\expandafter\xr@f}\def\xr@f[#1]{#1}
\def\refs#1{\count255=1[\r@fs #1{\hbox{}}]}
\def\r@fs#1{\ifx\UNd@FiNeD#1\message{reflabel \string#1 is undefined.}%
\nref#1{need to supply reference \string#1.}\fi%
\vphantom{\hphantom{#1}}{\let\hyperref=\relax\xdef\next{#1}}%
\ifx\next\em@rk\def\next{}%
\else\ifx\next#1\ifodd\count255\relax\xref#1\count255=0\fi%
\else#1\count255=1\fi\let\next=\r@fs\fi\next}
%

%
\newwrite\ffile\global\newcount\figno \global\figno=1
\def\fig{fig.~\hyperref{}{figure}{\the\figno}{\the\figno}\nfig}
\def\nfig#1{\DefWarn#1%
\xdef#1{fig.~\noexpand\hyperref{}{figure}{\the\figno}{\the\figno}}%
\writedef{#1\leftbracket fig.\noexpand~\xfig#1}%
\ifnum\figno=1\immediate\openout\ffile=\jobname.figs\fi\chardef\wfile=\ffile%
{\let\hyperref=\relax
\immediate\write\ffile{\noexpand\medskip\noexpand\item{Fig.\ %
\noexpand\hyperdef\noexpand\hypernoname{figure}{\the\figno}{\the\figno}. }
\reflabeL{#1\hskip.55in}\pctsign}}\global\advance\figno by1\findarg}
\def\listfigs{\vfill\eject\immediate\closeout\ffile{\parindent40pt
\baselineskip14pt\centerline{{\bf Figure Captions}}\nobreak\medskip
\escapechar=` \input \jobname.figs\vfill\eject}}
\def\xfig{\expandafter\xf@g}\def\xf@g fig.\penalty\@M\ {}
\def\figs#1{figs.~\f@gs #1{\hbox{}}}
\def\f@gs#1{{\let\hyperref=\relax\xdef\next{#1}}\ifx\next\em@rk\def\next{}\else
\ifx\next#1\xfig #1\else#1\fi\let\next=\f@gs\fi\next}
\def\figin{\epsfcheck\figin}\def\figins{\epsfcheck\figins}
\def\epsfcheck{\ifx\epsfbox\UNd@FiNeD
\message{(NO epsf.tex, FIGURES WILL BE IGNORED)}
\gdef\figin##1{\vskip2in}\gdef\figins##1{\hskip.5in}
\else\message{(FIGURES WILL BE INCLUDED)}%
\gdef\figin##1{##1}\gdef\figins##1{##1}\fi}
\def\DefWarn#1{}
\def\figinsert{\goodbreak\midinsert}
\def\ifig#1#2#3{\DefWarn#1\xdef#1{fig.~\noexpand\hyperref{}{figure}%
{\the\figno}{\the\figno}}\writedef{#1\leftbracket fig.\noexpand~\xfig#1}%
\figinsert\figin{\centerline{#3}}\medskip\centerline{\vbox{\baselineskip12pt
\advance\hsize by -1truein\noindent\wrlabeL{#1=#1}\footnotefont%
{\bf Fig.~\hyperdef\hypernoname{figure}{\the\figno}{\the\figno}:} #2}}
\bigskip\endinsert\global\advance\figno by1}
\newwrite\lfile
{\escapechar-1\xdef\pctsign{\string\%}\xdef\leftbracket{\string\{}
\xdef\rightbracket{\string\}}\xdef\numbersign{\string\#}}
\def\writedefs{\immediate\openout\lfile=\jobname.defs \def\writedef##1{%
{\let\hyperref=\relax\let\hyperdef=\relax\let\hypernoname=\relax
 \immediate\write\lfile{\string\def\string##1\rightbracket}}}}%
\def\writestop{\def\writestoppt{\immediate\write\lfile{\string\pageno
 \the\pageno\string\startrefs\leftbracket\the\refno\rightbracket
 \string\def\string\secsym\leftbracket\secsym\rightbracket
 \string\secno\the\secno\string\meqno\the\meqno}\immediate\closeout\lfile}}
\def\writestoppt{}\def\writedef#1{}
\def\seclab#1{\DefWarn#1%
\xdef #1{\noexpand\hyperref{}{section}{\the\secno}{\the\secno}}%
\writedef{#1\leftbracket#1}\wrlabeL{#1=#1}}
\def\subseclab#1{\DefWarn#1%
\xdef #1{\noexpand\hyperref{}{subsection}{\secn@m.\the\subsecno}%
{\secn@m.\the\subsecno}}\writedef{#1\leftbracket#1}\wrlabeL{#1=#1}}
\def\applab#1{\DefWarn#1%
\xdef #1{\noexpand\hyperref{}{appendix}{\secn@m}{\secn@m}}%
\writedef{#1\leftbracket#1}\wrlabeL{#1=#1}}
\newwrite\tfile \def\writetoca#1{}
\def\leaderfill{\leaders\hbox to 1em{\hss.\hss}\hfill}
\def\writetoc{\immediate\openout\tfile=\jobname.toc
   \def\writetoca##1{{\edef\next{\write\tfile{\noindent ##1
   \string\leaderfill {\string\hyperref{}{page}{\noexpand\number\pageno}%
                       {\noexpand\number\pageno}} \par}}\next}}}
\newread\ch@ckfile
\def\listtoc{\immediate\closeout\tfile\immediate\openin\ch@ckfile=\jobname.toc
\ifeof\ch@ckfile\message{no file \jobname.toc, no table of contents this pass}%
\else\closein\ch@ckfile\centerline{\bf Contents}\nobreak\medskip%
{\baselineskip=12pt\footnotefont\parskip=0pt\catcode`\@=11\input\jobname.toc
\catcode`\@=12\bigbreak\bigskip}\fi}
\catcode`\@=12 
%
\edef\tfontsize{\ifx\answ\bigans scaled\magstep3\else scaled\magstep4\fi}
\font\titlerm=cmr10 \tfontsize \font\titlerms=cmr7 \tfontsize
\font\titlermss=cmr5 \tfontsize \font\titlei=cmmi10 \tfontsize
\font\titleis=cmmi7 \tfontsize \font\titleiss=cmmi5 \tfontsize
\font\titlesy=cmsy10 \tfontsize \font\titlesys=cmsy7 \tfontsize
\font\titlesyss=cmsy5 \tfontsize \font\titleit=cmti10 \tfontsize
\skewchar\titlei='177 \skewchar\titleis='177 \skewchar\titleiss='177
\skewchar\titlesy='60 \skewchar\titlesys='60 \skewchar\titlesyss='60
\def\titlefont{\def\rm{\fam0\titlerm}
\textfont0=\titlerm \scriptfont0=\titlerms \scriptscriptfont0=\titlermss
\textfont1=\titlei \scriptfont1=\titleis \scriptscriptfont1=\titleiss
\textfont2=\titlesy \scriptfont2=\titlesys \scriptscriptfont2=\titlesyss
\textfont\itfam=\titleit \def\it{\fam\itfam\titleit}\rm}
 \ifx\answ\bigans\else scaled\magstep1\fi
\ifx\answ\bigans\def\abstractfont{\tenpoint}\else
\font\absit=cmti10 scaled \magstep1
\font\abssl=cmsl10 scaled \magstep1
\font\absrm=cmr10 scaled\magstep1 \font\absrms=cmr7 scaled\magstep1
\font\absrmss=cmr5 scaled\magstep1 \font\absi=cmmi10 scaled\magstep1
\font\absis=cmmi7 scaled\magstep1 \font\absiss=cmmi5 scaled\magstep1
\font\abssy=cmsy10 scaled\magstep1 \font\abssys=cmsy7 scaled\magstep1
\font\abssyss=cmsy5 scaled\magstep1 \font\absbf=cmbx10 scaled\magstep1
\skewchar\absi='177 \skewchar\absis='177 \skewchar\absiss='177
\skewchar\abssy='60 \skewchar\abssys='60 \skewchar\abssyss='60
\def\abstractfont{\def\rm{\fam0\absrm}
\textfont0=\absrm \scriptfont0=\absrms \scriptscriptfont0=\absrmss
\textfont1=\absi \scriptfont1=\absis \scriptscriptfont1=\absiss
\textfont2=\abssy \scriptfont2=\abssys \scriptscriptfont2=\abssyss
\textfont\itfam=\absit \def\it{\fam\itfam\absit}\def\footnotefont{\tenpoint}%
\textfont\slfam=\abssl \def\sl{\fam\slfam\abssl}%
\textfont\bffam=\absbf \def\bf{\fam\bffam\absbf}\rm}\fi
\def\tenpoint{\def\rm{\fam0\tenrm}
\textfont0=\tenrm \scriptfont0=\sevenrm \scriptscriptfont0=\fiverm
\textfont1=\teni  \scriptfont1=\seveni  \scriptscriptfont1=\fivei
\textfont2=\tensy \scriptfont2=\sevensy \scriptscriptfont2=\fivesy
\textfont\itfam=\tenit \def\it{\fam\itfam\tenit}\def\footnotefont{\ninepoint}%
\textfont\bffam=\tenbf \def\bf{\fam\bffam\tenbf}\def\sl{\fam\slfam\tensl}\rm}
\font\ninerm=cmr9 \font\sixrm=cmr6 \font\ninei=cmmi9 \font\sixi=cmmi6
\font\ninesy=cmsy9 \font\sixsy=cmsy6 \font\ninebf=cmbx9
\font\nineit=cmti9 \font\ninesl=cmsl9 \skewchar\ninei='177
\skewchar\sixi='177 \skewchar\ninesy='60 \skewchar\sixsy='60
\def\ninepoint{\def\rm{\fam0\ninerm}
\textfont0=\ninerm \scriptfont0=\sixrm \scriptscriptfont0=\fiverm
\textfont1=\ninei \scriptfont1=\sixi \scriptscriptfont1=\fivei
\textfont2=\ninesy \scriptfont2=\sixsy \scriptscriptfont2=\fivesy
\textfont\itfam=\ninei \def\it{\fam\itfam\nineit}\def\sl{\fam\slfam\ninesl}%
\textfont\bffam=\ninebf \def\bf{\fam\bffam\ninebf}\rm}
%
%

\hyphenation{anom-aly anom-alies coun-ter-term coun-ter-terms}
\def\inv{^{\raise.15ex\hbox{${\scriptscriptstyle -}$}\kern-.05em 1}}

\def\Dsl{\,\raise.15ex\hbox{/}\mkern-13.5mu D} 
\def\dsl{\raise.15ex\hbox{/}\kern-.57em\partial}

 \def\Tr{{\rm Tr}}
\def\lspace{\ifx\answ\bigans{}\else\qquad\fi}
\def\lbspace{\ifx\answ\bigans{}\else\hskip-.2in\fi} 
\def\boxeqn#1{\vcenter{\vbox{\hrule\hbox{\vrule\kern3pt\vbox{\kern3pt
	\hbox{${\displaystyle #1}$}\kern3pt}\kern3pt\vrule}\hrule}}}
\def\mbox#1#2{\vcenter{\hrule \hbox{\vrule height#2in
		\kern#1in \vrule} \hrule}}  
%

\def\darr#1{\raise1.5ex\hbox{$\leftrightarrow$}\mkern-16.5mu #1}

\def\roughly#1{\raise.3ex\hbox{$#1$\kern-.75em\lower1ex\hbox{$\sim$}}}

\def\smallfig#1#2#3{\DefWarn#1\xdef#1{fig.~\the\figno}
\writedef{#1\leftbracket fig.\noexpand~\the\figno}%
\figinsert\figin{\centerline{#3}}\medskip\centerline{\vbox{
\baselineskip12pt\advance\hsize by -1truein
\noindent\footnotefont{\bf Fig.~\the\figno:} #2}}
\endinsert\global\advance\figno by1}

\def\bb{
\font\tenmsb=msbm10
\font\sevenmsb=msbm7
\font\fivemsb=msbm5
\textfont1=\tenmsb
\scriptfont1=\sevenmsb
\scriptscriptfont1=\fivemsb
}

\input amssym

%
%
\ifx\pdfoutput\undefined
\input epsf
\def\fig#1{\epsfbox{#1.eps}}
\def\figscale#1#2{\epsfxsize=#2\epsfbox{#1.eps}}
%
%
\else
\def\fig#1{\pdfximage {#1.pdf}\pdfrefximage\pdflastximage}
\def\figscale#1#2{\pdfximage width#2 {#1.pdf}\pdfrefximage\pdflastximage}
\fi

\def\IZ{\relax\ifmmode\mathchoice
{\hbox{\cmss Z\kern-.4em Z}}{\hbox{\cmss Z\kern-.4em Z}} {\lower.9pt\hbox{\cmsss Z\kern-.4em Z}}
{\lower1.2pt\hbox{\cmsss Z\kern-.4em Z}}\else{\cmss Z\kern-.4em Z}\fi}

\newif\ifdraft\draftfalse
\newif\ifinter\interfalse
\ifdraft\draftmode\else\interfalse\fi
\def\journal#1&#2(#3){\unskip, \sl #1\ \bf #2 \rm(19#3) }
\def\andjournal#1&#2(#3){\sl #1~\bf #2 \rm (19#3) }

\def\frac#1#2{{#1\over#2}}

\def\inbar{\,\vrule height1.5ex width.4pt depth0pt}
\def\IC{\relax\hbox{$\inbar\kern-.3em{\rm C}$}}
\def\IR{\relax{\rm I\kern-.18em R}}
\def\IP{\relax{\rm I\kern-.18em P}}
\def\Z{{\bf Z}}

%
%


%
\catcode`\@=11
\def\slash#1{\mathord{\mathpalette\c@ncel{#1}}}
\overfullrule=0pt

\def\underrel#1\over#2{\mathrel{\mathop{\kern\z@#1}\limits_{#2}}}

\catcode`\@=12


%


\def\[{[}
\def\]{]}

\def\comment#1{ }

%
\def\draftnote#1{\ifdraft{\baselineskip2ex
                 \vbox{\kern1em\hrule\hbox{\vrule\kern1em\vbox{\kern1ex
                 \noindent \underbar{NOTE}: #1
             \vskip1ex}\kern1em\vrule}\hrule}}\fi}
\def\internote#1{\ifinter{\baselineskip2ex
                 \vbox{\kern1em\hrule\hbox{\vrule\kern1em\vbox{\kern1ex
                 \noindent \underbar{Internal Note}: #1
             \vskip1ex}\kern1em\vrule}\hrule}}\fi}

%
%



%
%
%
%

%

\def\inv{^{-1}}


\def\Tr{{\rm Tr}}

\def\1{{\ds 1}}

\def\C{\hbox{$\bb C$}}

\def\Z{\hbox{$\bb Z$}}

\def\S{\hbox{$\bb S$}}

\newfam\frakfam
\font\teneufm=eufm10
\font\seveneufm=eufm7
\font\fiveeufm=eufm5
\textfont\frakfam=\teneufm
\scriptfont\frakfam=\seveneufm
\scriptscriptfont\frakfam=\fiveeufm
\def\frak{\fam\frakfam \teneufm}

\lref\NiarchosAH{
  V.~Niarchos,
  ``Seiberg dualities and the 3d/4d connection,''
JHEP {\bf 1207}, 075 (2012).
[arXiv:1205.2086 [hep-th]].
}

\lref\AharonyGP{
  O.~Aharony,
  ``IR duality in d = 3 N=2 supersymmetric USp(2N(c)) and U(N(c)) gauge theories,''
Phys.\ Lett.\ B {\bf 404}, 71 (1997).
[hep-th/9703215].
}

\lref\AffleckAS{
  I.~Affleck, J.~A.~Harvey and E.~Witten,
  ``Instantons and (Super)Symmetry Breaking in (2+1)-Dimensions,''
Nucl.\ Phys.\ B {\bf 206}, 413 (1982)..
}

\lref\IntriligatorID{
  K.~A.~Intriligator and N.~Seiberg,
  ``Duality, monopoles, dyons, confinement and oblique confinement in supersymmetric SO(N(c)) gauge theories,''
Nucl.\ Phys.\ B {\bf 444}, 125 (1995).
[hep-th/9503179].
}

\lref\PasquettiFJ{
  S.~Pasquetti,
  ``Factorisation of N = 2 Theories on the Squashed 3-Sphere,''
JHEP {\bf 1204}, 120 (2012).
[arXiv:1111.6905 [hep-th]].
}

\lref\HarveyIT{
  J.~A.~Harvey,
 ``TASI 2003 lectures on anomalies,''
[hep-th/0509097].
}

\lref\BeemMB{
  C.~Beem, T.~Dimofte and S.~Pasquetti,
  ``Holomorphic Blocks in Three Dimensions,''
[arXiv:1211.1986 [hep-th]].
}

\lref\DiPietroBCA{
  L.~Di Pietro and Z.~Komargodski,
  ``Cardy formulae for SUSY theories in $d =$ 4 and $d =$ 6,''
JHEP {\bf 1412}, 031 (2014).
[arXiv:1407.6061 [hep-th]].
}

\lref\SeibergPQ{
  N.~Seiberg,
  ``Electric - magnetic duality in supersymmetric nonAbelian gauge theories,''
Nucl.\ Phys.\ B {\bf 435}, 129 (1995).
[hep-th/9411149].
}

\lref\AharonyBX{
  O.~Aharony, A.~Hanany, K.~A.~Intriligator, N.~Seiberg and M.~J.~Strassler,
  ``Aspects of N=2 supersymmetric gauge theories in three-dimensions,''
Nucl.\ Phys.\ B {\bf 499}, 67 (1997).
[hep-th/9703110].
}

\lref\IntriligatorNE{
  K.~A.~Intriligator and P.~Pouliot,
  ``Exact superpotentials, quantum vacua and duality in supersymmetric SP(N(c)) gauge theories,''
Phys.\ Lett.\ B {\bf 353}, 471 (1995).
[hep-th/9505006].
}

\lref\KarchUX{
  A.~Karch,
  ``Seiberg duality in three-dimensions,''
Phys.\ Lett.\ B {\bf 405}, 79 (1997).
[hep-th/9703172].
}

\lref\SafdiRE{
  B.~R.~Safdi, I.~R.~Klebanov and J.~Lee,
  ``A Crack in the Conformal Window,''
[arXiv:1212.4502 [hep-th]].
}

\lref\SchweigertTG{
  C.~Schweigert,
  ``On moduli spaces of flat connections with nonsimply connected structure group,''
Nucl.\ Phys.\ B {\bf 492}, 743 (1997).
[hep-th/9611092].
}

\lref\GiveonZN{
  A.~Giveon and D.~Kutasov,
  ``Seiberg Duality in Chern-Simons Theory,''
Nucl.\ Phys.\ B {\bf 812}, 1 (2009).
[arXiv:0808.0360 [hep-th]].
}

\lref\Spiridonov{
  Spiridonov, V.~P.,
  ``Aspects of elliptic hypergeometric functions,''
[arXiv:1307.2876 [math.CA]].
}

\lref\GaiottoBE{
  D.~Gaiotto, G.~W.~Moore and A.~Neitzke,
  ``Framed BPS States,''
[arXiv:1006.0146 [hep-th]].
}

\lref\AldayRS{
  L.~F.~Alday, M.~Bullimore and M.~Fluder,
  ``On S-duality of the Superconformal Index on Lens Spaces and 2d TQFT,''
JHEP {\bf 1305}, 122 (2013).
[arXiv:1301.7486 [hep-th]].
}

\lref\ArdehaliBLA{
  A.~Arabi Ardehali,
  ``High-temperature asymptotics of supersymmetric partition functions,''
JHEP {\bf 1607}, 025 (2016).
[arXiv:1512.03376 [hep-th]].
}

\lref\RazamatJXA{
  S.~S.~Razamat and M.~Yamazaki,
  ``S-duality and the N=2 Lens Space Index,''
[arXiv:1306.1543 [hep-th]].
}

\lref\NiarchosAH{
  V.~Niarchos,
  ``Seiberg dualities and the 3d/4d connection,''
JHEP {\bf 1207}, 075 (2012).
[arXiv:1205.2086 [hep-th]].
}

\lref\almost{
  A.~Borel, R.~Friedman, J.~W.~Morgan,
  ``Almost commuting elements in compact Lie groups,''
arXiv:math/9907007.
}

\lref\BobevKZA{
  N.~Bobev, M.~Bullimore and H.~C.~Kim,
  ``Supersymmetric Casimir Energy and the Anomaly Polynomial,''
JHEP {\bf 1509}, 142 (2015).
[arXiv:1507.08553 [hep-th]].
}

\lref\KapustinJM{
  A.~Kapustin and B.~Willett,
  ``Generalized Superconformal Index for Three Dimensional Field Theories,''
[arXiv:1106.2484 [hep-th]].
}

\lref\SeibergBD{
  N.~Seiberg,
  ``Five-dimensional SUSY field theories, nontrivial fixed points and string dynamics,''
Phys.\ Lett.\ B {\bf 388}, 753 (1996).
[hep-th/9608111].
}
\lref\ChanQC{
  C.~S.~Chan, O.~J.~Ganor and M.~Krogh,
  ``Chiral compactifications of 6-D conformal theories,''
Nucl.\ Phys.\ B {\bf 597}, 228 (2001).
[hep-th/0002097].
}

\lref\GanorPC{
  O.~J.~Ganor, D.~R.~Morrison and N.~Seiberg,
  ``Branes, Calabi-Yau spaces, and toroidal compactification of the N=1 six-dimensional E(8) theory,''
Nucl.\ Phys.\ B {\bf 487}, 93 (1997).
[hep-th/9610251].
}

\lref\AharonyGP{
  O.~Aharony,
  ``IR duality in d = 3 N=2 supersymmetric USp(2N(c)) and U(N(c)) gauge theories,''
Phys.\ Lett.\ B {\bf 404}, 71 (1997).
[hep-th/9703215].
}

\lref\FestucciaWS{
  G.~Festuccia and N.~Seiberg,
  ``Rigid Supersymmetric Theories in Curved Superspace,''
JHEP {\bf 1106}, 114 (2011).
[arXiv:1105.0689 [hep-th]].
}

\lref\RomelsbergerEG{
  C.~Romelsberger,
  ``Counting chiral primaries in N = 1, d=4 superconformal field theories,''
Nucl.\ Phys.\ B {\bf 747}, 329 (2006).
[hep-th/0510060].
}

\lref\KapustinKZ{
  A.~Kapustin, B.~Willett and I.~Yaakov,
  ``Exact Results for Wilson Loops in Superconformal Chern-Simons Theories with Matter,''
JHEP {\bf 1003}, 089 (2010).
[arXiv:0909.4559 [hep-th]].
}

\lref\DolanQI{
  F.~A.~Dolan and H.~Osborn,
  ``Applications of the Superconformal Index for Protected Operators and q-Hypergeometric Identities to N=1 Dual Theories,''
Nucl.\ Phys.\ B {\bf 818}, 137 (2009).
[arXiv:0801.4947 [hep-th]].
}

\lref\GaddeIA{
  A.~Gadde and W.~Yan,
  ``Reducing the 4d Index to the $S^3$ Partition Function,''
JHEP {\bf 1212}, 003 (2012).
[arXiv:1104.2592 [hep-th]].
}

\lref\DolanRP{
  F.~A.~H.~Dolan, V.~P.~Spiridonov and G.~S.~Vartanov,
  ``From 4d superconformal indices to 3d partition functions,''
Phys.\ Lett.\ B {\bf 704}, 234 (2011).
[arXiv:1104.1787 [hep-th]].
}

\lref\ImamuraUW{
  Y.~Imamura,
 ``Relation between the 4d superconformal index and the $S^3$ partition function,''
JHEP {\bf 1109}, 133 (2011).
[arXiv:1104.4482 [hep-th]].
}

\lref\BeemYN{
  C.~Beem and A.~Gadde,
  ``The $N=1$ superconformal index for class $S$ fixed points,''
JHEP {\bf 1404}, 036 (2014).
[arXiv:1212.1467 [hep-th]].
}

\lref\HamaEA{
  N.~Hama, K.~Hosomichi and S.~Lee,
  ``SUSY Gauge Theories on Squashed Three-Spheres,''
JHEP {\bf 1105}, 014 (2011).
[arXiv:1102.4716 [hep-th]].
}

\lref\MekareeyaOTA{
  N.~Mekareeya, K.~Ohmori, Y.~Tachikawa and G.~Zafrir,
  ``$E_8$ instantons on type-A ALE spaces and supersymmetric field theories,''
[arXiv:1707.04370 [hep-th]].
}

\lref\GaddeEN{
  A.~Gadde, L.~Rastelli, S.~S.~Razamat and W.~Yan,
  ``On the Superconformal Index of N=1 IR Fixed Points: A Holographic Check,''
JHEP {\bf 1103}, 041 (2011).
[arXiv:1011.5278 [hep-th]].
}

\lref\EagerHX{
  R.~Eager, J.~Schmude and Y.~Tachikawa,
  ``Superconformal Indices, Sasaki-Einstein Manifolds, and Cyclic Homologies,''
[arXiv:1207.0573 [hep-th]].
}

\lref\AffleckAS{
  I.~Affleck, J.~A.~Harvey and E.~Witten,
  ``Instantons and (Super)Symmetry Breaking in (2+1)-Dimensions,''
Nucl.\ Phys.\ B {\bf 206}, 413 (1982)..
}

\lref\SeibergPQ{
  N.~Seiberg,
  ``Electric - magnetic duality in supersymmetric nonAbelian gauge theories,''
Nucl.\ Phys.\ B {\bf 435}, 129 (1995).
[hep-th/9411149].
}

\lref\BahDG{
  I.~Bah, C.~Beem, N.~Bobev and B.~Wecht,
  ``Four-Dimensional SCFTs from M5-Branes,''
JHEP {\bf 1206}, 005 (2012).
[arXiv:1203.0303 [hep-th]].
}

\lref\debult{
  F.~van~de~Bult,
  ``Hyperbolic Hypergeometric Functions,''
University of Amsterdam Ph.D. thesis
}

\lref\OhmoriST{
  K.~Ohmori, H.~Shimizu and Y.~Tachikawa,
  ``Anomaly polynomial of E-string theories,''
JHEP {\bf 1408}, 002 (2014).
[arXiv:1404.3887 [hep-th]].}
  
\lref\OhmoriAMP{
  K.~Ohmori, H.~Shimizu, Y.~Tachikawa and K.~Yonekura,
  ``Anomaly polynomial of general $6d$ SCFTs,''
PTEP {\bf 2014}, 103B07 (2014).
[arXiv:1408.5572 [hep-th]].}

\lref\Shamirthesis{
  I.~Shamir,
  ``Aspects of three dimensional Seiberg duality,''
  M. Sc. thesis submitted to the Weizmann Institute of Science, April 2010.
  }

\lref\slthreeZ{
  J.~Felder, A.~Varchenko,
  ``The elliptic gamma function and $SL(3,Z) \times Z^3$,'' $\;\;$
[arXiv:math/0001184].
}

\lref\BeniniNC{
  F.~Benini, T.~Nishioka and M.~Yamazaki,
  ``4d Index to 3d Index and 2d TQFT,''
Phys.\ Rev.\ D {\bf 86}, 065015 (2012).
[arXiv:1109.0283 [hep-th]].
}

\lref\GaiottoWE{
  D.~Gaiotto,
  ``N=2 dualities,''
  JHEP {\bf 1208}, 034 (2012).
  [arXiv:0904.2715 [hep-th]].
}

\lref\SpiridonovZA{
  V.~P.~Spiridonov and G.~S.~Vartanov,
  ``Elliptic Hypergeometry of Supersymmetric Dualities,''
Commun.\ Math.\ Phys.\  {\bf 304}, 797 (2011).
[arXiv:0910.5944 [hep-th]].
}

\lref\BeniniMF{
  F.~Benini, C.~Closset and S.~Cremonesi,
  ``Comments on 3d Seiberg-like dualities,''
JHEP {\bf 1110}, 075 (2011).
[arXiv:1108.5373 [hep-th]].
}

\lref\BeniniGI{
  F.~Benini, S.~Benvenuti and Y.~Tachikawa,
  ``Webs of five-branes and N=2 superconformal field theories,''
JHEP {\bf 0909}, 052 (2009).
[arXiv:0906.0359 [hep-th]].
}

\lref\ClossetVP{
  C.~Closset, T.~T.~Dumitrescu, G.~Festuccia, Z.~Komargodski and N.~Seiberg,
  ``Comments on Chern-Simons Contact Terms in Three Dimensions,''
JHEP {\bf 1209}, 091 (2012).
[arXiv:1206.5218 [hep-th]].
}

\lref\SpiridonovHF{
  V.~P.~Spiridonov and G.~S.~Vartanov,
  ``Elliptic hypergeometry of supersymmetric dualities II. Orthogonal groups, knots, and vortices,''
[arXiv:1107.5788 [hep-th]].
}

\lref\RazamatQFA{
  S.~S.~Razamat,
  ``On the ${\cal N} =$ 2 superconformal index and eigenfunctions of the elliptic RS model,''
Lett.\ Math.\ Phys.\  {\bf 104}, 673 (2014).
[arXiv:1309.0278 [hep-th]].
}

\lref\SpiridonovWW{
  V.~P.~Spiridonov and G.~S.~Vartanov,
  ``Elliptic hypergeometric integrals and 't Hooft anomaly matching conditions,''
JHEP {\bf 1206}, 016 (2012).
[arXiv:1203.5677 [hep-th]].
}

\lref\HeckmanXDL{
  J.~J.~Heckman, P.~Jefferson, T.~Rudelius and C.~Vafa,
  ``Punctures for Theories of Class ${\cal{S}}_\Gamma$,''
[arXiv:1609.01281 [hep-th]].
}

\lref\DimoftePY{
  T.~Dimofte, D.~Gaiotto and S.~Gukov,
  ``3-Manifolds and 3d Indices,''
[arXiv:1112.5179 [hep-th]].
}

\lref\KimWB{
  S.~Kim,
  ``The Complete superconformal index for N=6 Chern-Simons theory,''
Nucl.\ Phys.\ B {\bf 821}, 241 (2009), [Erratum-ibid.\ B {\bf 864}, 884 (2012)].
[arXiv:0903.4172 [hep-th]].
}

\lref\WillettGP{
  B.~Willett and I.~Yaakov,
  ``N=2 Dualities and Z Extremization in Three Dimensions,''
[arXiv:1104.0487 [hep-th]].
}

\lref\ImamuraSU{
  Y.~Imamura and S.~Yokoyama,
  ``Index for three dimensional superconformal field theories with general R-charge assignments,''
JHEP {\bf 1104}, 007 (2011).
[arXiv:1101.0557 [hep-th]].
}

\lref\FreedYA{
  D.~S.~Freed, G.~W.~Moore and G.~Segal,
  ``The Uncertainty of Fluxes,''
Commun.\ Math.\ Phys.\  {\bf 271}, 247 (2007).
[hep-th/0605198].
}

\lref\HwangQT{
  C.~Hwang, H.~Kim, K.~-J.~Park and J.~Park,
  ``Index computation for 3d Chern-Simons matter theory: test of Seiberg-like duality,''
JHEP {\bf 1109}, 037 (2011).
[arXiv:1107.4942 [hep-th]].
}

\lref\GreenDA{
  D.~Green, Z.~Komargodski, N.~Seiberg, Y.~Tachikawa and B.~Wecht,
  ``Exactly Marginal Deformations and Global Symmetries,''
JHEP {\bf 1006}, 106 (2010).
[arXiv:1005.3546 [hep-th]].
}

\lref\IntriligatorJJ{
  K.~A.~Intriligator and B.~Wecht,
  ``The Exact superconformal R symmetry maximizes a,''
Nucl.\ Phys.\ B {\bf 667}, 183 (2003).
[hep-th/0304128].
}

\lref\IntriligatorID{
  K.~A.~Intriligator and N.~Seiberg,
  ``Duality, monopoles, dyons, confinement and oblique confinement in supersymmetric SO(N(c)) gauge theories,''
Nucl.\ Phys.\ B {\bf 444}, 125 (1995).
[hep-th/9503179].
}

\lref\HoravaQA{
  P.~Horava and E.~Witten,
  ``Heterotic and type I string dynamics from eleven-dimensions,''
Nucl.\ Phys.\ B {\bf 460}, 506 (1996).
[hep-th/9510209].
}

\lref\SeibergNZ{
  N.~Seiberg and E.~Witten,
  ``Gauge dynamics and compactification to three-dimensions,''
In *Saclay 1996, The mathematical beauty of physics* 333-366.
[hep-th/9607163].
}

\lref\HoravaMA{
  P.~Horava and E.~Witten,
  ``Eleven-dimensional supergravity on a manifold with boundary,''
Nucl.\ Phys.\ B {\bf 475}, 94 (1996).
[hep-th/9603142].
}

\lref\KinneyEJ{
  J.~Kinney, J.~M.~Maldacena, S.~Minwalla and S.~Raju,
  ``An Index for 4 dimensional super conformal theories,''
  Commun.\ Math.\ Phys.\  {\bf 275}, 209 (2007).
  [hep-th/0510251].
}

\lref\KT{
  S.~M.~Kuzenko and S.~Theisen,
  ``Correlation Functions of Conserved Currents in ${\cal N}=2$ Superconformal Theory,''
Class.\ Quant.\ Grav.\ {\bf 17}, 665 (2000).
[hep-th/9907107].
}

\lref\NakayamaUR{
  Y.~Nakayama,
  ``Index for supergravity on AdS(5) x T**1,1 and conifold gauge theory,''
Nucl.\ Phys.\ B {\bf 755}, 295 (2006).
[hep-th/0602284].
}

\lref\GaddeKB{
  A.~Gadde, E.~Pomoni, L.~Rastelli and S.~S.~Razamat,
  ``S-duality and 2d Topological QFT,''
JHEP {\bf 1003}, 032 (2010).
[arXiv:0910.2225 [hep-th]].
}

\lref\GaddeTE{
  A.~Gadde, L.~Rastelli, S.~S.~Razamat and W.~Yan,
  ``The Superconformal Index of the $E_6$ SCFT,''
JHEP {\bf 1008}, 107 (2010).
[arXiv:1003.4244 [hep-th]].
}

\lref\AharonyCI{
  O.~Aharony and I.~Shamir,
  ``On $O(N_c)$ d=3 N=2 supersymmetric QCD Theories,''
JHEP {\bf 1112}, 043 (2011).
[arXiv:1109.5081 [hep-th]].
}

\lref\GiveonSR{
  A.~Giveon and D.~Kutasov,
  ``Brane dynamics and gauge theory,''
Rev.\ Mod.\ Phys.\  {\bf 71}, 983 (1999).
[hep-th/9802067].
}

\lref\SpiridonovQV{
  V.~P.~Spiridonov and G.~S.~Vartanov,
  ``Superconformal indices of ${\cal N}=4$ SYM field theories,''
Lett.\ Math.\ Phys.\  {\bf 100}, 97 (2012).
[arXiv:1005.4196 [hep-th]].
}
\lref\GaddeUV{
  A.~Gadde, L.~Rastelli, S.~S.~Razamat and W.~Yan,
  ``Gauge Theories and Macdonald Polynomials,''
Commun.\ Math.\ Phys.\  {\bf 319}, 147 (2013).
[arXiv:1110.3740 [hep-th]].
}

\lref\AlvarezGaumeIG{
  L.~Alvarez-Gaume and E.~Witten,
 ``Gravitational Anomalies,''
Nucl.\ Phys.\ B {\bf 234}, 269 (1984).
}

\lref\KapustinGH{
  A.~Kapustin,
  ``Seiberg-like duality in three dimensions for orthogonal gauge groups,''
[arXiv:1104.0466 [hep-th]].
}

\lref\orthogpaper{O. Aharony, S. S. Razamat, N.~Seiberg and B.~Willett, 
``3d dualities from 4d dualities for orthogonal groups,''
[arXiv:1307.0511 [hep-th]].
}

\lref\readinglines{
  O.~Aharony, N.~Seiberg and Y.~Tachikawa,
  ``Reading between the lines of four-dimensional gauge theories,''
[arXiv:1305.0318 [hep-th]].
}

\lref\WittenNV{
  E.~Witten,
  ``Supersymmetric index in four-dimensional gauge theories,''
Adv.\ Theor.\ Math.\ Phys.\  {\bf 5}, 841 (2002).
[hep-th/0006010].
}

\lref\WittenNVT{
  E.~Witten,
  ``Supersymmetric index of three-dimensional gauge theory,''
[hep-th/9903005].
}

\lref\WittenNVS{
  E.~Witten,
  ``Toroidal compactification without vector structure,''
JHEP {\bf 9802}, 006 (1998).
[hep-th/9712028].
}

\lref\Kac{
  V.~Kac and A.~Smilga,
  ``Vacuum structure in supersymmetric Yang-Mills theories with any gauge group,''
[hep-th/9902029].
}

\lref\Keure{
  A.~Keurentjes,
  ``Nontrivial flat connections on the 3 torus I: G(2) and the orthogonal groups,''
JHEP {\bf 9905}, 001 (1999).
[hep-th/9901154].
}

\lref\Keura{
  A.~Keurentjes,
  ``Nontrivial flat connections on the three torus. 2. The Exceptional groups F4 and E6, E7, E8,''
JHEP {\bf 9905}, 014 (1999).
[hep-th/9902186].
}

\lref\tHooft{
G.~'t~Hooft, 
``{A Property of Electric and Magnetic Flux in Nonabelian Gauge Theories},''
  Nucl.\ Phys.\ B {\bf 153}, 141 (1979).
}

\lref\GaddeUV{
  A.~Gadde, L.~Rastelli, S.~S.~Razamat and W.~Yan,
  ``Gauge Theories and Macdonald Polynomials,''
Commun.\ Math.\ Phys.\  {\bf 319}, 147 (2013).
[arXiv:1110.3740 [hep-th]].
}

\lref\GaddeIK{
  A.~Gadde, L.~Rastelli, S.~S.~Razamat and W.~Yan,
  ``The 4d Superconformal Index from q-deformed 2d Yang-Mills,''
Phys.\ Rev.\ Lett.\  {\bf 106}, 241602 (2011).
[arXiv:1104.3850 [hep-th]].
}

\lref\GaiottoXA{
  D.~Gaiotto, L.~Rastelli and S.~S.~Razamat,
  ``Bootstrapping the superconformal index with surface defects,''
JHEP {\bf 1301}, 022 (2013).
[arXiv:1207.3577 [hep-th]].
}

\lref\GaiottoUQ{
  D.~Gaiotto and S.~S.~Razamat,
  ``Exceptional Indices,''
JHEP {\bf 1205}, 145 (2012).
[arXiv:1203.5517 [hep-th]].
}

\lref\RazamatUV{
  S.~S.~Razamat,
  ``On a modular property of N=2 superconformal theories in four dimensions,''
JHEP {\bf 1210}, 191 (2012).
[arXiv:1208.5056 [hep-th]].
}

\lref\noumi{
  Y.~Komori, M.~Noumi, J.~Shiraishi,
  ``Kernel Functions for Difference Operators of Ruijsenaars Type and Their Applications,''
SIGMA 5 (2009), 054.
[arXiv:0812.0279 [math.QA]].
}

\lref\SpirWarnaar{
  V.~P.~Spiridonov and S.~O.~Warnaar,
  ``Inversions of integral operators and elliptic beta integrals on root systems,''
Adv. Math. 207 (2006), 91-132
[arXiv:math/0411044].
}

\lref\RazamatJXA{
  S.~S.~Razamat and M.~Yamazaki,
  ``S-duality and the N=2 Lens Space Index,''
[arXiv:1306.1543 [hep-th]].
}

\lref\RazamatOPA{
  S.~S.~Razamat and B.~Willett,
  ``Global Properties of Supersymmetric Theories and the Lens Space,''
[arXiv:1307.4381 [hep-th]].
}

\lref\GaddeTE{
  A.~Gadde, L.~Rastelli, S.~S.~Razamat and W.~Yan,
  ``The Superconformal Index of the $E_6$ SCFT,''
JHEP {\bf 1008}, 107 (2010).
[arXiv:1003.4244 [hep-th]].
}

\lref\deBult{
  F.~J.~van~de~Bult,
  ``An elliptic hypergeometric integral with $W(F_4)$ symmetry,''
The Ramanujan Journal, Volume 25, Issue 1 (2011)
[arXiv:0909.4793[math.CA]].
}

\lref\MaruyoshiCAF{
  K.~Maruyoshi and J.~Yagi,
  ``Surface defects as transfer matrices,''
[arXiv:1606.01041 [hep-th]].
}

\lref\GaddeKB{
  A.~Gadde, E.~Pomoni, L.~Rastelli and S.~S.~Razamat,
  ``S-duality and 2d Topological QFT,''
JHEP {\bf 1003}, 032 (2010).
[arXiv:0910.2225 [hep-th]].
}

\lref\ArgyresCN{
  P.~C.~Argyres and N.~Seiberg,
  ``S-duality in N=2 supersymmetric gauge theories,''
JHEP {\bf 0712}, 088 (2007).
[arXiv:0711.0054 [hep-th]].
}

\lref\SpirWarnaar{
  V.~P.~Spiridonov and S.~O.~Warnaar,
  ``Inversions of integral operators and elliptic beta integrals on root systems,''
Adv. Math. 207 (2006), 91-132
[arXiv:math/0411044].
}

\lref\GaiottoHG{
  D.~Gaiotto, G.~W.~Moore and A.~Neitzke,
  ``Wall-crossing, Hitchin Systems, and the WKB Approximation,''
[arXiv:0907.3987 [hep-th]].
}

\lref\RuijsenaarsVQ{
  S.~N.~M.~Ruijsenaars and H.~Schneider,
  ``A New Class Of Integrable Systems And Its Relation To Solitons,''
Annals Phys.\  {\bf 170}, 370 (1986).
}

\lref\RuijsenaarsPP{
  S.~N.~M.~Ruijsenaars,
  ``Complete Integrability Of Relativistic Calogero-moser Systems And Elliptic Function Identities,''
Commun.\ Math.\ Phys.\  {\bf 110}, 191 (1987).
}

\lref\HallnasNB{
  M.~Hallnas and S.~Ruijsenaars,
  ``Kernel functions and Baecklund transformations for relativistic Calogero-Moser and Toda systems,''
J.\ Math.\ Phys.\  {\bf 53}, 123512 (2012).
}

\lref\ZafrirRGA{
  G.~Zafrir,
  ``Brane webs, $5d$ gauge theories and $6d$ ${\cal N}$$=(1,0)$ SCFT's,''
JHEP {\bf 1512}, 157 (2015).
[arXiv:1509.02016 [hep-th]].
}

\lref\OhmoriTKA{
  K.~Ohmori and H.~Shimizu,
 ``$S^1/T^2$ compactifications of 6d $ {\cal N}=\left(1,\;0\right) $ theories and brane webs,''
JHEP {\bf 1603}, 024 (2016).
[arXiv:1509.03195 [hep-th]].
}

\lref\kernelA{
S.~Ruijsenaars,
  ``Elliptic integrable systems of Calogero-Moser type: Some new results on joint eigenfunctions'', in Proceedings of the 2004 Kyoto Workshop on "Elliptic integrable systems", (M. Noumi, K. Takasaki, Eds.), Rokko Lectures in Math., no. 18, Dept. of Math., Kobe Univ.
}

\lref\ellRSreview{
Y.~Komori and S.~Ruijsenaars,
  ``Elliptic integrable systems of Calogero-Moser type: A survey'', in Proceedings of the 2004 Kyoto Workshop on "Elliptic integrable systems", (M. Noumi, K. Takasaki, Eds.), Rokko Lectures in Math., no. 18, Dept. of Math., Kobe Univ.
}

\lref\langmann{
E.~Langmann,
  ``An explicit solution of the (quantum) elliptic Calogero-Sutherland model'', [arXiv:math-ph/0407050].
}

\lref\TachikawaWI{
  Y.~Tachikawa,
  ``4d partition function on $S^1 \times S^3$ and 2d Yang-Mills with nonzero area,''
PTEP {\bf 2013}, 013B01 (2013).
[arXiv:1207.3497 [hep-th]].
}

\lref\TachikawaYMS{
  Y.~Tachikawa,
  ``On S-duality of 5d super Yang-Mills on $S^1$,''
JHEP {\bf 1111}, 123 (2011).
[arXiv:1110.0531 [hep-th]].
}

\lref\MinahanFG{
  J.~A.~Minahan and D.~Nemeschansky,
  ``An N=2 superconformal fixed point with E(6) global symmetry,''
Nucl.\ Phys.\ B {\bf 482}, 142 (1996).
[hep-th/9608047].
}

\lref\AldayKDA{
  L.~F.~Alday, M.~Bullimore, M.~Fluder and L.~Hollands,
  ``Surface defects, the superconformal index and q-deformed Yang-Mills,''
[arXiv:1303.4460 [hep-th]].
}

\lref\FukudaJR{
  Y.~Fukuda, T.~Kawano and N.~Matsumiya,
  ``5D SYM and 2D q-Deformed YM,''
Nucl.\ Phys.\ B {\bf 869}, 493 (2013).
[arXiv:1210.2855 [hep-th]].
}

\lref\XieHS{
  D.~Xie,
  ``General Argyres-Douglas Theory,''
JHEP {\bf 1301}, 100 (2013).
[arXiv:1204.2270 [hep-th]].
}

\lref\DrukkerSR{
  N.~Drukker, T.~Okuda and F.~Passerini,
  ``Exact results for vortex loop operators in 3d supersymmetric theories,''
[arXiv:1211.3409 [hep-th]].
}
\lref\XieGMA{
  D.~Xie,
  ``M5 brane and four dimensional N = 1 theories I,''
JHEP {\bf 1404}, 154 (2014).
[arXiv:1307.5877 [hep-th]].
}

\lref\qinteg{
  M.~Rahman, A.~Verma,
  ``A q-integral representation of Rogers' q-ultraspherical polynomials and some applications,''
Constructive Approximation
1986, Volume 2, Issue 1.
}

\lref\qintegOK{
  A.~Okounkov,
  ``(Shifted) Macdonald Polynomials: q-Integral Representation and Combinatorial Formula,''
Compositio Mathematica
June 1998, Volume 112, Issue 2. 
[arXiv:q-alg/9605013].
}

\lref\macNest{
 H.~Awata, S.~Odake, J.~Shiraishi,
  ``Integral Representations of the Macdonald Symmetric Functions,''
Commun. Math. Phys. 179 (1996) 647.
[arXiv:q-alg/9506006].
}

\lref\MitevJQJ{
  V.~Mitev and E.~Pomoni,
  ``2D CFT blocks for the 4D class $\mathcal{S}_k$ theories,''
JHEP {\bf 1708}, 009 (2017).
[arXiv:1703.00736 [hep-th]].
}

\lref\deBult{
  F.~J.~van~de~Bult,
  ``An elliptic hypergeometric integral with $W(F_4)$ symmetry,''
The Ramanujan Journal, Volume 25, Issue 1 (2011)
[arXiv:0909.4793[math.CA]].
}

\lref\Rains{
E.~M.~Rains,
  ``Transformations of elliptic hypergometric integrals,''
Annals of Mathematics, Volume  171, Issue 1 (2010)
[arXiv:math/0309252].
}

\lref\ItoFPL{
  Y.~Ito and Y.~Yoshida,
  ``Superconformal index with surface defects for class ${\cal S}_k$,''
[arXiv:1606.01653 [hep-th]].
}

\lref\BeniniMZ{
  F.~Benini, Y.~Tachikawa and B.~Wecht,
  ``Sicilian gauge theories and N=1 dualities,''
JHEP {\bf 1001}, 088 (2010).
[arXiv:0909.1327 [hep-th]].
}

\lref\DimoftePD{
  T.~Dimofte and D.~Gaiotto,
  ``An E7 Surprise,''
JHEP {\bf 1210}, 129 (2012).
[arXiv:1209.1404 [hep-th]].
}

\lref\GaddeFMA{
  A.~Gadde, K.~Maruyoshi, Y.~Tachikawa and W.~Yan,
  ``New N=1 Dualities,''
JHEP {\bf 1306}, 056 (2013).
[arXiv:1303.0836 [hep-th]].
}

\lref\AgarwalVLA{
  P.~Agarwal, K.~Intriligator and J.~Song,
  ``Infinitely many ${\cal N}=1 $ dualities from m + 1 ? m = 1,''
JHEP {\bf 1510}, 035 (2015).
[arXiv:1505.00255 [hep-th]].
}

\lref\MinahanCJ{
  J.~A.~Minahan and D.~Nemeschansky,
  ``Superconformal fixed points with E(n) global symmetry,''
Nucl.\ Phys.\ B {\bf 489}, 24 (1997).
[hep-th/9610076].
}

\lref\GorskyTN{
  A.~Gorsky,
  ``Dualities in integrable systems and N=2 SUSY theories,''
J.\ Phys.\ A {\bf 34}, 2389 (2001).
[hep-th/9911037].
}

\lref\FockAE{
  V.~Fock, A.~Gorsky, N.~Nekrasov and V.~Rubtsov,
  ``Duality in integrable systems and gauge theories,''
JHEP {\bf 0007}, 028 (2000).
[hep-th/9906235].
}

\lref\RazamatPTA{
  S.~S.~Razamat and B.~Willett,
  ``Down the rabbit hole with theories of class $ \cal S $,''
JHEP {\bf 1410}, 99 (2014).
[arXiv:1403.6107 [hep-th]].
}

\lref\RazamatL{ A.~Gadde, S.~S.~Razamat, and  B.~Willett,
  ``A ``Lagrangian'' for a non-Lagrangian theory,''
  Phys. Rev. Lett. {\bf 115}, 171604 (2015).
  [arXiv:1505.05834 [hep-th]].
  }
  
  \lref\OhmoriPUA{
  K.~Ohmori, H.~Shimizu, Y.~Tachikawa and K.~Yonekura,
  ``6d ${\cal N}=(1,0)$ theories on $T^2$ and class S theories: Part I,''
JHEP {\bf 1507}, 014 (2015).
[arXiv:1503.06217 [hep-th]].
}

\lref\OhmoriPIA{
  K.~Ohmori, H.~Shimizu, Y.~Tachikawa and K.~Yonekura,
  ``6d ${\cal N}=(1,\;0) $ theories on S$^{1}$ /T$^{2}$ and class S theories: part II,''
JHEP {\bf 1512}, 131 (2015).
[arXiv:1508.00915 [hep-th]].
}

\lref\RastelliTBZ{
  L.~Rastelli and S.~S.~Razamat,
  ``The supersymmetric index in four dimensions,''
[arXiv:1608.02965 [hep-th]].
}

\lref\vandrue{
J. F. van Diejen, 
``Integrability of difference Calogero--Moser systems'', J. Math. Phys. 35 (1994)}

\lref\DelZottoRCA{
  M.~Del Zotto, C.~Vafa and D.~Xie,
  ``Geometric engineering, mirror symmetry and $ 6{{d}}_{\left(1,0\right)}\to 4{{d}}_{\left({\cal N}=2\right)} $,''
JHEP {\bf 1511}, 123 (2015).
[arXiv:1504.08348 [hep-th]].
}

\lref\HananyPFA{
  A.~Hanany and K.~Maruyoshi,
  ``Chiral theories of class $ {\cal S} $,''
JHEP {\bf 1512}, 080 (2015).
[arXiv:1505.05053 [hep-th]].
}

\lref\GaiottoLCA{
  D.~Gaiotto and A.~Tomasiello,
  ``Holography for (1,0) theories in six dimensions,''
JHEP {\bf 1412}, 003 (2014).
[arXiv:1404.0711 [hep-th]].
}

\lref\ApruzziZNA{
  F.~Apruzzi, M.~Fazzi, A.~Passias and A.~Tomasiello,
  ``Supersymmetric AdS$_{5}$ solutions of massive IIA supergravity,''
JHEP {\bf 1506}, 195 (2015).
[arXiv:1502.06620 [hep-th]].
}

\lref\ComanBQQ{
  I.~Coman, E.~Pomoni, M.~Taki and F.~Yagi,
  ``Spectral curves of ${\cal N}=1$ theories of class ${\cal S}_k$,''
[arXiv:1512.06079 [hep-th]].
}

\lref\FrancoJNA{
  S.~Franco, H.~Hayashi and A.~Uranga,
  ``Charting Class ${\cal S}_k$ Territory,''
Phys.\ Rev.\ D {\bf 92}, no. 4, 045004 (2015).
[arXiv:1504.05988 [hep-th]].
}

\lref\SpiridonovZR{
  V.~P.~Spiridonov and G.~S.~Vartanov,
 ``Superconformal indices for N = 1 theories with multiple duals,''
Nucl.\ Phys.\ B {\bf 824}, 192 (2010).
[arXiv:0811.1909 [hep-th]].
}

\lref\AharonyDHA{
  O.~Aharony, S.~S.~Razamat, N.~Seiberg and B.~Willett,
  ``3d dualities from 4d dualities,''
JHEP {\bf 1307}, 149 (2013).
[arXiv:1305.3924 [hep-th]].}

\lref\HeckmanPVA{
  J.~J.~Heckman, D.~R.~Morrison and C.~Vafa,
  ``On the Classification of 6D SCFTs and Generalized ADE Orbifolds,''
JHEP {\bf 1405}, 028 (2014), Erratum: [JHEP {\bf 1506}, 017 (2015)].
[arXiv:1312.5746 [hep-th]].
}

\lref\RazamatDPL{
  S.~S.~Razamat, C.~Vafa and G.~Zafrir,
  ``4d ${\cal N}=1 $ from 6d (1, 0),''
JHEP {\bf 1704}, 064 (2017).
[arXiv:1610.09178 [hep-th]].}

\lref\HeckmanBFA{
  J.~J.~Heckman, D.~R.~Morrison, T.~Rudelius and C.~Vafa,
  ``Atomic Classification of 6D SCFTs,''
Fortsch.\ Phys.\  {\bf 63}, 468 (2015).
[arXiv:1502.05405 [hep-th]].
}

\lref\CsakiCU{
  C.~Csaki, M.~Schmaltz, W.~Skiba and J.~Terning,
  ``Selfdual N=1 SUSY gauge theories,''
Phys.\ Rev.\ D {\bf 56}, 1228 (1997).
[hep-th/9701191].
}

\lref\BahGPH{
  I.~Bah, A.~Hanany, K.~Maruyoshi, S.~S.~Razamat, Y.~Tachikawa and G.~Zafrir,
  ``4d ${\cal N}=1 $ from 6d $ {\cal N}=(1,0) $ on a torus with fluxes,''
JHEP {\bf 1706}, 022 (2017).
[arXiv:1702.04740 [hep-th]].
}

\lref\DelZottoHPA{
  M.~Del Zotto, J.~J.~Heckman, A.~Tomasiello and C.~Vafa,
  ``6d Conformal Matter,''
JHEP {\bf 1502}, 054 (2015).
[arXiv:1407.6359 [hep-th]].
}

\lref\Brz{
C. Beem, S. S. Razamat, G. Zafrir,
``Atiyah-Bott and deformations of ${\cal N}=1$ CFTs,'' work in progress
}

\lref\BhardwajXXA{
  L.~Bhardwaj,
  ``Classification of 6d  N=(1,0) gauge theories,''
JHEP {\bf 1511}, 002 (2015).
[arXiv:1502.06594 [hep-th]].
}

\lref\HeeKiZVS{
H.-C.~Kim, S.~S. Razamat, C. Vafa, and G. Zafrir,
work in progress}

\lref\JKRV{
P. Jefferson, H.-C.~Kim, S.~S. Razamat, and C. Vafa,
work in progress}

\lref\GaiottoUSA{
  D.~Gaiotto and S.~S.~Razamat,
  ``${\cal N}=1 $ theories of class $ {\cal S}_k $,''
JHEP {\bf 1507}, 073 (2015).
[arXiv:1503.05159 [hep-th]].
}

\lref\PestunZXK{
  V.~Pestun {\it et al.},
  ``Localization techniques in quantum field theories,''
[arXiv:1608.02952 [hep-th]].
}

\lref\WittenGX{
  E.~Witten,
  ``Small instantons in string theory,''
Nucl.\ Phys.\ B {\bf 460}, 541 (1996).
[hep-th/9511030].
}

\lref\dieskla{
S. Ruijsenaars, ``Hilbert-Schmidt operators vs. integrable systems of elliptic Calogero-Moser type. IV. The relativistic Heun (van Diejen) case,'' SIGMA 11 (2015), 004}

\lref\diesklb{
E. Rains and S. Ruijsenaars, ``Difference operators of Sklyanin and van Diejen type,'' Comm. Math. Phys. 320 (2013)}

\lref\GanorMU{
  O.~J.~Ganor and A.~Hanany,
  ``Small E(8) instantons and tensionless noncritical strings,''
Nucl.\ Phys.\ B {\bf 474}, 122 (1996).
[hep-th/9602120].
}
\lref\SeibergVS{
  N.~Seiberg and E.~Witten,
  ``Comments on string dynamics in six-dimensions,''
Nucl.\ Phys.\ B {\bf 471}, 121 (1996).
[hep-th/9603003].
}

\lref\FazziEEC{
  M.~Fazzi and S.~Giacomelli,
  ``${\cal N} = 1$ superconformal theories with $D_N$ blocks,''
Phys.\ Rev.\ D {\bf 95}, no. 8, 085010 (2017).
[arXiv:1609.08156 [hep-th]].
}

\lref\MorrisonPP{
  D.~R.~Morrison and C.~Vafa,
  ``Compactifications of F theory on Calabi-Yau threefolds. 2.,''
Nucl.\ Phys.\ B {\bf 476}, 437 (1996).
[hep-th/9603161].
}
\lref\WittenQB{
  E.~Witten,
  ``Phase transitions in M theory and F theory,''
Nucl.\ Phys.\ B {\bf 471}, 195 (1996).
[hep-th/9603150].
}

\lref\MinahanVR{
  J.~A.~Minahan, D.~Nemeschansky, C.~Vafa and N.~P.~Warner,
  ``E strings and N=4 topological Yang-Mills theories,''
Nucl.\ Phys.\ B {\bf 527}, 581 (1998).
[hep-th/9802168].
}

\lref\SeibergBD{
  N.~Seiberg,
  ``Five-dimensional SUSY field theories, nontrivial fixed points and string dynamics,''
Phys.\ Lett.\ B {\bf 388}, 753 (1996).
[hep-th/9608111].
}
\lref\GaiottoLCA{
  D.~Gaiotto and A.~Tomasiello,
  ``Holography for (1,0) theories in six dimensions,''
JHEP {\bf 1412}, 003 (2014).
[arXiv:1404.0711 [hep-th]].
}

\lref\GaiottoUNA{
  D.~Gaiotto and H.~C.~Kim,
  ``Duality walls and defects in 5d $ {\cal N}=1 $ theories,''
JHEP {\bf 1701}, 019 (2017).
[arXiv:1506.03871 [hep-th]].
}
\lref\MorrisonNP{
  D.~R.~Morrison and W.~Taylor,
  ``Classifying bases for 6D F-theory models,''
Central Eur.\ J.\ Phys.\  {\bf 10}, 1072 (2012).
[arXiv:1201.1943 [hep-th]].
}

\Title{\vbox{\baselineskip12pt
}}
{\vbox{
\centerline{E-String Theory on Riemann Surfaces}
\vskip7pt 
\centerline{}
}
}

\centerline{Hee-Cheol Kim,$^a$ Shlomo S. Razamat,$^b$ Cumrun Vafa,$^a$ and Gabi Zafrir$^{c}$}
\bigskip
\centerline{{\it ${}^a$ Jefferson Physical Laboratory, Harvard University, Cambridge, MA 02138, USA}}
\centerline{{\it ${}^b$ Physics Department, Technion, Haifa, Israel 32000}}
\centerline{{\it ${}^c$ Kavli IPMU (WPI), UTIAS, the University of Tokyo, Kashiwa, Chiba 277-8583, Japan}}
\bigskip
\vskip.1in \vskip.2in \centerline{\bf Abstract}

We study  compactifications of the 6d E-string theory, the theory of a small $E_8$ instanton, to four dimensions. In particular we identify ${\cal N}=1$ field theories in four dimensions corresponding to compactifications on arbitrary Riemann surfaces with punctures and with arbitrary non-abelian flat connections as well as fluxes for the abelian sub-groups of the $E_8$ flavor symmetry.   This sheds light on emergent symmetries in a number of 4d ${\cal N}=1$ SCFTs (including the `E7 surprise' theory) as well as leads to new predictions for a large number of 4-dimensional exceptional dualities and symmetries.

\vskip.2in

\noindent  

\vfill

\Date{September 2017}

\newsec{Introduction}

In recent decades we have learnt a lot about the dynamics of supersymmetric quantum field theories in four dimensions. These models often exhibit properties which are hard to explain from first principles. One example of such a property is duality, either exact equivalence of different CFTs or, more ubiquitously, different UV models flowing to the same IR SCFT. Another, less well studied phenomenon, is appearance of symmetries at IR fixed points that are not manifest in the UV description. An interesting question about such phenomena is whether there is any organizing principle responsible for their existence and whether there is a systematic way to discover examples of models possessing such surprising properties. 

Recently, mainly due to proliferation of exact non-perturbative techniques \PestunZXK, on the one hand we are able to relatively easily produce, or more precisely conjecture, many examples of surprising properties of QFTs. On the other hand, many such properties can be fit in a geometric construction realizing the theories of interest as dimensional reduction of some six dimensional supersymmetric model on a two dimensional surface.  Such geometric construction gives precisely the desired organizing principle both giving arguments to why one should have models exhibiting dualities and certain symmetries already observed, and more importantly predicting existence of many new examples.    

The geometric constructions of SCFTs in four dimensions start from a choice of a six dimensional $(1,0)$ supersymmetric model. A vast variety of such models is believed to exist (see, e.g. \refs{\MorrisonNP,\GaiottoLCA,\DelZottoHPA}) and a classification of them has been proposed in \refs{\HeckmanPVA,\HeckmanBFA,\BhardwajXXA}.   Once we compactify these theories on a Riemann surface, it is difficult to ascertain detailed properties of the resulting theories.   In such compactifications for a general choice of the setup one can derive predictions for existence of four dimensional models exhibiting certain duality and symmetry properties \RazamatDPL .  However, for special cases one can say more.
An important case of 6d supersymmetric theories which has been widely studied \refs{\GaiottoWE,\BeniniMZ,\BahDG} and one can say much more about is the $(2,0)$ supersymmetric theory living on a stack of M5 branes compactified on a Riemann surface.  Here many of the compactifications give rise to CFTs in four dimensions with extended supersymmetry, a fact which allows to perform more computations (Seiberg-Witten curves, $\S^4$ partition functions) testing a conjectured map between compactifications and four dimensional constructions. 
 Another example studied recently,  now with ${\cal N}=1$ supersymmetry, is that of M5 branes probing A-type singularity \GaiottoUSA\ (see also  \refs{\HananyPFA,\FrancoJNA,\ComanBQQ}).  Here beyond special cases (for example two M5 branes probing $\Z_2$ singularity \refs{\GaiottoUSA,\RazamatDPL} on general surface, or $N$ M5 branes on a torus with fluxes for global symmetry \refs{\GaiottoUSA,\BahGPH})\foot{One can also understand compactifications of more general $(1,0)$ theories on a torus  with no fluxes, which have extended supersymmetry,  by relating them to the better studied compactifications of the $(2,0)$ theory \refs{\OhmoriPUA,\OhmoriPIA,\DelZottoRCA,\ZafrirRGA,\OhmoriTKA,\MekareeyaOTA}.} an explicit map between  predicted models and 4d field theoretic constructions is hard to derive. When a convenient 4d field theory is identified for a 6d theory on a particular Riemann surface, it might lead  to a stepping stone  which can be used to unravel the whole map for an arbitrary Riemann surface.
 
In this paper we study in detail yet another example of such geometric constructions.  In particular we study Riemann surface compactifications of perhaps the most `minimal' 6d (1,0) theory:   the 6d theory of a small $E_8$ instanton \WittenGX. This model has a variety of other string/M/F-theoretic constructions. It can also be viewed as the theory on an M5 brane probing the Horava-Witten $E_8$-wall \refs{\GanorMU,\SeibergVS}, as  the theory obtained by blowing up a point in the $\C^2$ base of F-theory \refs{\WittenQB,\MorrisonPP}, or as the theory on an M5 brane probing a $D_4$ singularity.  This 6d theory is often referred to as  the E-string model as the corresponding tensionless string enjoys $E_8$ symmetry \MinahanVR.  One can also consider higher rank E-string theories, corresponding to having more than one $M5$ brane probing the $E_8$-wall.   We will construct 4d SCFTs corresponding to compactifications of the rank one E-string model on a general Riemann surface with general values of the fluxes and holonomies for the global symmetry in six dimensions.    As we will discuss, the resulting four dimensional models for certain choices of compactification parameters, should exhibit exceptional symmetry ($E_8$, $E_7\times U(1)$, $E_6\times SU(2)\times U(1)$, and so on).  A stepping stone for our derivation of the map between four and six dimensions will be a particular case of a four dimensional model for which it is believed that exceptional enhancement of symmetry happens \DimoftePD, the {\it $E_7$ surprise model}.    Here the apparent $SU(  8    )$ symmetry of the Lagrangian enhances to $E_7$ at some point on the conformal manifold of the IR SCFT.     In fact, we derive this field theory using known facts about compactifications of E-string theory on a circle (which leads to a description with $SU(2)$ gauge theory with 8 flavors \SeibergBD ).  In particular our construction demystifies the {\it E7-surprise}.  This single entry on the map will allow us to chart the whole correspondence between six and four dimensions in this case. In particular we will derive a large variety of quiver theories for which we conjecture the symmetry of the IR fixed points is enhanced to various sub-groups of $E_8$. We will also construct models which have $E_8$ itself as the symmetry group of the fixed point.

The paper is organized as follows. In section two we discuss the general predictions for four dimensional theories derived from six dimensions starting from the E string model. We compute the anomalies of the theories in four dimensions and the expected flavor symmetry. In section three we discuss the field theories corresponding to compactifications on a torus with fluxes preserving $E_7\times U(1)$ symmetry. In particular we develop the basic entry, a sphere with two punctures, on the correspondence map which will be utilized to bootstrap it in what follows. In section four we consider a five dimensional perspective from which this correspondence can be deduced.  Moreover we use the $5d$ picture to derive the resulting $4d$ theory for a sphere with two punctures and arbitrary fluxes.   In section five we present several checks of this prediction deriving theories corresponding to toroidal compactification and also sphere with two punctures with flux breaking the symmetry of the four dimensional models to subgroups of $E_8$. In section six we study the procedure of closing punctures and in particular discuss spheres with one puncture. In section seven we propose a model corresponding to a sphere with three maximal punctures.  From this theory we can then construct models corresponding to general Riemann surfaces with punctures and general values of the flux. In section eight we summarize the results.  Several appendices complement the text with additional details and computations. In particular, Appendix B includes comments on generalizations of our results to higher rank E-string  theories.

\

\newsec{E-string}

For a general 6d $(1,0)$ theory compactified on a Riemann surface we would expect to obtain an ${\cal N}=1$ supersymmetric theory in 4d.  To preserve the supersymmetry we embed the $U(1)$ holonomy of the surface in the $SU(2)$ R-symmetry of the 6d theory.  We can also turn on supersymmetry preserving flat connections for the flavor symmetries of the 6d theory, as well as turn on  fluxes in an abelian subgroup of the flavor group \RazamatDPL.  One can then predict the symmetries, the dimension of the conformal manifold \RazamatDPL, the number of and charges of certain relevant deformations \Brz,  as well as the 't Hooft anomalies (from which, assuming there are no accidental abelian symmetries in four dimensions,  $a$ and $c$ central charges can be computed) for the resulting 4d theories.  In this section we shall discuss the compactification of the rank $Q$ E-string theory to $4d$ with fluxes under its  flavor symmetry. 
The E-string theory has flavor symmetry $E_8$ for rank one and $SU(2)\times E_8$ for rank higher than one.  In what follows we will concentrate mostly on the rank one case, however the six dimensional analysis can be easily done for the general case and we will keep the rank as a parameter here.  

We start from the computation of the anomalies of the  $4d$ models resulting from the compactification of the mother $6d$ theory. 
For that we require the anomaly polynomial of the rank $Q$ E-string theory. This was computed in \refs{\OhmoriST,\OhmoriAMP}, who found it to be,

\eqn\APEstring{\eqalign{
&I_{E-string}  =\cr &\qquad  \frac{Q(4Q^2 + 6Q +3)}{24}C^2_2 (R) + \frac{(Q-1)(4Q^2 - 2Q +1)}{24}C^2_2 (L) - \frac{Q(Q^2 - 1)}{3} C_2 (R) C_2 (L)  \cr & +  \frac{(Q-1)(6Q+1)}{48} C_2 (L) p_1 (T) - \frac{Q(6Q+5)}{48} C_2 (R) p_1 (T) + \frac{Q(Q-1)}{120} C_2 (L) C_2(E_8)_{\bf{248}}  \cr  &\quad -  \frac{Q(Q+1)}{120} C_2 (R) C_2(E_8)_{\bf{248}} + \frac{Q}{240} p_1 (T) C_2(E_8)_{\bf{248}} + \frac{Q}{7200} C^2_2(E_8)_{\bf{248}} \cr & +  (30Q-1)\frac{7p_1 (T) - 4p_2 (T)}{5760}\,. }
}
We use the notation $C_2 (R), C_2 (L)$ for the second Chern classes in the fundamental representation of the $SU(2)_R$ and $SU(2)_L$ symmetries, respectively.  Here $SU(2)_R$ denotes the R-symmetry and $SU(2)_L$ denotes the global symmetry of the higher rank E-string theory.
We also employ the notation $C_2(G)_{\bf{R}}$ for the second Chern class of the global symmetry $G$, evaluated in the representation $\bf{R}$, and $p_1 (T), p_2 (T)$ for the first and second Pontryagin classes respectively.

Next we consider compactifying the theory on a torus with fluxes under $U(1)$ subgroups of $E_8$.  
We shall first consider the case of a single $U(1)$ and then remark about more general cases.

\

\subsec{Some properties of $E_8$}

There are eight convenient generators of $U(1)$'s inside $E_8$. These are just given by the Cartan subalgebra of $E_8$. To each $U(1)$ we can associate a node in the Dynkin diagram of $E_8$. Then for each node we get a different embedding of a $U(1)$ inside $E_8$ where the commutant of the $U(1)$ in $E_8$ is given by the Dynkin diagram one is left with after removing that node. The Dynkin diagram of $E_8$, in a standard numbering, is given in figure 1. In Table 2.2 
we have provided the commutant of the $U(1)$ inside $E_8$ for each node, as well as data that is useful when performing  calculation of the anomaly. The branching rules for the adjoint of $E_8$ are given in  Appendix A for all of the eight choices. These then also serve to define the normalizations we are using for the various $U(1)$'s.\foot{In this paper we will, unless otherwise stated, be cavalier with global properties of the groups.}

\

\centerline{\figscale{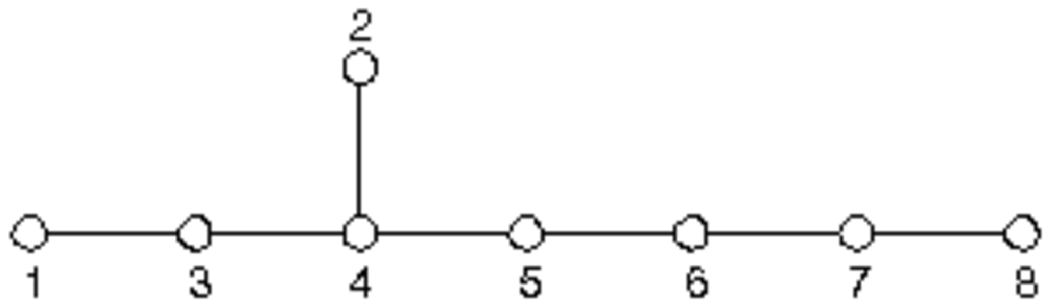}{1.9in}}
\medskip\centerline{\vbox{
\baselineskip12pt\advance\hsize by -1truein
\centerline{\noindent\footnotefont{\bf Fig.~1:} The Dynkin diagram of $E_8$.}}}

\eqn\FluxesinEstring{
\vbox{\offinterlineskip\tabskip=0pt
\halign{\strut\vrule#
\vrule
&~#~\hfil\vrule\vrule
&~#~\hfil\vrule
&~#\hfil\vrule
&\vrule#
\cr
\noalign{\hrule}
Node Number & Associated representation & Commutant in $E_8$ & $\xi$ & \cr\noalign{\hrule}
 $\;\;\;8$ & $\bf{248}$ & $U(1)\times E_7$  & $1$& \cr\noalign{\hrule} 
  $\;\;\;1$ & $\bf{3875}$ & $U(1)\times SO(14)$ & $2$ & \cr\noalign{\hrule} 
   $\;\;\;7$ & $\bf{30380}$ & $U(1)\times SU(2) \times E_6$  & $3$& \cr\noalign{\hrule} 
  $\;\;\;2$ & $\bf{147250}$ & $U(1)\times SU(  8    )$  & $4$ & \cr\noalign{\hrule} 
 $\;\;\;6$ & $\bf{2450240}$ & $U(1)\times SU(3)\times SO(10)$  & $6$ & \cr\noalign{\hrule} 
$\;\;\;3$ & $\bf{6696000}$ & $U(1)\times SU(2)\times SU(7)$  & $7$ & \cr\noalign{\hrule} 
 $\;\;\;5$ & $\bf{146325270}$ & $U(1)\times SU(4)\times SU(5)$  & $10$ & \cr\noalign{\hrule} 
 $\;\;\;4$ & $\bf{6899079264}$ & $U(1)\times SU(2)\times SU(3)\times SU(5)$  & $15$ & \cr\noalign{\hrule}
}
\hrule}}

\

Specifically, we shall need to decompose the second Chern class of $E_8$ into the second Chern classes of the commutant and the first Chern class of the $U(1)$. In this case we find that,

\eqn\CCDecom{
C_2(E_8) = - 2 \xi C^2_1 (U(1)) + \sum_j C_2(G_j) .  }
Here $C_1 (U(1))$ is the first Chern class of the $U(1)$, normalized so that the minimal charge is $1$, and $\xi$ is a $U(1)$ dependent integer which values for the various $U(1)$'s are given in Table \FluxesinEstring. The sum $j$ is over all  simple groups that are commutants of the $U(1)$ in $E_8$. Here we adopted a representation independent normalization of the second Chern class defined as: $C_2(G)_{\bf{R}} = T(G)_{\bf{R}} C_2(G)$, where $T(G)_{\bf{R}}$ is the Dynkin index of the representation.\foot{For $SU(N)$ and $USp(2N)$ groups, the Dynkin index of the fundamental representation is $\frac{1}{2}$, for $SO(N)$ groups it is $1$, and for $E_6, E_7$ and $E_8$ it is $3, 6$ and $30$ respectively.}

The generalization for fluxes in say $n$ $U(1)$'s is straightforward. The decomposition can now be written as,

\eqn\CCDecomGen{
C_2(E_8) = - 2 \sum^n_{i,j=1} \Xi^{i j} C_1 (U(1)_i) C_1 (U(1)_j) + \sum_j C_2(G_j) \,    ,  }
where $\Xi$ is an $n \times n$ real symmetric matrix. Thus there is a basis in which the matrix $\Xi$ is diagonal. In this diagonal basis the decomposition becomes:

\eqn\CCDecomDiag{
C_2(E_8) = - 2 \sum^n_{i=1} \xi_{i} C^2_1 (U(1)_i) + \sum_j C_2(G_j)\,       , }   
where  $\xi_{i}$ are the ones given in Table \FluxesinEstring\ for each $U(1)$.\foot{
For example consider the $n=2$ cases whose branching rules are given in  Appendix A. For all three cases appearing appearing in Appendix A the basis used is diagonal and we can immediately write the decomposition. }

Using the group theory information discussed here we can next compute the anomalies of the resulting $4d$ theory. We shall first deal with the case of flux in a single $U(1)$, and after that go on to discuss more general cases.

\subsec{Anomalies of the E-string theory with flux in a single $U(1)$}

We can consider compactifying the $6d$ theory on a Riemann surface $\Sigma$ with flux under a $U(1)$, that is $\int_{\Sigma} C_1 (U(1)) = -z$ where $z$ is an integer. First let us concentrate on the case where $\Sigma$ is a torus. As the torus is flat we do not need to twist to preserve SUSY. However, SUSY is still broken down to ${\cal N}=1$ in $4d$ by the flux. The $4d$ theory inherits a natural $U(1)_R$ R-symmetry from the Cartan of the $6d$ $SU(2)_R$ though this in general is not the superconformal R-symmetry. Under the embedding of $U(1)_R \subset SU(2)_R$, the characteristic classes decompose as, $C_2 (R) = -C^2_1 (U(1)_R)$. 

Next we need to decompose $E_8$ to the subgroup preserved by the flux as is done in \CCDecom. Finally we set: $C_1 (U(1)) = -z t + \epsilon C_1 (U(1)_R) + C_1 (U(1)_F)$. The first term is the flux on the Riemann surface, where we use $t$ for a unit flux two form on $\Sigma$, that is $\int_{\Sigma} t = 1$. The second term takes into account possible mixing of the $4d$ global $U(1)$ with the superconformal R-symmetry, where $\epsilon$ is a parameter to be determined via a-maximization \IntriligatorJJ. For a-maximization one has to be careful that accidental $U(1)$ symmetries do not appear in the IR.    Finally, the third term is the $4d$ curvature of the $U(1)$.   
Next, we plug these decompositions into \APEstring\ and integrate over the Riemann surface. This yields the $4d$ anomaly polynomial six -- form. From this we can evaluate $a$ and determine $\epsilon$. We find that,

\eqn\mi{
\epsilon = sign(z) \sqrt{\frac{3Q+5}{18 \xi}}
}
Inserting this into the $4d$ anomaly polynomial we find,

\eqn\aci{
a= \frac{\sqrt{2 \xi}Q(3Q+5)^{\frac{3}{2}}|z|}{16}, \qquad\, c = \frac{Q\sqrt{2\xi(3Q+5)}(3Q+7)|z|}{16},}
\eqn\fli{
Tr(U(1)_F) = -12 \xi Q z , \qquad \, Tr(U(1)^3_F) = -12 \xi^2 Q z,}
\eqn\riu{
Tr(U(1)_R U(1)^2_F) = -(2\xi)^{\frac{3}{2}} Q \sqrt{3Q+5}|z|, \qquad \, Tr(U(1)_F U(1)^2_R) = -\frac{4 \xi Q z}{3}}
\eqn\ngi{
Tr(U(1)_R G^2) = - \frac{Q \sqrt{\xi(3Q+5)}|z|}{3 \sqrt{2}}, \qquad \, Tr(U(1)_F G^2) = - Q \xi z, }
\eqn\ggy{
Tr(U(1)_R SU(2)^2_L) = - \frac{Q(Q-1) \sqrt{2\xi(3Q+5)}|z|}{12}, \qquad \,  Tr(U(1)_F SU(2)^2_L) = - \frac{1}{2}Q(Q-1) \xi z,  }
We can package the anomalies in a trial $a$ and $c$ function. Define $R=R'-\frac{s}2 F- h\, T$ where $T$ is the Cartan of $SU(2)$ and $F$ is the $U(1)$ generator. The generator $R'$ is the six dimensional R symmetry before we extremize the trial $a$. The trial conformal anomalies for rank $Q$ E-string on torus with   flux $z$ for single $U(1)$ are then, 

\eqn\ans{\eqalign{&a=
\frac{9}{128} z \xi s Q \left(6 s^2 \xi+3 h^2 (Q-1)-4 (3 Q+5)\right)\,,\cr&
c=\frac{3}{128}\xi z  s Q \left(18 s^2 \xi+9 h^2 (Q-1)-4 (9 Q+19)\right)\,.
}}

\

\subsec{Symmetry and flux quantization}

We have chosen to normalize the $U(1)$'s so that the minimal charge is $1$. This   implies that the flux is quantized so as to be an integer. The global symmetry in 4d is then the commutant of the flux inside $E_8$.
However, as detailed in  Appendix C, fractional fluxes may still be consistent if they are accompanied by flux in the center of the non-abelian symmetry\foot{Specifically, the flux is generated by two holonomies that do not commute up to an element of the center. As such, it breaks the global symmetry to a smaller group. It can also be regarded as a nonzero Stiefel-Whitney class for the global symmetry bundle preserved by the $U(1)$ fluxes. Again we refer the reader to Appendix C for details.}. The possible spectrum can be inferred by studying the branching rules in Appendix A, and looking for combined transformations that act trivially. We won't give here a full classification rather mention a few cases that will play a role later.

Consider the breaking of $E_8 \rightarrow U(1) \times E_7$. In this case we can support also half-integer fluxes. Under this choice of fluxes only the ${\bf 56}^{\pm 1}$ will transform non-trivially, which can be canceled by turning on a flux in the center of $E_7$. However, this will break $E_7$ to a smaller group. The maximal subgroup one can preserve is $F_4$. It should be noted though that the commutant depends on the choice of elements used to implement this flux, and in particular different choices can lead to different symmetries though the rank remains invariant.

As a more complicated example, consider the breaking of $E_8 \rightarrow U(1) \times SO(14)$. In this case we can also support fluxes of the form $\frac{n}{4}$ where $n$ is an integer. This follows as $SO(14)$ has a $\Z_4$ center which we can turn on flux in to compensate for the incomplete transformation generated by the $U(1)$ flux. In the case of half-integer flux, the element in the center that is used is exactly the one corresponding to a $2\pi$ rotation in the $SO$ group. In this case the maximal commutant group is $SO(11)$.

As a final example consider the case of $E_8 \rightarrow U(1) \times SU(2) \times E_6$. Now we can incorporate fluxes quantized as $\frac{n}{6}$ where $n$ is an integer. In this we use the $\Z_2$ center of $SU(2)$ and the $\Z_3$ center of $E_6$. In the specific case of half-integer flux, we rely only on flux in the $\Z_2$ center of the $SU(2)$. This breaks completely the $SU(2)$. 

\subsec{Anomalies of the E-string theory with fluxes in more than one $U(1)$}

It is straightforward to generalize to flux in more $U(1)$'s. At the level of the anomaly polynomial this implies we  take $\int_{\Sigma} C_1 (U(1)_i) = -z_i$, where $z$ is a vector of fluxes. This means that we  use the decomposition \CCDecomGen, and take $C_1 (U(1)_i) = -z_i t + \epsilon_i C_1 (U(1)_R) + C_1 (U(1)_{F_i})$, but otherwise proceed as before. Therefore, after integrating the $6d$ anomaly polynomial we get the $4d$ one.

The $4d$ anomaly polynomial has increasingly more terms, the more $U(1)$'s we turn on flux in. Similarly, to get $a$ and $c$ we will need to determine all $\epsilon$'s by performing a-maximization.

As an example let's consider the case of two $U(1)$'s. We shall assume that these can be written in a diagonal basis. In this case we can show that,

\eqn\mitwo{
\epsilon_i = z_i \sqrt{\frac{3Q+5}{18 (\xi_1 z^2_1 + \xi_2 z^2_2)}}
}     
From this we can evaluate various anomalies. For instance for $a$ and $c$ we find,

\eqn\actwoi{
a= \frac{\sqrt{2 (\xi_1 z^2_1 + \xi_2 z^2_2)}Q(3Q+5)^{\frac{3}{2}}}{16}\,\qquad c = \frac{Q\sqrt{2(\xi_1 z^2_1 + \xi_2 z^2_2)(3Q+5)}(3Q+7)}{16}\,.}
Finally we can deal with the case of arbitrary flux. Assuming again that we are using a diagonal basis, then it is possible to show that we now get,

\eqn\acgeni{
a= \frac{\sqrt{2 (\sum^8_{i=1} \xi_i z^2_i)}Q(3Q+5)^{\frac{3}{2}}}{16},\,\qquad  c = \frac{Q\sqrt{2(\sum^8_{i=1} \xi_i z^2_i)(3Q+5)}(3Q+7)}{16}\,.}
This leaves the issue of finding possible diagonal bases. Recall that abelian fluxes can be identified with points in the root lattice so the problem can be reduced to finding a diagonal basis for the root lattice. For this it is convenient to use a basis of roots given by the $SO(16)\subset E_8$. The roots can be represented by their charges under the eight Cartans. As the adjoint of $E_8$ decomposes as: $\bf{248} \rightarrow \bf{120} + \bf{128}$, the roots are given by ,

\eqn\eeightroots{\eqalign{&
(\pm 2 , \pm 2, 0 , 0 , 0 , 0 , 0 , 0 ) +\, permutations,\cr& (\pm 1 , \pm 1, \pm 1 , \pm 1 , \pm 1 , \pm 1 , \pm 1 , \pm 1 )\; with\, even\, number\, of\, minus\, signs,}}
where the first term gives the $112$ roots in $\bf{120}$, and the second term the $\bf{128}$. Here we have normalized the $U(1)$'s so that the minimal charge is $1$ as we have used in this article.

A convenient choice of basis then is the $SO(16)$ one: $(4 , 0 , 0 , 0 , 0 , 0 , 0 , 0 ) +$ permutations. This is a diagonal basis where each $U(1)$ preserves $U(1)\times SO(14)\subset E_8$. Thus, in this basis $\xi_i=2$. One can choose other bases. For instance the basic roots each preserve a $U(1)\times E_7\subset E_8$ and a diagonal basis can be made just from them. For instance the choice: $( 2 , 2 , 0 , 0 , 0 , 0 , 0 , 0 ) + ( 2 , -2 , 0 , 0 , 0 , 0 , 0 , 0 ) + $ $\Z_4$ cyclic permutations, is a diagonal basis with $\xi_i=1$. Finally in Table (2.17) we have written several vector choices for each basic subgroup. 

\eqn\FluxesandRoots{
\vbox{\offinterlineskip\tabskip=0pt
\halign{\strut\vrule#
\vrule
&~#~\hfil\vrule\vrule
&~#~\hfil\vrule
&~#\hfil\vrule
&\vrule#
\cr
\noalign{\hrule}
Node & Associated vectors & Commutant in $E_8$ &  $\|vector\|^2$ & \cr\noalign{\hrule}
 $\;\;\;8$ & $(2 , 2, 0 , 0 , 0 , 0 , 0 , 0 ), (1 , 1, 1 , 1 , 1 , 1 , 1 , 1 )$ & $U(1)\times E_7$  & $8$& \cr\noalign{\hrule} 
  $\;\;\;1$ & $(4 , 0, 0 , 0 , 0 , 0 , 0 , 0 ), (2 , 2, 2 , 2 , 0 , 0 , 0 , 0 )$ & $U(1)\times SO(14)$ & $16$ & \cr\noalign{\hrule} 
   $\;\;\;7$ & $(2 , 2, 2 , 2 , 2 , 2 , 0 , 0 ), (3 , 3, 1 , 1 , 1 , 1 , 1 , 1 )$ & $U(1)\times SU(2) \times E_6$  & $24$& \cr\noalign{\hrule} 
  $\;\;\;2$ & $(4 , 2, 2 , 2 , 2 , 0 , 0 , 0 ), (5 , 1, 1 , 1 , 1 , 1 , 1 , 1 )$ & $U(1)\times SU(  8    )$  & $32$ & \cr\noalign{\hrule} 
 $\;\;\;6$ & $(4 , 4, 2 , 2 , 2 , 2 , 0 , 0 ), (4 , 4, 4 , 0 , 0 , 0 , 0 , 0 )$ & $U(1)\times SU(3)\times SO(10)$  & $48$ & \cr\noalign{\hrule} 
$\;\;\;3$ & $(6 , 2, 2 , 2 , 2 , 2 , 0 , 0 )$ & $U(1)\times SU(2)\times SU(7)$  & $56$ & \cr\noalign{\hrule} 
 $\;\;\;5$ & $(4 , 4, 4 , 4 , 4 , 0 , 0 , 0 ), (5 , 5, 5 , 1 , 1 , 1 , 1 , 1 )$ & $U(1)\times SU(4)\times SU(5)$  & $80$ & \cr\noalign{\hrule} 
 $\;\;\;4$ & $(9 , 3, 3 , 3 , 3 , 1 , 1 , 1 )$ & $U(1)\times SU(2)\times SU(3)\times SU(5)$  & $120$ & \cr\noalign{\hrule}
}
\hrule}} 
{{\it Basic vectors on the root lattice preserving a given subgroup inside $E_8$. Here we show the minimal choice where others can be generated by multiplying the vector. Also each choice has several possibilities that are related by Weyl transformations, and we have only given some of them. We have also given the length of the vector, which is a Weyl invariant. } }

\

While one can choose any basis to work with, when studying the theories that appear in $4d$ a convenient basis presents itself. This basis uses the $U(1) \times SU(  8    )$ subgroup embedded as: $U(1) \times SU(  8    ) \subset U(1) \times E_7 \subset E_8$. For this we can introduce the flux vector $(n_t; n_i)$, where $n_t$ is the flux under the $U(1)$ and $n_i$ are the fluxes under the $SU(  8    )$, as such they obey $\sum_i n_i = 0$. The normalization of the $U(1)$ is as in Appendix A, and $n_i$ are normalized such that: $\bf 8 = \sum_i a_i$.

The flux vector as given is overcomplete. One can combine the fluxes in $n_t$ and $n_i$ to form the flux vector $(f_i)$ where $f_i = 2n_i + n_t$. This leads to an $SO(14)$ basis which is exactly the one we introduced before. This is a convenient basis as the fluxes precisely match with points in the root lattice of $E_8$ in the $SO(16)$ basis we introduced, which can be used to infer the global symmetry preserved by the flux. For example, the flux $f_i = 1$ preserves $E_7$ while the one $f_1 = f_2 = f_3 = f_4 = 1, f_5 = f_6 = f_7 = f_8 = 0$ preserves $SO(14)$, and likewise for other fluxes appearing in Table \FluxesandRoots.

Another convenient property of these bases is that they are diagonal. In terms of the $(n_t; n_i)$ presentation then $n_t$ is an $E_7$ preserving $U(1)$ so $\xi_t = 1$. One cannot give a flux to just one $n_i$ yet for anomaly calculations one can associate with them the unphysical $\xi_{n_i} = \frac{1}{2}$. Then in this basis we have:

\eqn\fluxbasiso{
\sum_i \xi_i z^2_i = n^2_t + \frac{1}{2} \sum_i n^2_i.}

The combined basis, $(f_i)$, is an $SO(14)$ one, but we have chosen to normalize it so that the $SO(14)$ preserving roots are $(1,0,0,0,0,0,0,0)$ instead of $(4,0,0,0,0,0,0,0)$. Thus, in this basis we have:  

\eqn\fluxbasist{
\sum_i \xi_i z^2_i = 2 \sum_i \biggl(\frac{f_i}{4}\biggr)^2 =\frac{1}{8} \sum_i f^2_i.}

\

\subsec{Anomalies of the E-string theory with fluxes on a closed Riemann surface}

We consider the case of a generic closed Riemann surface of genus $g$. This differs from the previous case as Riemann surfaces are generically curved and so supersymmetry is completely broken. We can preserve supersymmetry by twisting the $SO(2)$ acting on the tangent space of the Riemann surface with the Cartan of $SU(2)_R$. At the level of the anomaly polynomial, this changes the decomposition of $C_2(R)$: $C_2(R) = - C_1(R)^2 + 2(1-g) t C_1(R) + O(t^2)$. The rest proceed exactly as before. It is convenient in this case to normalize the flux with respect to the genus. For that we define $\tilde{z} = \frac{z}{2-2g}$.    

The simplest case is the compactification with no flux for which we find,

 \eqn\noflux{
a= \frac{75}{16} (g-1)\,,\,\qquad  c = \frac{43}{8} (g-1) .}
This only makes sense for $g>1$. The case of $g=1$ is known to give the Minahan-Nemeschansky \MinahanCJ\ $E_8$ theory \refs{\SeibergBD,\GanorPC}. This case has ${\cal N}=2$ supersymmetry, where the ${\cal N}=2$ $U(1)_R$ is an accidental symmetry from the 6d view point. As a result we can not compute the anomalies involving this symmetry. As for ${\cal N}=2$ superconformal theories all non-vanishing anomalies must involve this symmetry \KT, we cannot determine them using this method. In that light the result of \noflux\  is consistent though un-informative. Finally when $g=0$ we do not expect a $4d$ superconformal theory. It should be noted though that with sufficiently high flux, even sphere compactifications can lead to interesting $4d$ models \FazziEEC.

For completeness we shall next write the anomalies for general compactifications,

\eqn\sdrgen{
Tr(U(1)^3_R)= (g-1)Q(4Q^2 + 6Q + 3), \qquad \, Tr(U(1)_R)= -(g-1)Q(6Q + 5),}
\eqn\fli{
Tr(U(1)_{F_i}) = 24(g-1) Q \xi_i \tilde{z}_i ,  \qquad \, Tr(U(1)^3_{F_i}) = 24(g-1) Q \xi^2_i \tilde{z}_i,}
\eqn\riu{
Tr(U(1)_R U(1)^2_{F_i}) = -2 Q (Q+1)(g-1)\xi_i,   \qquad \, Tr(U(1)_{F_i} U(1)^2_R) = -4 Q (Q+1)(g-1)\xi_i \tilde{z}_i}
\eqn\ngi{
Tr(U(1)_{F_j} U(1)^2_{F_i}) = 8 Q(g-1)\xi_i \xi_j \tilde{z}_j,  \qquad \,  Tr(U(1)_{F_k} U(1)_{F_j} U(1)_{F_i}) = Tr(U(1)_R U(1)_{F_j} U(1)_{F_i})=0, }
\eqn\ggy{
Tr(U(1)_R SU(2)^2_L) = - \frac{Q(Q^2-1) (g-1)}{3}, \qquad \,  Tr(U(1)_{F_i} SU(2)^2_L) = Q(Q-1)(g-1) \xi_i \tilde{z}_i,  }
where here we use the $6d$ R-symmetry.

\subsec{Anomalies of a puncture}

We are now interested in the anomalies of a generic Riemann surface with genus $g$ and $s$ punctures. The anomalies without punctures, as we discussed, can be obtained from the anomaly polynomial of the E-string theory by integrating it on the Riemann surface. This means that the anomalies of the genus $g$ Riemann surface are determined by the topology of the Riemann surface including the $U(1)$ fluxes as well as the anomalies of the original 6d E-string theory. However, when we add a number of punctures, the symmetries and the anomalies assigned to the punctures are not fully captured by the topological data. These properties are associated to boundary conditions of the E-string theory around the punctures. It is difficult to study the boundary conditions of the E-string theory directly in six-dimensions. Instead, we can use a circle reduction of the E-string theory for this purpose.

We first elongate the geometry around a puncture as a thin and long tube with a boundary. The puncture corresponds to the boundary condition of the tube. The 4d theory of the deformed Riemann surface remains the same as the 4d theory of the original surface since the theory depends only on the topology of the Riemann surface other than the puncture data. Now the appropriate theory living on the thin long tube is the 5d theory of the E-string theory compactified on a small circle. The puncture data of the 4d theory such as symmetries and anomalies are encoded in the boundary condition of the 5d theory at the tip of the tube, say $x^5 = 0$ where the tube stretches along the $x^5$ direction in 5d.
This 5d theory is a well-known theory \SeibergBD. It is the ${\cal N}=1$ $SU(2)$ gauge theory with $8$ fundamental hypermultiplets, often called 5d E-string theory.
The classical gauge theory preserves $SO(16)$ global symmetry acting on the 8 fundamental hypers and $U(1)$ topological symmetry whose charge is carried by non-perturbative instanton particles. It is expected that this 5d theory at strong coupling uplifts to the 6d E-string theory.

\

\centerline{\figscale{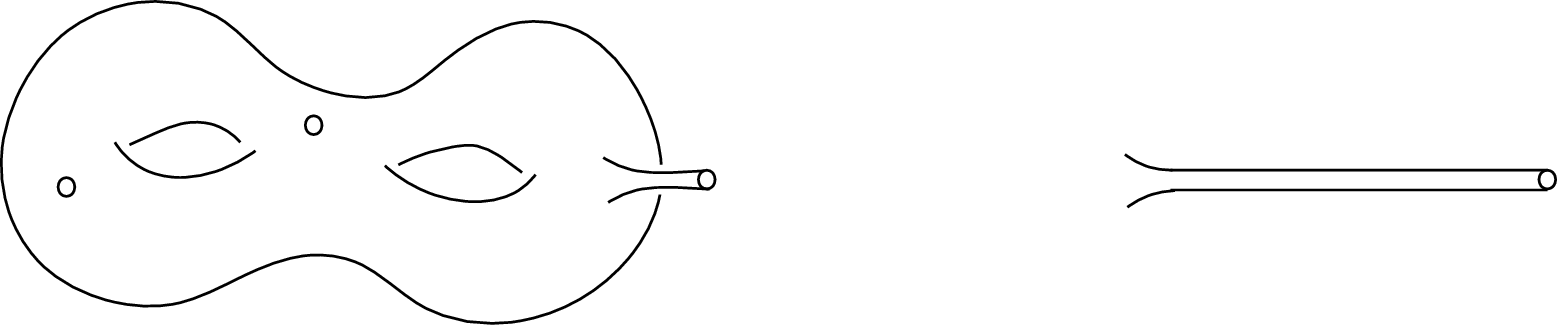}{4in}}
\medskip\centerline{\vbox{
\baselineskip12pt\advance\hsize by -1truein
\noindent\footnotefont{\bf Fig.~2:} Geometry near a puncture in Riemann surface (left) can be deformed as a long thin tube in the right.  Boundary condition at the end of the tube determines type of the puncture.}
}
\

The puncture preserves four supersymmetries, so it is related to a boundary condition preserving four supersymmetries of the 5d ${\cal N}=1$ supersymmetry. There is a simple 1/2 BPS boundary condition. We give Dirichlet boundary conditions to the $SU(2)$ vector multiplet of the 5d theory at the boundary. For the hypermultiplets, we first split them into two sets of eight  chiral multiplets so as to be compatible with 4d boundary ${\cal N}=1$ supersymmetry, and we choose Neumann boundary conditions for one set and Dirichlet boundary conditions for the other. This is the simplest boundary condition preserving four supersymmetries. We will refer to this boundary condition as `{\it maximal} boundary condition' as it maximally preserves the symmetry of the E-string theory with boundaries.

The $E_8$ global symmetry of the 6d E-string theory will be broken to $U(8)$ or $U(1)\times SU(  8    )$ global symmetry because of the splitting of the eight hypermultiplets. The 4d theory involving punctures from this maximal boundary condition is expected to have $U(8)$ global symmetry or its subgroup depending on the bulk topology of the Riemann surface and the fluxes and also other punctures. This classical global symmetry sometimes enhances to a bigger symmetry in special points in the marginal deformations by quantum effects. The bulk $SU(2)$ gauge symmetry  becomes an $SU(2)$ global symmetry due to the Dirichlet boundary condition of the vector multiplet, which lead to an additional $SU(2)$ global symmetry for each puncture.

We claim that the maximal boundary condition of the 5d E-string theory gives rise to the punctures in the 4d theories in the following sections.  We can of course in principle try to construct other 1/2 BPS boundary conditions by coupling some additional 4d ${\cal N}=1$ degrees of freedom to this simplest boundary condition or we may be able to find new 1/2 BPS boundary conditions with same or different global symmetries. More punctures and boundary conditions associated to 6d theories will be studied in a separate paper \JKRV.
We will focus on the maximal boundary condition and associated punctures in this work and will not attempt a classification of the punctures.

We can study many important properties of punctures by using the 5d boundary condition analysis. For example, we have already identified the global symmetries related to the puncture.  We will now compute the 't Hooft anomalies assigned to the puncture. The anomalies of punctures have two distinct contributions. One is the geometric contribution which we can compute by integrating the 6d anomaly polynomial of the E-string theory around the puncture with fluxes. Another contribution comes from the 5d boundary conditions. The 5d fermions with Neumann boundary condition generates anomaly inflows toward the boundary and it induces non-trivial 't Hooft anomalies for the puncture. This can be interpreted as the anomaly inflows from the effective Chern-Simons (CS) term in the $5d$ E-string theory  in the presence of the boundary where the effective CS term is induced by the fermion loops with Neumann boundary condition.  The combination of the geometric contribution and the inflow contribution of the $5d$ boundary condition determines the total anomalies of the puncture.

The geometric contributions to the puncture anomalies from the $6d$ anomaly polynomial depends on the Riemann surface and fluxes. For the two punctured sphere, the full geometric anomalies including two puncture contributions are
\eqn\hfyui{\eqalign{
	& Tr(U(1)_{F_i}) = -12\xi_i z_i \ , \quad Tr(U(1)_{F_i}^3) = -12 \xi_i^2 z_i \ ,  \cr
	& Tr(U(1)_{F_i}U(1)_R^2)=4\xi_iz_i  \ , \quad Tr(U(1)_{F_j} U(1)_{F_j}^2) = -4\xi_i\xi_j z_j \ .}
}
Other anomalies are zero. Here the $U(1)_R$ is the Cartan of the $6d$ $SU(2)_R$ R-symmetry before mixed with other abelian global symmetries. The geometric anomalies of a generic Riemann surface including  $s$ punctures can be easily computed from the anomalies with no punctures given in \sdrgen, \fli, \riu, \ngi, \ggy \ by replacing $g \rightarrow g+s/2$.

Now let us compute the anomaly inflow arising from the $5d$ boundary condition.
First, the $SU(2)$ vector multiplet is in the Dirichlet boundary condition. This kills a chiral half of the gaugino $\lambda$ and leaves an anti-chiral gaugino at the boundary. Namely, the anti-chiral gaugino, i.e. $\gamma^5\lambda=-\lambda$, satisfies Neumann boundary condition. Note that the gauginos are in the adjoint representation of $SU(2)$ and they are subject to the $5d$ symplectic-Majorana condition. Under this condition, this anti-chiral gaugino satisfying Neumann boundary condition with $U(1)_R$ R-charge "+1"  is identified with a chiral gaugino with $U(1)_R$ R-charge $-1$ which contributes to an anomaly inflow toward the 4d boundary. Therefore the anomaly inflow from the vector multiplet in the Dirichlet boundary condition is
\eqn\inflowvvector{
	Tr(U(1)_R^3) = -\frac{3}{2} \ , \quad Tr(U(1)_R) = -\frac{3}{2} \ , \quad Tr(U(1)_R SU(2)^2) = -1 \ . 
}
We remark here that the anomaly inflow induced by an chiral fermion coming from a 5d chiral fermion is half of the anomalies from a $4d$ chiral fermion carrying the exactly same charge \refs{\HoravaQA,\HoravaMA,\GaiottoUNA}.\foot{For example, a single $5d$ hypermultiplet in a segment with a small length $L\ll 1$ becomes a $4d$ chiral multiplet including one chiral fermion when it satisfies Neumann boundary condition at both ends. The anomalies of the $4d$ chiral multiplet can be interpreted as the sum of the $5d$ anomaly inflows toward two boundaries. This means that the anomaly inflow at each $4d$ boundary is half of the anomalies of the $4d$ chiral multiplet since two boundary contributions should be the same.}  Regarding this fact, we have  multiplied by a factor of  $\frac{1}{2}$  in the above anomaly results.

Next, a singlet hypermultiplet in the maximal boundary condition leaves a chiral fermion of $\gamma^5\psi = \psi$. This chiral fermion is a singlet under the $SU(2)_R$ symmetry. The flavor charges of the chiral fermions depend on the choice of chiral half of the scalar fields in the hypermultiplet.
When $a$-th chiral scalar (of eight hypermultiplets) with the $U(1)_{F_i}$ charge $q_{ai}$  satisfies Neumann boundary condition, the anomaly inflow contributions coming from its fermionic partner are
\eqn\inflowwhyper{\eqalign{
	&Tr(U(1)_{F_i}^3) =  \sum_{a=1}^8 q_{ai}^3 \ , \quad Tr(U(1)_{F_i}) = \sum_{i=1}^8 q_{ai} \ , \quad Tr(U(1)_{F_i} SU(2)^2) = \frac{1}{4}\sum_{a=1}^8 q_{ai} \ .
}}
Therefore, the anomaly inflow contribution for a puncture is given by the sum of \inflowvvector\   
 and \inflowwhyper.

\

\

\newsec{Rank one E-string on a torus: $E_8$$\to E_7\times U(1)$ }

We now construct the four dimensional field theories resulting in compactification of rank one E-string on torus with flux preserving $E_7\times U(1)$ subgroup. We will present a more systematic construction going through five dimensions in   section four. Here we will argue for the model obtained in such a compactification directly in four dimensions and will be guided by anomaly and symmetry considerations.

\

\centerline{\figscale{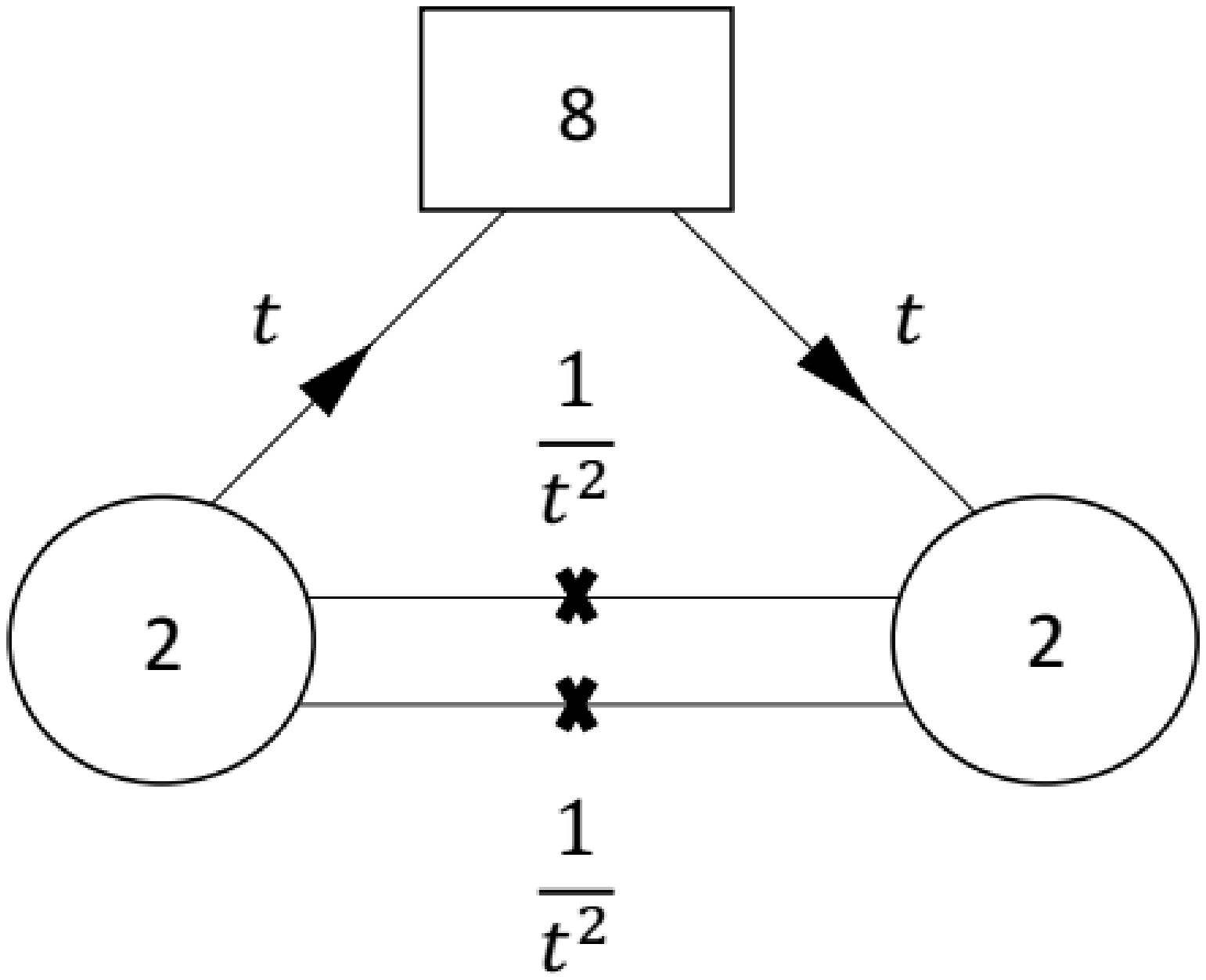}{1.7in}}
\medskip\centerline{\vbox{
\baselineskip12pt\advance\hsize by -1truein
{\noindent\footnotefont{\bf Fig.~3:} The basic theory. Circles represent gauge groups while squares represent global symmetries. There is a cubic superpotential for the two triangles. Also there are two singlets flipping the fields marked with an X. There is a global $U(1)$ whose charges are written using the fugacity $t$. Also all fields have superconformal R-charge $\frac23$. With six dimensional R-symmetry the fields charged under $SU(8)$ have R-charge one, bifundamental of the gauge symmetry have R charge zero, and finally the flip fields have R-charge two.}}} 

\

We expect the theory to have $E_7\times U(1)$ symmetry, in particular all the protected states to fall in $E_7$ representations. There is a natural candidate to be related to such a model, the $E_7$ surprise theory of Dimofte and Gaiotto~\DimoftePD\
 (see \refs{\Rains,\SpiridonovZR} for precursor observations).  This model is two copies of $SU(2)$ SQCD with four flavors with the bilinear gauge invariants of the copies coupled through a quartic superpotential. The Lagrangian of this model shows $SU(  8    )$ symmetry, and by studying supersymmetric spectrum of the model one can argue that it is reasonable that somewhere on the conformal manifold of the IR SCFT the symmetry enhances to (at least) $E_7$.
However, it is easy to check that the anomalies of this theory do not match the anomalies predicted from six dimensions for any simple choice of flux, punctures, and genus.  A small variation of this theory, such that the surprise theory is a relevant deformation of it, has actually all the  needed properties.   A conjecture for the theory with minimal flux for the $U(1)$ is depicted in Fig. 3. It consists of two $SU(2)$ gauge nodes with two copies of bi-fundamental chiral fields and each node has additional eight chiral fields. 
We have a superpotential for each triangle in the quiver and also we have two gauge singlet fields flipping the gauge invariant mesons built from the bifundamentals. As aside comment let us say that this theory, without the flip fields, is related to a $\Z_2$ orbifold of $SU(2)$ ${\cal N}=2$ SYM with four flavors, which also appears as a trinion for two M5 branes probing $\Z_2$ singularity with flux breaking the $SO(7)$ symmetry of that setup to $SO( 5  )\times U(1)$. 
We will actually derive this model from first principles based on compactifictions of E-string theory on a circle.  This derivation will be postponed to the next section.
 
We can compute the anomalies of the model. In particular the superconformal R-charge is the free one. The $a$ and $c$ anomalies are,

\eqn\anoms{
a=2\,,\qquad c=\frac52\,.} These anomalies match precisely the ones predicted for rank one E string with one unit of flux on a torus. At this stage the flux can be either positive or negative, yet, for reasons of concreteness and convenience, we shall associate a flux of $-1$ with this theory. All the other anomalies can be also computed and match six dimensions, where $U(1)_t$ is related to the $6d$ one by a factor of $-\frac{1}{2}$. Thus, under the $6d$ $U(1)$ symmetry the fundamentals are charged minus half, the bifundamentals  one, and the flippers minus two.
 Then for example,

\eqn\anomsd{\eqalign{&
\Tr F R^2 =32(\frac23-1)^2(-\frac12)+8(\frac23-1)^2(1)+2(\frac23-1)^2 (-2)= -\frac43\,,\cr &\qquad \Tr R F^2 =32(\frac23-1)(-\frac12)^2+8(\frac23-1)(1)^2+2(\frac23-1)(-2)^2=-8\,,\cr&
\Tr F^3 = 32(-\frac12)^3+8(1)^3+2\, (-2)^3= -12\,,\qquad \Tr F = 32(-\frac12)+8(1)+2\, (-2)^3= -12\,.
}} 
This theory has manifestly $SU(  8    )\times U(1)$ flavor symmetry. One can compute the dimension of the conformal manifold to be six \RazamatDPL\ and show that the symmetry at a general locus is broken to $U(1)^8$. The $SU(  8    )$ symmetry, in fact $SU(  8    )\times SU(  8    )\times SU(2)\times U(1)$, is recovered at zero coupling. Interestingly, computing the index one can find that the protected spectrum organizes in representations of $E_7\times U(1)$ with the $SU(  8    )\times U(1)$ being the maximal subgroup. For example, one has fundamental fields in ${\bf 8}$ 
and $\bar {\bf 8}$ for the two gauge nodes. The gauge invariants then are in ${\bf 28}$ and $\overline   {\bf 28}$ which combine to form ${\bf 56}$ of $E_7$. The complete index can be formed in $E_7$ characters where it reads\foot{In case the reader is not familiar with suprsymmetric index \refs{\RomelsbergerEG,\KinneyEJ} nomenclature we recommend \DolanQI\ for beautiful exposition and \RastelliTBZ\ for a review, and we will use the notations of the latter.}

\eqn\indexEseven{\eqalign{
I_{E_7} &= 1 + (p q)^{\frac{2}{3}} (\frac{3}{t^4}  + t^2\chi[{\bf 56}]) - 2 p q + (p q)^{\frac{2}{3}}(p+q)(\frac{2}{t^4} + t^2 \chi[{\bf 56}]) \cr &+ (p q)^{\frac{4}{3}} (\frac{6}{t^8} + \frac{1}{t^2} \chi[{\bf 56}] + z^4 (\chi[{\bf 1463 }]-\chi[{\bf 133}]-1)) + ...}} 
Here we have ignored the singlets as these are just free fields.

So we  have seen that many things are consistent with the $6d$ interpretation. It is thus natural to conjecture that there is a locus on the conformal manifold where the symmetry enhances to $E_7\times U(1)$ or larger. There is one problem with this conjecture. The index at order $p q$ is given by $-2$. At this order the index captures the number of marginal deformations minus the conserved currents \BeemYN. Assuming that somewhere the symmetry enhances to $E_7$ we thus should write ${\bf 133}-{\bf 133}-1-1=-2$. The $-{\bf 133}-1$ is the conserved current of $E_7\times u (1)$. The additional $-1$ is to be interpreted as a conserved current of an accidental $U(1)$ at that point, and ${\bf 133}$ is the marginal deformation. However, this implies that the dimension of the conformal manifold is the number of independent invariants \GreenDA\  of the adjoint representation of $E_7$ which is seven. This does not agree with the computation at the free point. We thus deduce that although the index  (and one can check other partition functions) are consistent with $SU(  8    )$ symmetry enhancing  to  $E_7$, there is no point on the conformal manifold where this actually happens. From six dimensional point of view if the theory is to be associated with the compactification of E-string preserving $E_7\times U(1)$ symmetry, this implies that there is a holonomy breaking the $E_7$ which cannot be turned to zero. Note that the naive dimension of the conformal manifold is nine, one for complex structure and eight for holonomies, however as it is usual for the torus with no punctures and low value of flux, the actual conformal manifold is different. 

The $E_7$ symmetry can be obtained if we give a vacuum expectation value to flipper fields which will provide  a mass  to the bifundamental chirals. One obtains the $E_7$ surprise of \DimoftePD.   Such a vacuum expectation value deformation breaks the $U(1)$ symmetry and might have the effect of switching off the holonomies breaking $E_7$.  Although we do not have a point with $E_7$ symmetry for the  model discussed here, we have seen that the assertion that this theory corresponds to $E_7$ compactifications is consistent with numerous non trivial computations.   In fact, given the derivation we present for this theory in the next section, this demystifies the $E_7$ surprise.

As we have mentioned the flux here is the one breaking $E_8$ to $U(1) \times E_7$. We identify this $U(1)$ with the $U(1)_t$ we introduced previously, and associate with this theory the flux $(-1; 0 , 0 , 0 , 0 , 0 , 0 , 0 , 0 )$, where we note that $z=1$ correspond to a flux of $-1$ on the torus. In terms of the complete basis the flux associated is: $(-1, -1 , -1 , -1 , -1 , -1 , -1 , -1 )$.

\

This construction has a generalization. Consider the quiver diagram of Fig. 4. This is a triangulation of a circle with $2z$ triangles (with $z=2$ for Fig. 4. (a)). We again add extra singlet fields.\foot{We comment again that up to the flip fields this model is related to $\Z_{2z}$ orbifold of $SU(2)$ ${\cal N}=2 $ SYM with four flavors.} The anomalies are given by, 

\eqn\anomss{
a=2 z\,,\qquad c=\frac52 z\,.} Which matches the six dimensions with flux $\mp z$, and here we will make the choice to associate for concreteness this model with $-z$. The supersymmetric states  again fall in representations of $E_7\times U(1)$.

\ 

\centerline{\figscale{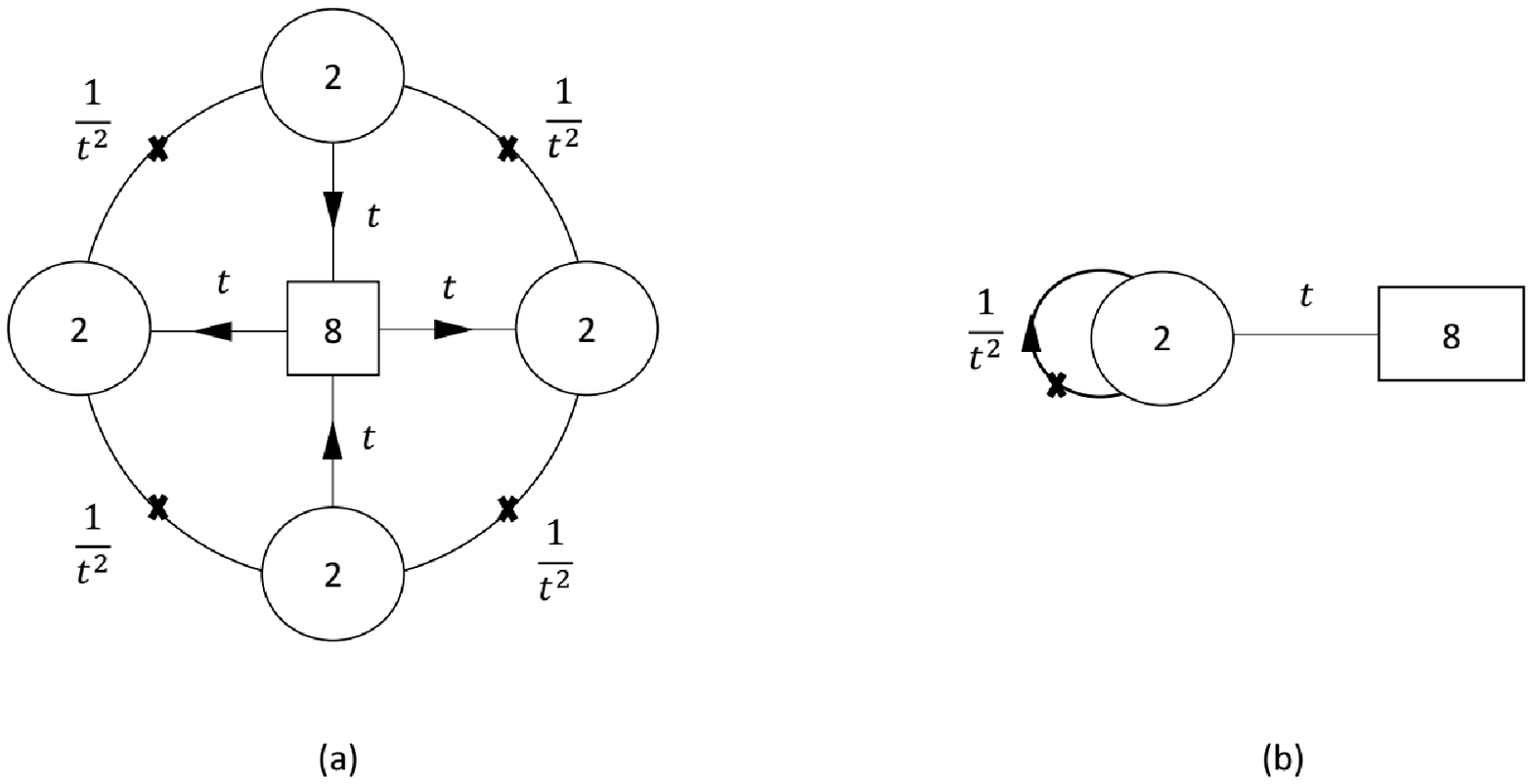}{3.7in}}
\medskip\centerline{\vbox{
\baselineskip12pt\advance\hsize by -1truein
\noindent\footnotefont{\bf Fig.~4:} (a) Theory with two units of flux. (b) Theory with half a unit of flux. Here the $SU(  8    )$ is broken to $SO(  8    )$ by the superpotential. Note that the line from the $SU(2)$ to itself stands for an adjoint plus a singlet.}}

\

Note that with odd number of triangles the group is broken from $SU(  8    )$ to $SO(  8    )$. In particular one triangle is just the ${\cal N}=2$ case with a flip. The flux here is $-\frac{1}{2}$. It will be interesting to study some aspects of these theories. Let's first consider the ${\cal N}=2$  (with the flip field) case shown in figure 4 (b).  Due to the fractional flux the compactification involves also a center flux breaking $E_7$ which explains the breaking of the $SU(  8    )$ in field theory.
The remaining global symmetry depends on the choice of holonomies used to generate the flux. Arbitrary holonomies are expected to break $E_7$ down to $U(1)^4$ which for special choices will be enhanced to various groups, the largest of which being $F_4$. Turning on holonomies on a Riemann surface are usually mapped to marginal operators in the $4d$ theory. Thus we expect there to be a conformal manifold on special points of which the symmetry enhances to various groups including $F_4$. 

We can try and test this using the superconformal index. Evaluating the index we find:

 \eqn\indexEsevenhalf{\eqalign{
I_{E_7} &= 1 + (p q)^{\frac{2}{3}} (\frac{1}{t^4} + t^2\chi[{\bf 28}]) - \chi[{\bf 28}] p q - \frac{(p q)^{\frac{1}{3}}(p+q)}{t^2} + ...\; ,}}
where we have again ignored the singlet fields. It was noted that this index in fact forms characters of $F_4$   \deBult. This works as one can reinterpret the ${\bf 28}$ of $SO(  8    )$ as the ${\bf 26} + 1 + 1$ of $F_4$. However, there are several problems with this. First the ${\bf 28}$ contribute negatively to the $p q$ order. This fits with the conserved current of $SO(  8    )$, but not with an $F_4$ interpretation as the ${\bf 26}$ is not the adjoint of $F_4$. Another issue is that the only marginal operator here is the $SU(2)$ gauge coupling which does not break $SO(  8    )$.
Therefore, this case bares similarities to the  case of minimal integer flux. Particularly, we have some expectations for symmetry enhancement on the conformal manifold. These expectations are supported by the index forming characters of the desired symmetry. However, the enhanced symmetry point does not exist. In both cases we note that the conformal manifold is smaller than predicted from $6d$. This can be explained by postulating that there is some holonomy in these cases that we cannot turn off. This then may also explain why certain symmetries are not realized in $4d$ despite the $6d$ expectations. 

Finally we remark a bit on the general case. Again we can formulate the same expectations where for integer $z$ we expect a point with $E_7$ symmetry while $z$ half integer an $F_4$ point is expected. We can again test this by evaluating the superconformal index. Ignoring the singlets, we find:

\eqn\indexEsevenhalf{\eqalign{
I_{E_7} &= 1 + (p q)^{\frac{2}{3}} (\frac{2 z}{t^4} +z t^2(\chi[{\bf 28}] + \chi[{\bf \bar{28}}])) + ... ,}}
where we note that the $p q$ order vanishes. When $z<2$ then there are additional terms owing to the existence of extra marginal operators or symmetries at the free point. 

We now note several observations regarding the index. For $z$ integer it forms characters of $E_7$, at least to the order we evaluated it. Assuming there is a point with $E_7$ global symmetry, we expect there to be an $8$ dimensional conformal manifold on a generic point of which the symmetry is broken to $U(1)^8$. Now there is no contradiction with the free point.

\centerline{\figscale{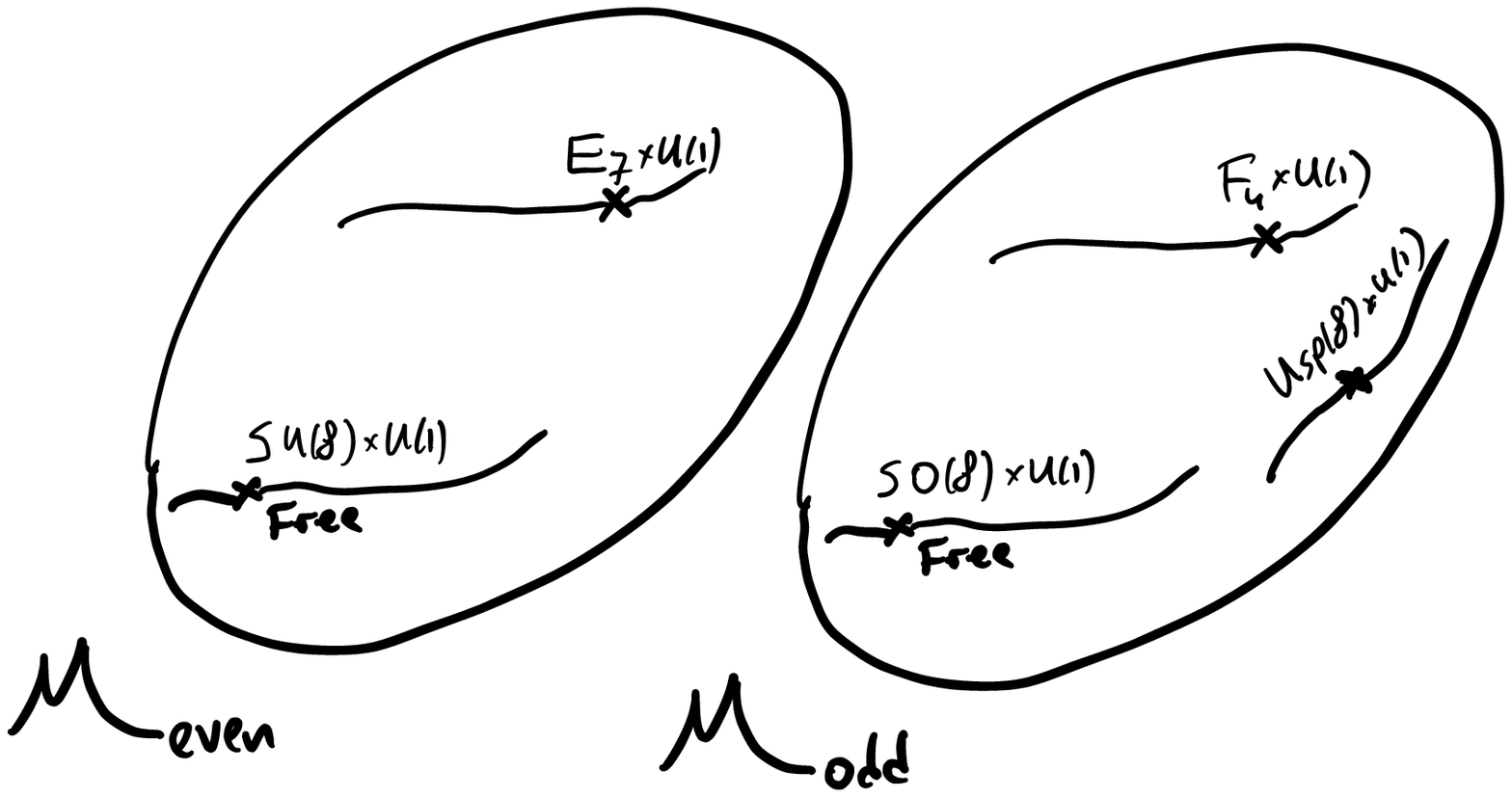}{2.9in}}
\medskip\centerline{\vbox{
\baselineskip12pt\advance\hsize by -1truein
\noindent\footnotefont{\bf Fig.~5:} On the left we have a drawing of the conformal manifold of the model with  integer flux. We have $U(1)^8$ symmetry on a general point of the conformal  manifold. The symmetry enhances to $SU(  8    )\times  U(1)$ on a line passing through free point and we conjecture that there is another line on which the symmetry is $E_7\times U(1)$ passing through strong coupling. For half integer flux we have a line with $SO(  8    )\times U(1)$ symmetry passing through the free point and $U(1)^5$ symmetry at a general point. We conjecture that there exist additional lines on which symmetry enhances with the maximal enhancement being $F_4\times U(1)$.}}

\

\

For $z$ half integer the index forms characters of $SO(  8    )$, but these can be reinterpreted as characters of $F_4$, $USp(8)$ and a variety of other symmetries. As the $p q$ order vanishes, there is no contradiction with interpreting them as global symmetries. Assuming such points exist, we expect there to be a $5$ dimensional conformal manifold on a generic point of which the symmetry is broken to $U(1)^5$. We see no contradiction with this from the free point. This structure is consistent with what we expect from $6d$.

It is illuminating to also consider the index including the singlets:

\eqn\indexEsevenhalf{\eqalign{
I_{E_7} &= 1 + 2 z t^4 (p q)^{\frac{1}{3}} + (p q)^{\frac{2}{3}} (z (2z+1) t^8 + z t^2(\chi[{\bf 28}] + \chi[{\bf \bar{28}}])) \cr &+ 2 z t^4 (p q)^{\frac{1}{3}}(p + q) + p q (\frac{2}{3} t^{12} z (z+1)(2z+1)+2 z^2 t^6(\chi[{\bf 28}] + \chi[{\bf \bar{28}}]))) + ... ,}}
where we again assume $z\geq 2$. The interesting thing here is that we can identify some of the contributions as coming from the $6d$ conserved current multiplet of the $E_8$ global symmetry. Particularly, the contributions $2 z t^4 (p q)^{\frac{1}{3}}$ and $z t^2(\chi[{\bf 28}] + \chi[{\bf \bar{28}}])$ have the same R-charge as marginal operators under the $6d$ R-symmetry. Furthermore the representations they carry exactly match those required to complete $E_7$ to $E_8$ (see the branching rule in Appendix A).

Another interesting thing is that the number of such operators is exactly as expected from the reasoning of \refs{\Brz} (see Appendix E for brief summary). The marginal operators expected from holonomies also work as we expect, $g-1=0$ ones in the adjoint of $E_7\times U(1)$. This is especially interesting as such simple 
reasonings are known to be unreliable for the torus. We also expect one more marginal deformation related to the complex structure moduli of the torus, which is absent. Absence of such exactly marginal deformation appears already in the case with no flux, that is the MN $E_8$ theory.  

\

\subsec{Sphere with two punctures and gaugings}

The theories corresponding to the torus can be constructed in a rather natural way by gluing together theories we would correspond to a sphere with two maximal punctures, which have $SU(2)$ symmetry in our case, and flux value of $-\frac{1}{2}$. The field theory is drawn in Fig. 6. This is a Wess-Zumino model of a collection of chiral fields. In terms of the flux basis it is associated with $(-\frac{1}{2}; 0 , 0 , 0 , 0 , 0 , 0 , 0 , 0 )$ in the overcomplete basis and $(-\frac{1}{2}, -\frac{1}{2} , -\frac{1}{2} , -\frac{1}{2} , -\frac{1}{2} , -\frac{1}{2} , -\frac{1}{2} , -\frac{1}{2} )$ in the complete basis. 

\

\centerline{\figscale{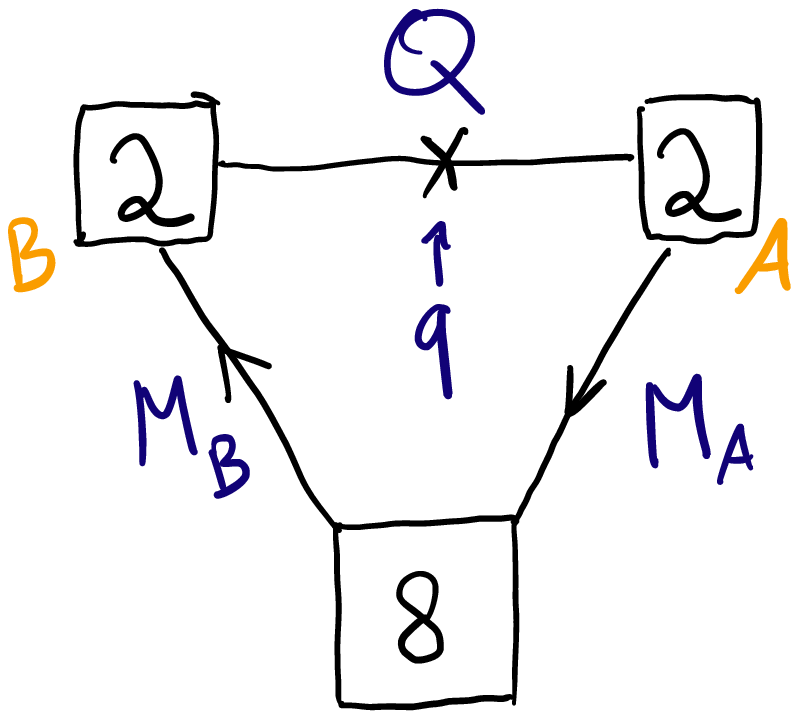}{1.7in}}
\medskip\centerline{\vbox{
\baselineskip12pt\advance\hsize by -1truein
\noindent\footnotefont{\bf Fig.~6:} Sphere with two maximal punctures and half a unit of flux. The six dimensional R-charge of the fields $M_A$ and $M_B$ is one, of bifundamentals is zero, and the flip fields have R-charge two.} }

\

Note that each $SU(2)$ flavor symmetry, that we associate to a puncture, has an operator in the fundamental of $SU(2)$ and ${\bf 8}$ or $\bar {\bf 8}$ of the $SU(  8    )$. We denote these operators, which are fields in this theory, by $M$. We think of the punctures as having a color label depending on the embedding of $SU(  8    )$ in $E_8$. Here we have fixed $E_7$ in $E_8$ and have two choices,  depending on what representation of $SU(  8    )$ $M$ is in, and we denote the choices by plus and minus. 
When we glue punctures of color $+$ we introduce a bifundamental field of $SU(2)\times SU(  8    )$ in $\bar{ \bf 8}$, call it $\Phi$, and couple it through the superpotential $W=M_A\Phi -M_B\Phi$. We then gauge the $SU(2)$ symmetry. The $-$ color is glued in a similar manner. Note that the chiral fields also have $U(1)$ charge, with $M$ charged one and the bifundamentals of two $SU(2)$ groups minus two. Combining two theories we can obtain a sphere with two maximal punctures of same color and one unit of flux, and we depict it in Fig. 7.  We can next glue two maximal punctures together to get our torus theory.

\

\centerline{\figscale{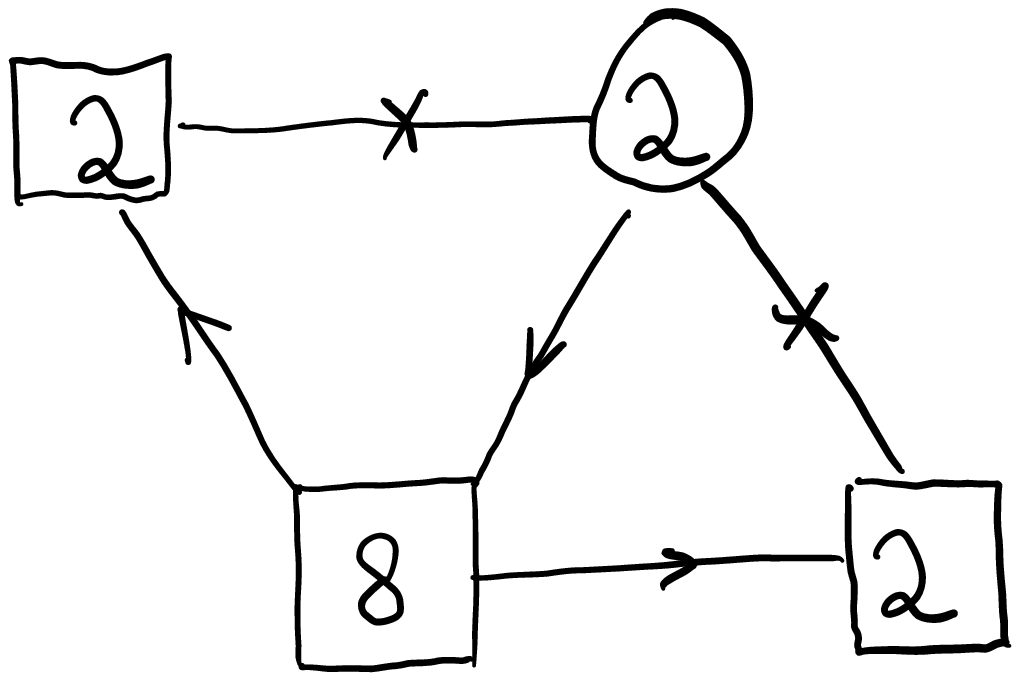}{1.7in}}
\medskip\centerline{\vbox{
\baselineskip12pt\advance\hsize by -1truein
\centerline{\noindent\footnotefont{\bf Fig.~7:} Sphere with two punctures and one unit of flux.}} }

\

The anomalies of the Wess-Zumino model discussed in this section  match (see equation (4.13) in the next section) the anomalies computed from six dimensions for theory with half a unit of flux on a sphere with two punctures. In particular,

\eqn\fueyuoyo{\eqalign{
& Tr R =  4(-1)+(2-1)=-3\,,\qquad Tr R^3 = 4(-1)^3+ (2-1)^3=-3\,,    
\cr & Tr U(1) R^2 = 4 (1) (-1)^2+(-2)(2-1)^2=2\,, Tr U(1) = 16\times 2 \times (-\frac12) +4 (1) + (-2)=-14\,, \qquad\, \cr & Tr U(1)^3=16\times 2\times(-\frac12)^3+4(1)^3+(-2)^3=-8\,,}
} 
where the $U(1)$ is normalized such that the bifundamental has charge $1$.

We also have an analogue of S-gluing of \refs{\GaiottoUSA,\RazamatDPL}. We note first that if we conjugate the representations under all the symmetries we will get a theory which we will associate to a compactification with opposite values of the fluxes. In particular we will assign an additional label, call this sign in analogy to \GaiottoUSA, to punctures depending on the charge of the $M$ operators under the $U(1)$.  Consider gluing together two theories along punctures of  opposite signs. As when we change the sign we conjugate all representations, the gluing is obtained without additional fields by gauging $SU(2)$ and adding the supepotential $W=M_AM_B$.  As the operators $M$ here are fields the superpotential gives them mass and they disappear in the IR. The gauge group is then $SU(2)$ with four chirals. In particular the dynamics in the IR identifies the  flavor group with the gauge group connected to the group we gauge now, that is Higgsing it due to the deformed quantum  moduli space. We then obtain a theory with half a unit less flux and with maximal puncture with opposite sign. See Fig 8.

\centerline{\figscale{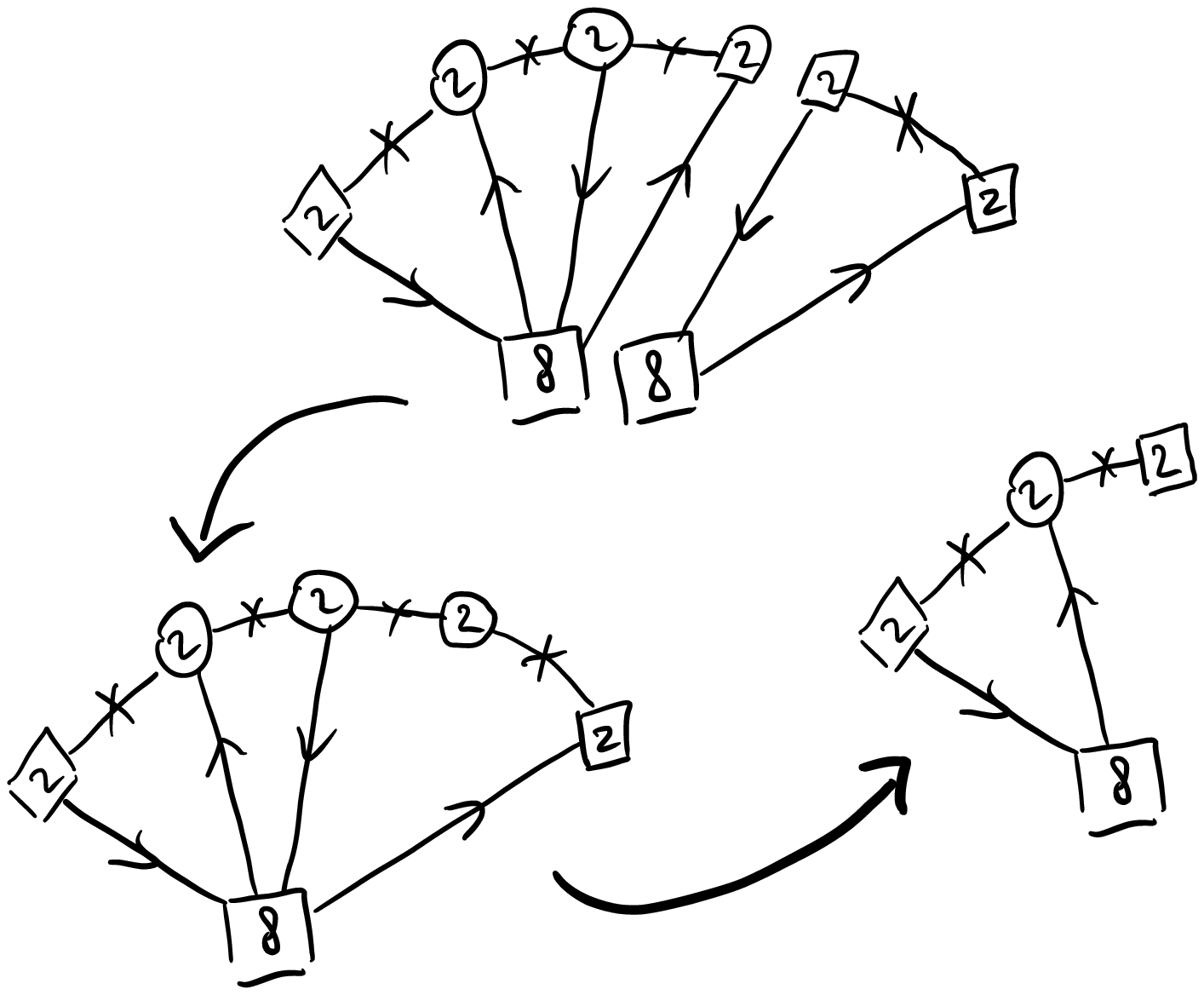}{2.7in}}
\medskip\centerline{\vbox{
\baselineskip12pt\advance\hsize by -1truein
\noindent\footnotefont{\bf Fig.~8:} S gluing of two punctures. The final theory is sphere with half unit of flux less than the original one and two punctures of opposite sign. Recall that the sign of puncture can be changed by flipping the ``moment map'' operator charged under it. In this case this operator is field $M$ and thus it becomes massive. }}

\

\

\newsec{Five dimensions, domain walls, and tube models}

In this section, we will derive our models of two punctured spheres (or tubes) from the $5d$ E-string theory and domain walls in it. We will first review  boundary conditions and the duality domain walls in 5d gauge theories studied in \GaiottoUNA. It turns out that the duality domain wall induces nonzero flux for a $U(1)$ global symmetry associated to the duality in the $5d$ theory.  The domain wall studied in \GaiottoUNA \ connects two different gauge theories that are dual to each other by a Weyl reflection in an $SU(2)$ subgroup of the $5d$ global symmetry, which flips the sign of the mass parameter of the $U(1)\subset SU(2)$.  This domain wall was called the duality domain wall in \GaiottoUNA\ because in that context the $SU(2)$ global symmetry was part of the emergent duality symmetry group.  In our context the $U(1)$ will be part of the $E_8$ global symmetry of the E-string theory.  We shall see that our tube models with fluxes in this paper can be interpreted as particular concatenations of these domain walls with suitable boundary conditions at both ends of the $5d$ E-string theory on a segment.

\subsec{Domain walls and fluxes}
We start with reviewing the construction of the 5d duality domain wall of \GaiottoUNA. Let us consider a domain wall inserted at $x^5=0$ in a $5d$ $SU(2)_G$ gauge symmetry with some number of fundamental hypermultiplets and a flavor symmetry which includes a $U(1)_F$.
The domain wall splits the $5d$ theory into the left and the right chambers. Each chamber now has its own $SU(2)_G$ gauge group.
We choose Neumann boundary condition for the $SU(2)_G$ gauge multiplets on both sides of the wall, which preserves half of the $5d$ ${\cal N}=1$ supersymmetries. This boundary condition introduces a $4d$ bi-fundamental chiral multiplet, say $q$, stuck at the $4d$ domain wall \foot{Bi-fundamental fields of the same kind appear  in systems with multiple D-branes divided by NS5-branes.}. For the fundamental hypermultiplets, we choose the maximal boundary condition we discussed in section 2.6. Let us call a chiral half of the hypermultiplets as $X_i$ and another chiral half as $Y_i$ in one chamber.   Here $i$ is a flavor index for the hypermultiplets.  Then we will choose Neumann boundary condition for $X_i$ in the left chamber and also for $Y'_i$ in the right chamber. So $Y_i$ and $X'_i$ are given Dirichlet boundary condition. The `duality' domain wall of \GaiottoUNA
\ leads to the following $4d$ superpotential coupling at the interface:
\eqn\dompo{
	{\cal W}|_{x^5=0} = b\,{\rm det} q + \sum_i Y'_iqX_i \ .
}
Here we added an extra $4d$ chiral multiplet  $b$ which is neutral under the bulk gauge symmetries. This singlet field $b$ will be later identified with the flipping fields in the $4d$ tube models.

This system has two anomaly free $U(1)$ global symmetries and we can choose a basis for them such that the $4d$ fields $b$ and $q$ transform only by one $U(1)$, which we will call $U(1)_F$.  The fundamental chiral fields $X_i$ and $Y_i'$ carry $-\frac{1}{2}$ charge and the $4d$ fields $b$ and $q$ carry $-2$ and $+1$ charges, respectively, under this $U(1)_F$ symmetry. Furthermore, 5d bulk topological symmetries of the instanton number current $J_I=\frac{1}{8\pi^2}Tr(F\wedge F)$ mix with this symmetry. A neutral instanton with instanton number `$+1$' in the left chamber carries $U(1)_F$  charge $-1+\frac{N_f}{8}$ and that in the right chamber has the  $U(1)_F$  charge $1-\frac{N_f}{8}$.  We note that the $U(1)_F$ symmetry does not mix with the topological symmetries when $N_f=8$.

This domain wall flips the sign of $U(1)_F$ charges and the sign of the corresponding mass parameter \GaiottoUNA. The chiral fields $X$ and $X'$ are both in the fundamental representation of $SU(  N_f   )$, but they carry opposite $U(1)_F$ charges. This means that the $U(1)_F$ charge on the left chamber flips its sign on the right chamber after crossing the domain wall.
 Accordingly, the mass parameter of $U(1)_F$ changes along the $x^5$ coordinate from $-m|_{x^5\rightarrow -\infty}$ to $m|_{x^5\rightarrow +\infty}$. 

We now consider this domain wall in the context of E-string theory compactified to $5$ dimensions.  Classically, this system has $SU(  8    )\times U(1)_F\times U(1)_I$ symmetry where $SU(  8    )\times U(1)_F$ is a subgroup of $E_8$ symmetry of the $6d$ E-string theory. 
The chiral multiplets $X_i$ and $X'_i$ are fundamentals of $SU(  8    )$.
The $U(1)_F$ is the Cartan of $SU(2)_F$ in $SU(2)_F\times E_7\subset E_8$. We will use this $U(1)_F$ in the construction of the domain wall.    This coincides with the $U(1)_F$ discussed above \GaiottoUNA .  
 $U(1)_I$ is another gauge anomaly free $U(1)$ which acts on the $5d$ instanton particles.  This $U(1)_I$ is associated to the Kaluza-Klein momentum of the $6d$ E-string theory. Since the $6d$ Kaluza-Klein states are truncated in the $4d$ limit, the $4d$ theories cannot see this $U(1)_I$.
Since $U(1)_F$ has no mixing with the $U(1)_I$ symmetry in the E-string theory, the domain wall action on $U(1)_F$ is independent of the $6d$ Kaluza-Klein momentum. Therefore the $4d$ limit is well-defined in the presence of the domain walls.

Recall that $U(1)_F$ is also a symmetry of the $6d$ theory, and in this context we now explain why the mass flipping  by the domain wall is related to $6d$ $U(1)_F$ flux.\foot{A similar construction relating $6d$ flux and $5d$ domain walls has been studied in \ChanQC .}
Suppose that the $U(1)_F$ flux is localized along a circle in  the location of the domain wall as drawn in the right-handed side of Fig. 9. The flux acts on $6d$ fermions $\Psi$ as
\eqn\ferflux{
	\Psi_R = g\, \Psi_L \ , \quad g= e^{2\pi i Q F_{56}\Gamma^{56}} \ ,
}
where $Q$ is the $U(1)_F$ charge of the fermion $\Psi_L$ and $F_{56}$ is the flux on the tube. $\Gamma^{56}$ is a $6d$ gamma matrix bi-linear and it reduces to $\gamma^5$ in the $5d$ reduction along the $x^6$ direction. The mass sign appears in front of the fermion mass terms in the Lagrangian of the $5d$ E-string theory. In the presence of the $6d$ $U(1)$ flux acting as (4.2), the $5d$ fermion mass term on the right-handed side of  the flux wall becomes
\eqn\fermass{
	m \bar\Psi_R \Psi_R= m \Psi_L^\dagger  g^\dagger\gamma^0g\Psi_L = m\bar\Psi_L e^{4\pi i QF_{56}\gamma^5}\Psi_L \ .
}
Here we have assumed that the fermions in the $5d$ E-string theory transforms in the same manner as the 6d fermions under $U(1)$ flux.

The mass sign will be flipped by the flux when $|QF_{56}|=\frac{1}{4}$.  In our domain wall construction, the fermion fields $\Psi_i$ carry the $U(1)_F$ charge $-\frac{1}{2}$. This means that an half-unit flux, i.e. $F_{56} = -\frac{1}{2}$, changes the sign of the mass parameter for this $U(1)_F$.  Therefore, we can say that the $5d$ domain wall flipping the mass of $U(1)_F$ corresponds to half a unit flux of $U(1)_F$ of the E-string theory in $6d$.
 In other words we have learned that
{\it the $5d$ `duality domain wall' of \GaiottoUNA\ is the same as half a unit of $U(1)_F$ flux from the $6d$ perspective}.

\

\centerline{\figscale{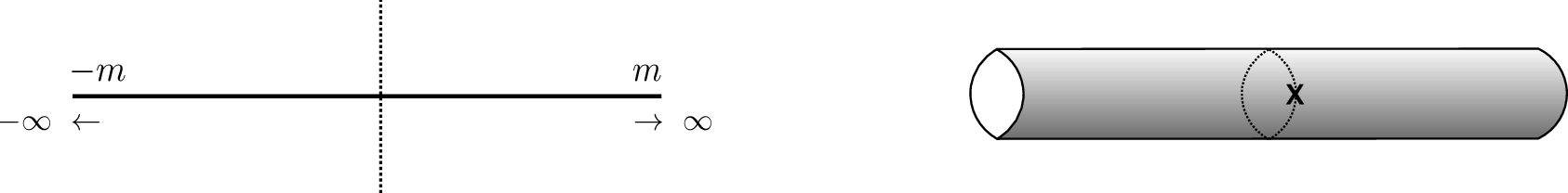}{5in}}
\medskip\centerline{\vbox{
\baselineskip12pt\advance\hsize by -1truein
\noindent\footnotefont{\bf Fig.~9:} Duality domain wall (dotted line) flips mass sign of  $U(1)_F$ symmetry. $5d$ E-string theory with duality wall corresponds to $6d$ E-string theory on a tube with half a unit $U(1)_F$ flux.}} 

\

\subsec{$4d$ models from tubes}
We can derive all $4d$ models of tubes with fluxes directly from the $5d$ E-string theory with domain walls on a segment with finite length $L$. According to our earlier discussion in section 2.6, a puncture with $SU(2)$ global symmetry at one end of the segment is defined by the maximal boundary condition. Following this, we set the maximal boundary condition on both ends of the segment. In low energy less than $L^{-1}$,  this leads to a new $4d$ theory corresponding to the E-string theory on a tube with fluxes.

Let us first consider the case with a single domain wall. There are two vector multiplets on the left and the right chambers. They satisfy Neumann boundary condition at the domain wall. However, they are given Dirichlet boundary condition at the other ends of the chambers. This implies that the vector multiplets in both chambers are truncated  in the $4d$ limit and thus two $SU(2)_G$ gauge symmetries simply become $4d$ global symmetries $SU(2)\times SU(2)$. For the hypermultiplets, the  maximal boundary condition sets Neumann boundary condition on $X$ and $Y'$ and Dirichlet boundary condition on $Y$ and $X'$. The fields $X$ and $Y'$ satisfy Neumann boundary condition on both ends of the segment, thus they become $4d$ chiral multiplets. Therefore, the E-string theory with a single duality domain wall on a segment with the maximal boundary condition gives rise to a $4d$ Lagrangian theory with the chiral multiplets $X$, $Y'$ coming from the $5d$ theory coupled to the additional $4d$ fields $q,b$ through the superpotential \dompo.
This $4d$ theory is precisely our $E_7$ model of a tube with half a unit $U(1)_F$ flux in Fig. 6. We here derived the $E_7$ model of the E-string theory on a tube directly using its $5d$ reduction dressed by the duality domain wall. We note that the $U(1)_F$ flux of this theory is $-\frac{1}{2}$ which precisely agrees with our claim that $U(1)_F$ flux introduced by a single domain wall is `one-half'.

We can also consider more complicated configurations with multiple domain walls on the segment. We expect that domain wall configurations giving different fluxes lead to different $4d$ Lagrangian theories. Let us now study how to connect two or more domain walls basically following the discussions in \GaiottoUNA.

Suppose that we attach two domain walls and  the first domain wall turns on half a unit of $U(1)_F$ flux. Two domain walls divide the $5d$ theory into three chambers. For the second domain wall, we have many different choices of $U(1)$ flux. Let us first focus on the second flux $\pm\frac{1}{2}$ for the same $U(1)_F$ symmetry.

Remember that the $U(1)_F$ flux is correlated to the choice of the $U(1)$ charges rotating the $4d$ fields $q'$ and $b'$ in the second domain wall as well as the $5d$ fields with Neumann boundary condition in the second and the third chambers.
If we turn on the second flux $-\frac{1}{2}$, since this flux is in the same $U(1)_F$ as the first flux, the $4d$ fields $q$ and $b$ at the first domain wall and $q'$ and $b'$ at the second domain wall should carry the same $U(1)$ charges. Accordingly, a chiral half $Y'$ of the hypermultiplets in the second chamber and another chiral half $X''$ in the third chamber should obey Neumann boundary condition. The superpotential of this domain wall system is given by
\eqn\multidoms{
	{\cal W}_{4d} = {\cal W}_{x^5=t_1} + {\cal W}_{x^5=t_2} \ , \quad {\cal W}_{x^5=t_1} = b \, {\rm det}q + {\rm Tr}\,Y'qX \ , \quad {\cal W}_{x^5=t_2} = b' \, {\rm det}q' + {\rm Tr}\, Y'q'X'' \ ,
}
where $t_1,t_2$ are locations of two domain walls. 
 The net $U(1)_F$ flux of this system along the tube becomes $-\frac{1}{2}-\frac{1}{2}=-1$. Similarly, we can concatenate a number of duality domain walls using the same $U(1)_F$ flux and construct a system of the E-string on a tube with generic flux. The net flux will become $n$ when the number of domain walls is $2n$.
 We can put this $5d$ system on a finite segment and give the maximal boundary conditions at both ends. In the $4d$ limit, this will yield the $4d$ $E_7$ model of two punctured sphere with $n$ flux. For example, $n=1$ case leads to the $4d$ $E_7$ model in Fig. 7.

On the other hand, if we choose $\frac{1}{2}$ as the flux in the second domain wall, the $4d$ fields in the first domain wall and those in the second domain wall will have opposite $U(1)$ charges.
In this case, the chiral fields in Neumann boundary condition are $X$ and $Y'$ at $x^5=t_1$, and $X'$ and $Y''$ at $x^5=t_2$, and other chiral fields satisfy Dirichlet boundary condition. Since $X'$ and $Y'$ have opposite boundary conditions at both ends in the second chamber, all the chiral fields in the second chamber become massive and we can integrate them out. Integrating out these massive fields leaves a quartic superpotential in $4d$ limit as follows  \GaiottoUNA \ :
\eqn\nuldoms{
	{\cal W}_{4s} = b \, {\rm det}q+b' \, {\rm det}q' + Y''q'qX \ .
}
One can now show that this system reduces to an `empty' domain wall using Seiberg duality on the $SU(2)$ gauge theory in the second chamber. Namely, a combination of two duality domain walls with opposite $U(1)$ fluxes is equivalent to a system with no domain wall \GaiottoUNA, which is consistent with $-\frac{1}{2}+\frac{1}{2}=0$ flux. For this property of the duality domain wall, the $4d$ singlet field $b$, which is called `flipping field', is necessary.

From these examples we find a simple algorithm to construct domain wall configurations giving generic $U(1)$ fluxes. For this it is convenient to use fluxes in the complete basis as each element there corresponds to the flux felt by a single $5d$ hypermultiplet. Then ,for a given flux of a single $U(1)$ symmetry, we first decompose it into  a combination of half unit fluxes. For instance, consider the flux vector $(0,0,0,0,-\!1,-\!1,-\!1,-\!1)$ which describes an $SO(14)$ preserving flux of strength $-1$. We can construct it using a half-unit $U(1)$ flux as
\eqn\uflxso{
	 {\textstyle (0,0,0,0,-\!1,-\!1,-\!1,-\!1) = (-\!\frac{1}{2},-\!\frac{1}{2},-\!\frac{1}{2},-\!\frac{1}{2},-\!\frac{1}{2},-\!\frac{1}{2},-\!\frac{1}{2},-\!\frac{1}{2}) +(\frac{1}{2},\frac{1}{2},\frac{1}{2},\frac{1}{2},-\!\frac{1}{2},-\!\frac{1}{2},-\!\frac{1}{2},-\!\frac{1}{2}) \ .}
}

Next, we introduce a duality domain wall for each half a unit flux and connect all of them together with suitable  boundary conditions. Boundary condition at the domain wall depends on the flux associated to the domain wall. Each element in the vector of the half-unit flux determines the boundary condition of the corresponding $5d$ hypermultiplet. When $i$-th element is $-\frac{1}{2}$ (or $+\frac{1}{2}$), the chiral field $X_i$ (or $Y_i$) obeys Neumann boundary condition and thus can couple to the $4d$ degrees of freedom in the domain wall. When two domain walls are glued, a chiral field satisfying Neumann boundary condition at both domain walls becomes a $4d$ chiral field. This $4d$ chiral field couples to chiral fields in the adjacent chambers as well as the $4d$ fields $q$ and $b$ through the superpotential \dompo. On the other hand, when the boundary conditions at the two boundaries are different, the corresponding hypermultiplet is truncated and it will generate a quartic superpotential, like the last term in \nuldoms, connecting chiral fields in the two adjacent chambers. This will determine the boundary condition and the superpotentials for the hypermultiplets in the middle chamber. We can repeat these procedures for each domain wall and fields in the corresponding chambers.
We remark here that the number of different elements between two flux vectors of two adjacent domain walls should be even. Otherwise there will be $\Z_2$ gauge anomaly for the $SU(2)$ gauge symmetry in the middle chamber. This is manifested in the $6d$ side by flux consistency. Recall that fluxes in the complete basis are vectors on the $E_8$ root lattice. Then this follows from the structure of the $E_8$ root lattice. 

\centerline{\figscale{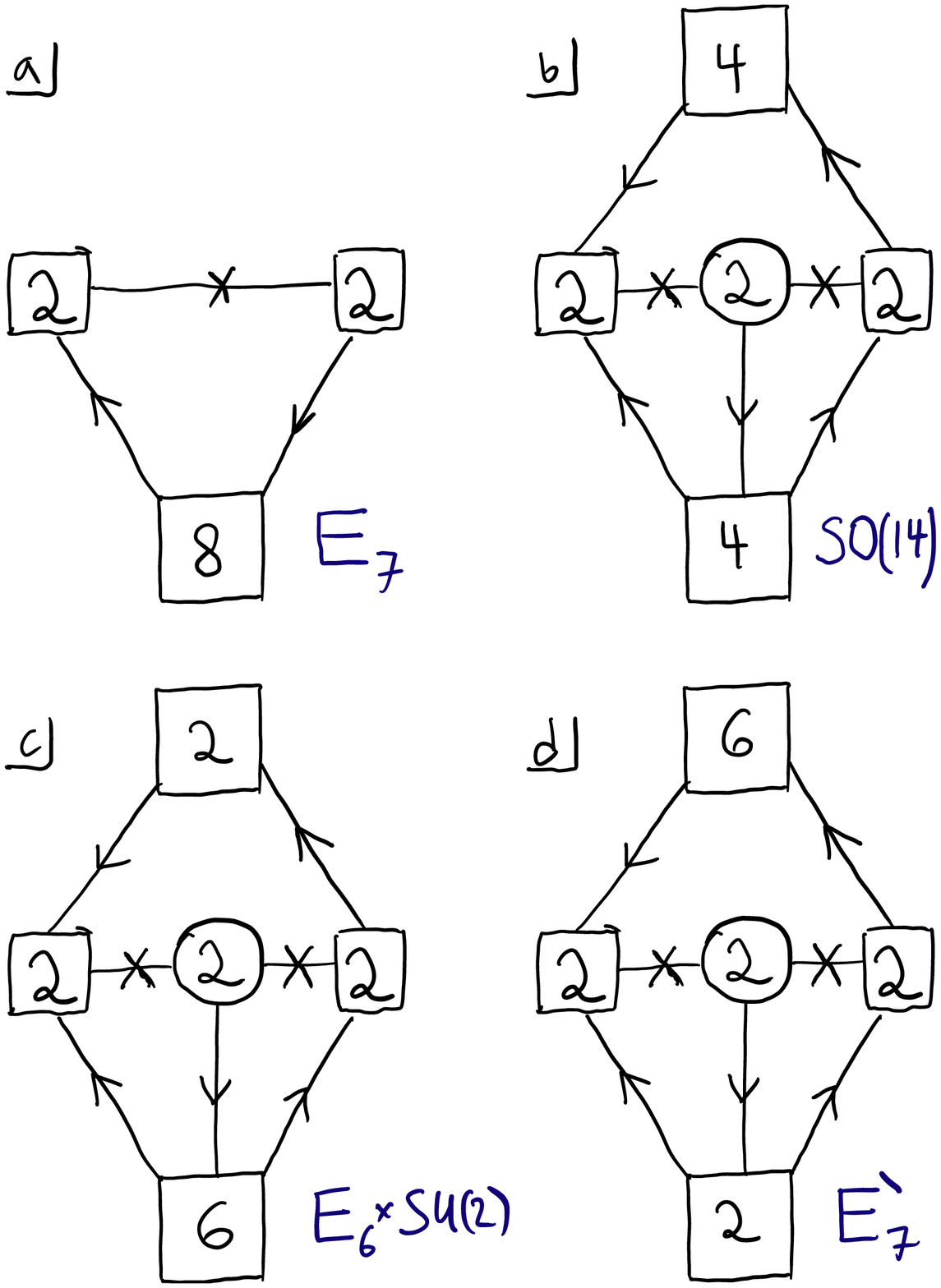}{2.7in}}
\medskip\centerline{\vbox{
\baselineskip12pt\advance\hsize by -1truein
\noindent\footnotefont{\bf Fig.~10:} a. The $E_7\times U(1)$ sphere with two maximal punctures and half unit of flux. b. The $SO(14)\times U(1)$ theory with two punctures and half unit of flux. c. The $E_6\times SU(2)\times U(1)$ theory with two maximal punctures and half unit of flux. d. Another representation of $E_7\times U(1)$ theory with two punctures and half unit of flux. Gluing  copies of this two a torus gives a  Seiberg dual to models constructed from the the first theory.}
}

\

For example, the $SO(14)$ model with the flux decomposition in \uflxso\ can be constructed by a concatenation of two domain walls with flux $(-\!\frac{1}{2},-\!\frac{1}{2},-\!\frac{1}{2},-\!\frac{1}{2},-\!\frac{1}{2},-\!\frac{1}{2},-\!\frac{1}{2},-\!\frac{1}{2}) $ and $(\frac{1}{2},\frac{1}{2},\frac{1}{2},\frac{1}{2},-\!\frac{1}{2},-\!\frac{1}{2},-\!\frac{1}{2},-\!\frac{1}{2})$ respectively. In the first domain wall, the chiral fields $X_{1,2,\cdots,8}$ in the first chamber and $Y_{1,2,\cdots,8}'$ in the second chamber satisfy Neumann boundary condition, while in the second domain wall the chiral fields $Y_{5,6,7,8}'$ and $X_{1,2,3,4}'$ in the second chamber, and $X_{5,6,7,8}''$ and $Y_{1,2,3,4}''$ in the third chamber satisfy Neumann boundary condition. This means the fields $X_{5,6,7,8}'$ and $Y_{5,6,7,8}'$ in the second chamber are truncated and the system will have the superpotential,
\eqn\sohalfpo{
	{\cal W} = b\ {\rm det}q + b'\ {\rm det}q' + \sum_{i=5}^8(Y'_iqX_i + Y'_iq'X''_i) + \sum_{i=1}^4 Y''_iqq'X_i \ .
}
When we put this system on a segment and impose the maximal boundary condition at both ends, we will obtain the $SO(14)$ model with flux \uflxso, which is drawn in Fig. 10 (b).

\

The $E_6$ model of half a unit flux can be constructed by domain walls using the following flux decomposition
\eqn\uflxe{
	 {\textstyle (0,0,-\!1,-\!1,-\!1,-\!1,-\!1,-\!1) = (-\!\frac{1}{2},-\!\frac{1}{2},-\!\frac{1}{2},-\!\frac{1}{2},-\!\frac{1}{2},-\!\frac{1}{2},-\!\frac{1}{2},-\!\frac{1}{2}) +(\frac{1}{2},\frac{1}{2},-\!\frac{1}{2},-\!\frac{1}{2},-\!\frac{1}{2},-\!\frac{1}{2},-\!\frac{1}{2},-\!\frac{1}{2}) \ .}
}
This leads to a domain wall configuration of  the $4d$ superpotential
\eqn\sohalfpo{
	{\cal W} = b\ {\rm det}q + b'\ {\rm det}q' + \sum_{i=3}^8(Y'_iqX_i + Y'_iq'X''_i) + \sum_{i=1}^2 Y''_iqq'X_i \ .
}
The chiral fields $X_i,Y_i,X'_i,Y'_i,X_i'',Y_i''$ appearing in this superpotential satisfy Neumann boundary condition at the boundaries. This $5d$ system on a finite segment yields the $4d$ $E_6$ model in Fig. 10 (c).
Similarly, we can construct the $E_7'$ model in Fig. 10 (d) from the flux decomposition
\eqn\epmdl{
	 {\textstyle(0,0,0,0,0,0,-\!1,-\!1) = (-\!\frac{1}{2},-\!\frac{1}{2},-\!\frac{1}{2},-\!\frac{1}{2},-\!\frac{1}{2},-\!\frac{1}{2},-\!\frac{1}{2},-\!\frac{1}{2})  +(\frac{1}{2},\frac{1}{2},\frac{1}{2},\frac{1}{2},\frac{1}{2},\frac{1}{2},-\!\frac{1}{2},-\!\frac{1}{2}) \ .}
}

\

\centerline{\figscale{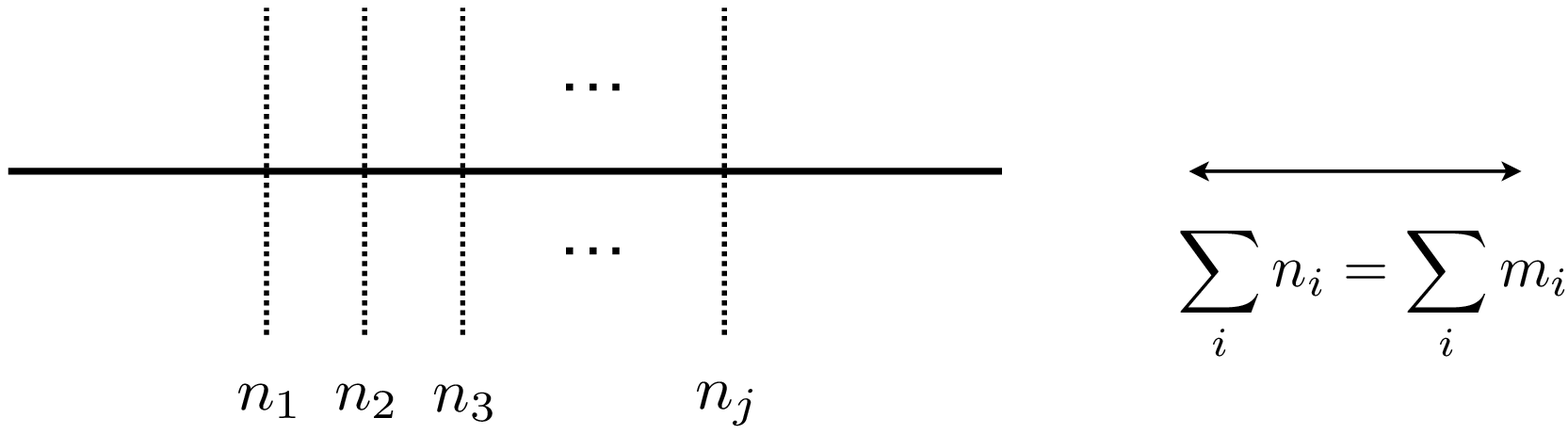}{5in}}
\medskip\centerline{\vbox{
\baselineskip12pt\advance\hsize by -1truein
\noindent\footnotefont{\bf Fig.~11:} There is a duality between different domain wall configurations obtained by  different decompositions of flux. $n_i$ and $m_i$ are $U(1)$ fluxes at each domain wall. If $\sum_in_i=\sum_im_i$, then two configurations describe the same physics.}} 

\

We claim that all tube models with generic $U(1)$ fluxes can be obtained by a combination of the duality domain walls. The number of domain walls and $U(1)$ charge assignments for the fields on each wall depends on the decomposition of $U(1)$ fluxes into a combination of half fluxes. Different decompositions give rise to different domain wall configurations, and therefore to different $4d$ Lagrangian tube models. However, as long as their net fluxes are the same (up to $E_8$ Weyl symmetry), all the different $4d$ models should describe the same physics at their conformal points. Therefore, we expect that our $5d$ domain wall picture can lead to a huge number of $4d$ (and also $5d$) dualities between tube models.

The simplest duality example of this type is the duality between the $E_7$ and the $E_7'$ models in Fig. 10 (a) and 10 (d). They are constructed by using different number of domain walls. However, their net fluxes are the same up to the Weyl transformation, thus we can expect a duality between these two theories. Indeed, they are Seiberg-dual to each other.

\

\

\noindent{\it Anomalies of two punctured spheres}

Let us now compute the anomalies of some of the spheres with two punctures. 
Knowing the anomaly polynomial and the $5d$ gauge theory description of a puncture of the $6d$ E-string theory discussed in section 2  allows us to compute the anomalies of $4d$ theories of punctured Riemann surfaces. The $4d$ anomalies are obtained, as we discussed above, by adding the geometric contributions and the $5d$ anomaly inflow contribution. We can compare the anomalies of two punctured spheres, which we compute directly from the $6d$ anomaly polynomial and the $5d$ boundary conditions, with the anomalies of the corresponding $4d$ models we will discuss later on.
This will provide another strong evidence that our $4d$ models of punctured Riemann surfaces are consistent with compactifications of the $6d$ E-string theory. Here we will compute the anomalies for some simple test cases and compare them with expectations based on corresponding $4d$ field theories proposed in the following sections.

Let us first compute the anomalies of the $E_7$ model from the two punctured sphere with flux $-z$ using the $6d$ anomaly polynomial and the $5d$ boundary condition. The geometric contribution is given by
\eqn\fghyuijo{\eqalign{
	&Tr(U(1)_F^3) = -12z\xi^2 \ , \quad Tr(U(1)_F) = -12z\xi \ , \cr
	&Tr(U(1)_F U(1)^2_R) = 4z\xi \ , \quad Tr(U(1)_R^3) = Tr(U(1)_R) = 0 \ ,
}}
with $z=\frac{1}{2}$ and $\xi=1$.
The $U(1)_F$ with flux $-\frac{1}{2}$ acts on the eight hypermultiplets in the $5d$ E-string theory with the same charge $-\frac{1}{2}$. Thus the anomaly inflow contribution from a single puncture is
\eqn\tyeuio{\eqalign{
	&Tr(U(1)_R^3) = - \frac{3}{2} \ , \quad Tr(U(1)_R) = - \frac{3}{2} \ , \quad
	Tr(U(1)_RSU(2)^2) = -1 \ , \cr
	& Tr(U(1)_F^3) = -1 \ , \quad Tr(U(1)_F) = -4 \ , \quad Tr(U(1)_FSU(2)^2) = -1 \ .
}}
The total anomalies are given by the sum of the geometric contribution and the anomaly inflows from two punctures which we find
\eqn\wreyuue{\eqalign{
	&Tr(U(1)_R^3) = -3 \ , \quad Tr(U(1)_R) = - 3  \ ,  \cr
	&Tr(U(1)_F^3) = -2-12z \ , \quad Tr(U(1)_F) = -8-12z \ , \quad Tr(U(1)_F U(1)_R^2) = 4z \ , \cr
	&Tr(U(1)_R SU(2)_{1,2}^2) = -1 \ , \quad Tr(U(1)_FSU(2)_{1,2}^2)  = -1 \ , \cr
}}
where $SU(2)_1$ and $SU(2)_2$ are puncture symmetries of two punctures respectively. As previously mentioned,
this result agrees with the anomalies of the WZ model, introduced in the previous section, which corresponds to  $E_7$ compactification on a sphere with two punctures and a unit of flux,  shown in Fig. 6.

We can also compute the anomalies of the two punctured sphere with $E_6\times SU(2)$ symmetries. The hypermultiplets in the $5d$ E-string theory transform under the $U(1)_F$, which is identified with $\frac{1}{2}U(1)_m$ in Fig. 14, with charges $\left(-\!\frac{3}{2},-\!\frac{3}{2},-\!\frac{1}{2},-\!\frac{1}{2},-\!\frac{1}{2},-\!\frac{1}{2},-\!\frac{1}{2},-\!\frac{1}{2}\right)$. This theory has the following anomalies.
\eqn\wfheuui{\eqalign{
	& Tr(U(1)_F^3) = \left(-\!108z\right)_{geo} + 2\!\times\!\left(\!-\!\frac{15}{2}\!\right)_{inf} = -\!15\!-\!108z \ , \quad Tr(U(1)^3_R) = 2\!\times\! \left(\!-\!\frac{3}{2}\!\right)_{inf} = -3 \ ,\cr
	& Tr(U(1)_F) = \left(-36z\right)_{geo}+2\times\left(-6\right)_{inf} = -12-36z \ , \quad Tr(U(1)_R) = 2\times \left(\!-\frac{3}{2}\!\right)_{inf} = -3 \ ,\cr 
	& Tr(U(1)_FU(1)_R^2) = \left(12z\right)_{geo}=12z \ , \quad Tr(U(1)_FSU(2)_{1,2}^2) = \left(-\frac{3}{2}\right)_{inf} = -\frac{3}{2} \ .
}}
Here, the subscripts $geo$ and $inf$ stand for the geometric contribution and the anomaly inflow contribution respectively. This result perfectly equals the $4d$ anomalies of our $E_6\times SU(2)$ model in Fig. 11c.

Similarly, the anomalies for the $SO(14)$ theory from two punctured sphere can be easily computed. The $5d$ hypermultiplets carry $(0,0,0,0,-\!1,-\!1,-\!1,-\!1)$ charge for the $U(1)_F$ (or $U(1)_m$ in Fig. 12). Regarding this, the anomalies are given by
\eqn\sfihuie{\eqalign{
	& Tr(U(1)_F^3) = \left(-48z\right)_{geo} + 2\times\left(-4\right)_{inf} = -8-48z \ , \quad Tr(U(1)^3_R) = 2\times \left(-\!\frac{3}{2}\right)_{inf} = -3 \ ,\cr 
	& Tr(U(1)_F) = \left(-24z\right)_{geo}+2\times\left(-4\right)_{inf} = -24z-8 \ , \quad Tr(U(1)_R) = 2\times \left(-\!\frac{3}{2}\right)_{inf} = -3 \ , \cr 
	& Tr(U(1)_FU(1)_R^2) = \left(8z\right)_{geo}=8z \ , \quad Tr(U(1)_FSU(2)_{1,2}^2) = \left(-1\right)_{inf} = -1 \ ,
}}
which are the same as the anomalies of the $SO(14)$ model in Fig. 10b.

\

\

\newsec{Tori and spheres with general fluxes}

We have constructed the theory associated to two punctured sphere for flux preserving $E_7\times U(1)$ matching anomalies of six dimensions and four dimensional construction in section three and understood how to derive two punctured spheres from this theory for more general fluxes in the previous section. Here  we will discuss the field theory constructions in detail for several different compactifications. In particular we will discuss how the anomalies and the symmetries of the quiver theories at hand exhibit the properties expected from six dimensional computations.

\

\subsec{Rank one E-string on a torus: $E_8\to$ $G\times U(1)$}

The most simple examples of tubes with flux in one $U(1)$ are depicted in Fig. 10 and we will discuss them in detail next.

\

\noindent{\bf \it $SO(14)\times U(1)$}

We have already argued that the theory corresponding to a sphere with two punctures and half a unit of flux breaking $E_8$ to $SO(14)\times U(1)$ is depicted in Fig. 10 (b). In terms of the flux basis we associate to it the flux $(-\frac{1}{2}; \frac{1}{4} , \frac{1}{4} , \frac{1}{4} , \frac{1}{4} , -\frac{1}{4} , -\frac{1}{4} , -\frac{1}{4} , -\frac{1}{4} )$ in the overcomplete basis and $( 0 , 0 , 0 , 0 , -1 , -1 , -1 , -1 )$ in the complete basis, where the last four $U(1)$'s are associated with the $SU(4)$ global symmetry seeing more flavors.
We can verify that the anomalies of the sphere with two punctures match the computation in six dimensions.

 Gluing two such spheres together into a torus we obtain the theory shown in Fig. 12. This theory then corresponds to an $SO(14)$ preserving torus compactification with unit flux. Next we shall analyze it in detail. 
First consider the case without the flipping fields. We inquire as to what is the superconformal R-symmetry, where the $6d$ R-symmetry can mix with the two $U(1)$'s. However we note that the charges under $U(1)_y$ are balanced so there is no mixing involving it. Thus the superconformal R-symmetry will be: $U(1)^{SC}_R = U(1)^{6d}_R + \alpha U(1)_m$. Performing a-maximization we find $\alpha = \frac{\sqrt{19}-3}{3}$. We find that all gauge invariant operators have dimension above the unitary bound so it is plausible that this theory flows to an interacting IR SCFT.

\

\centerline{\figscale{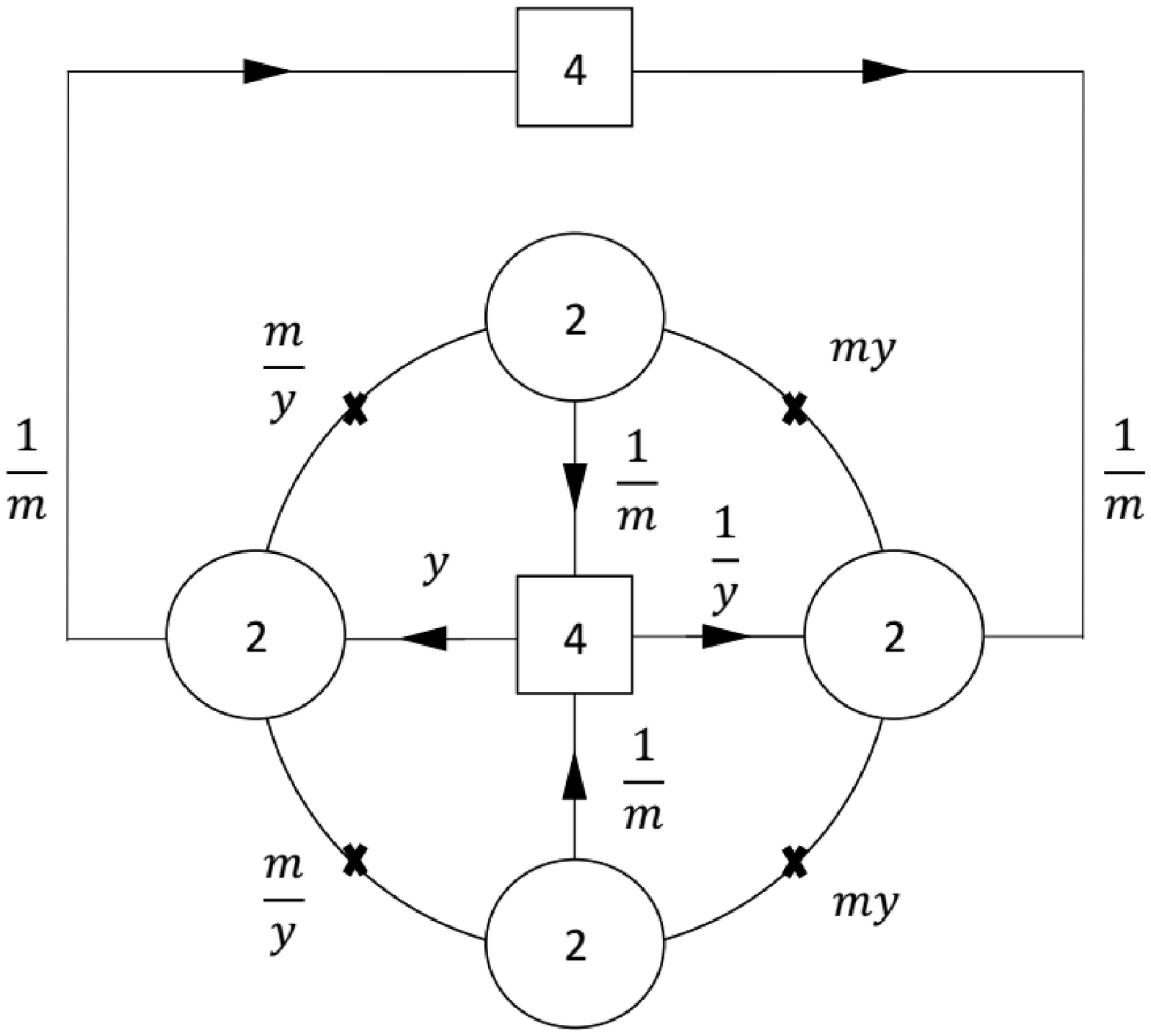}{2.1in}}
\medskip\centerline{\vbox{
\baselineskip12pt\advance\hsize by -1truein
\noindent\footnotefont{\bf Fig.~12:} The $U(1)\times SO(14)$ model from torus with one unit of flux. There is a cubic superpotential for each one of the four internal triangles, and two quartic ones for the bifundamentals of the external $SU(4)$ with the upper and lower half-circles. There is a natural R-symmetry, which is the one the theory inherits from $6d$, under which the gauge bifundamental have R-charge $0$, the flippers R-charge $2$, and the rest R-charge $1$. Besides the two $SU(4)$ global symmetries there are also two non-anomalous $U(1)$'s which we denote as $U(1)_m$ and $U(1)_y$. The charges of all the fields under these $U(1)$'s are represented by fugacities.}
}

\

Now we add $4$ singlets and couple them through the flipping superpotential. We find that this superpotential is relevant compared to the SCFT point so the theory will flow to a new theory in the IR. We can repeat the a-maximization for this case, finding: $\alpha = \frac{\sqrt{2}}{3}$. Using this values we obtain for the conformal anomalies,

\eqn\sofyo{
c=\frac52\sqrt{2}\,,\qquad \; a= 2\sqrt{2}\,.
} This agrees with the six dimensional computation noting that for $SO(14)$ $\xi=2$. Also we find that all gauge invariant operators are above the unitary bound so it is again plausible that this theory flows to an interacting SCFT in the IR. Note that the singlets do not have free R-charge in the SCFT and are thus an inseparable part of it. 

The $6d$ construction suggests that this theory has an $SO(14)$ global symmetry somewhere on its conformal manifold. This is definitely not visible from the Lagrangian so to test this we wish to evaluate the superconformal index. For this it is convenient to work with the non-superconformal R-symmetry: $U(1)^{'}_R = U(1)^{6d}_R + \frac{1}{2} U(1)_m$. Note that $\frac{\sqrt{2}}{3} - \frac{1}{2} \approx -0.03$, so this R-symmetry is very close to the true SC R-symmetry. Using this R-symmetry we indeed find that the index can be written in characters of $SO(14)$ at least to the order we evaluated. Particularly, the first terms in the supersymmetric index are,

\eqn\insuo{\eqalign{ {\cal I} &= 1 + \frac{2}{m^2} \chi[{\bf 14}] (p q)^{\frac{1}{2}} + \frac{1}{m} \chi[{\bf 64}] (p q)^{\frac{3}{4}} + \frac{2}{m^2} \chi[{\bf 14}] (p q)^{\frac{1}{2}}(p+q)\cr & + p q (m^4 + \frac{1}{m^4} (3 \chi[{\bf 104}] + \chi[{\bf 91}] -1) ) + ...
}}
where

\eqn\sodef{\eqalign{ \chi[{\bf 14}] &= y^2 + \frac{1}{y^2} + \chi[{\bf 6}, {\bf 1}] + \chi[{\bf 1}, {\bf 6}], \cr \chi[{\bf 64}]  &
 = y (\chi[{\bf 4}, {\bf \overline{4}}]+\chi[{\bf \overline{4}}, {\bf 4}]) + \frac{1}{y}(\chi[{\bf 4}, {\bf 4}]+\chi[{\bf \overline{4}}, {\bf \overline{4}}]).
}}

We next note several observations regarding the index. First it indeed forms characters of $SO(14)$ where $SU(4) \times SU(4) \times U(1)_y$ is enhanced to this symmetry. It is interesting to note that the two $SU(4)$ appear asymmetrically in the Lagrangian, but are symmetric in $SO(14)$. All the anomalies are consistent with the enhancement and with the $6d$ result. 

 The order $p q$ terms indicate that there are no marginal operators as all operators appearing at that order are charged under $U(1)_m$. Specifically, the $m^4$ state is relevant while the ones proportional to $\frac{1}{m^4}$ are irrelevant. Note that these contain negative contributions and so the fact that they are irrelevant avoids a contradiction with the results of \BeemYN .
If we wish to interpret the $SO(14)$ as the global symmetry somewhere on the conformal manifold, then we must view this term as $\chi[{\bf 91}] + 1 - \chi[{\bf 91}] -1$. This leads to an $8$ dimensional conformal manifold on a generic point of which the symmetry is broken to $U(1)^8$. This picture is consistent with the $6d$ conformal manifold generated via holonomies. We do not observe the complex structure moduli of the torus though it does not appear also in the case without the flux. In this case the theory is strongly coupled and we do not have a weak coupling point to compare against. 

We can again identify some of the contributions as coming from the $6d$ conserved current multiplet of the $E_8$ global symmetry. Particularly, the contributions $\frac{2}{m^2} \chi[{\bf 14}] (p q)^{\frac{1}{2}}$ and $\frac{1}{m} \chi[{\bf 64}] (p q)^{\frac{3}{4}}$ have the same R-charge as marginal operators under the 6d R-symmetry. Furthermore the representations they carry exactly match those required to complete $SO(14)$ to $E_8$ (again we refer the reader to  Appendix A for the branching rule). Interestingly, their number is again exactly as expected from the reasoning of \Brz. As previously mentioned, the marginal operators also behave as expected save for the absence of the marginal deformation expected from the complex structure moduli of the torus.

We can combine several spheres to form a torus with any value of the flux. When combining an odd number of spheres though some of the global symmetry is broken. For example consider closing the basic tube by gluing the two punctures. This results in the theory shown in Fig.   13. In the gluing we are forced to break $U(1)_y$ and also one of the $SU(4)$ groups to $USp(4)$. 

\

\centerline{\figscale{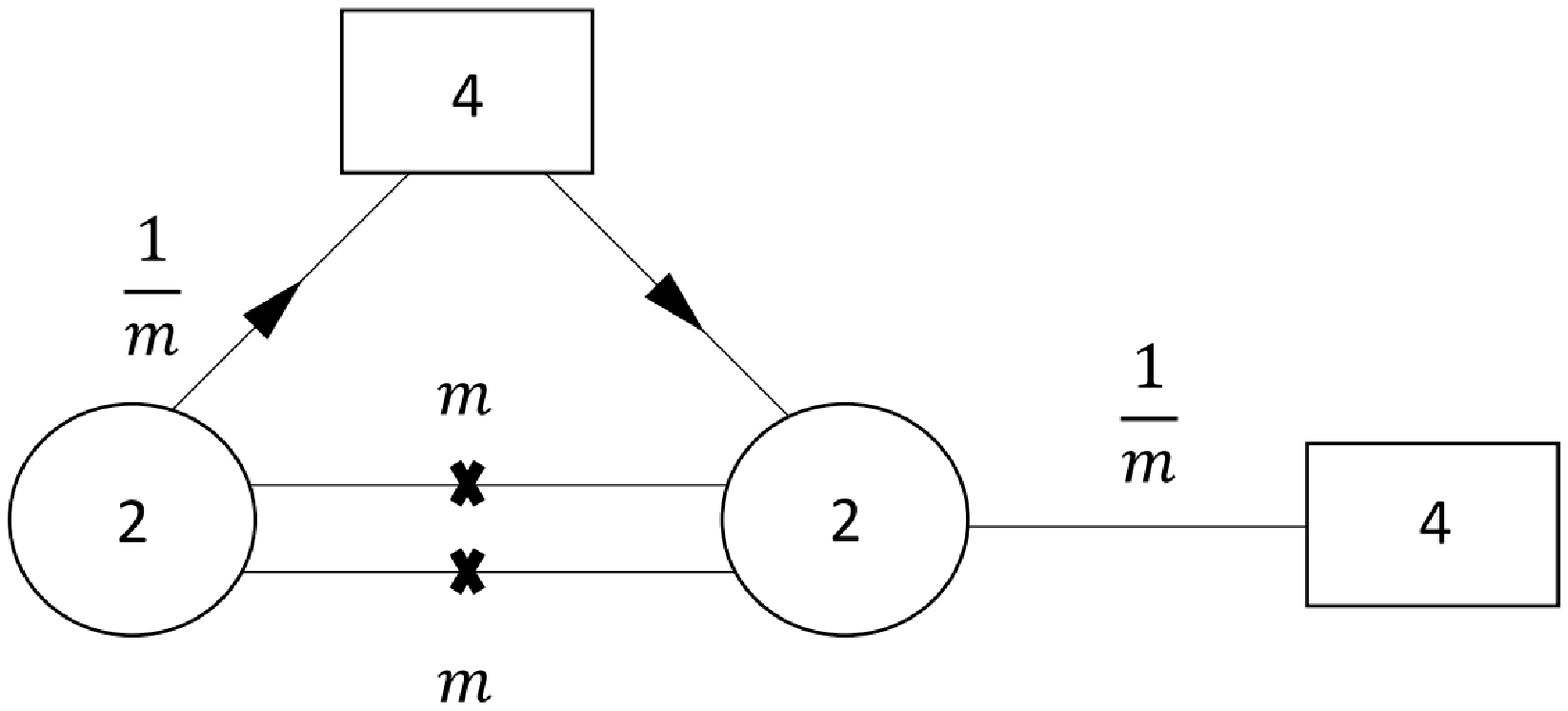}{2.1in}}
\medskip\centerline{\vbox{
\baselineskip12pt\advance\hsize by -1truein
\noindent\footnotefont{\bf Fig.~13:} The $SO(14)\times  U(1)$ model from torus with half a unit of flux. There is a cubic superpotential for the two triangles, and a quartic one for the bifundamentals of the lower right $SU(4)$ with the two gauge bifundamentals. Said quartic superpotential actually breaks the $SU(4)$ down to $USp(4)$. There is a natural R-symmetry, which is the one the theory inherits from $6d$, under which the gauge bifundamentals have R-charge $0$, the flippers R-charge $2$, and the rest R-charge $1$. Besides the $SU(4)$ and $USp(4)$ global symmetries there is now only one non-anomalous $U(1)$ corresponding to $U(1)_m$. Its charges are represented in the figure by fugacities.}
}

\

 Besides these points the dynamics of the theory are very similar to the previous case. Specifically, since the matter content is exactly half the one in the previous case, and as there was no mixing under $U(1)_y$, the R-symmetry maximizing a will be the same. The dimensions of all the operators are again the same so there are no violations of the unitary bound. Thus we expect again that this theory flows to an interacting SCFT. The anomalies are half those of the previous case which will therefore agree with the $6d$ analysis.

The interesting feature in this case is the breakdown of part of the global symmetry. From the $6d$ view point the breaking is done due to the center flux necessary for consistency of the compactification. From group theory the maximal global symmetry one can preserve is $SO(11)$, and so we expect the index of the theory to form characters of $SO(11)$. To check this we computed the index finding:

\eqn\insuo{\eqalign{ {\cal I} &= 1 + (\frac{3}{m^2} + m^2 + \frac{1}{m^2} \chi[{\bf 11}]) (p q)^{\frac{1}{2}} + \frac{1}{m} \chi[{\bf 32}] (p q)^{\frac{3}{4}} + (\frac{3}{m^2} + \frac{1}{m^2} \chi[{\bf 11}]) (p q)^{\frac{1}{2}}(p+q)\cr & + p q (m^4 + \frac{1}{m^4} (\chi[{\bf 65}] + 3\chi[{\bf 11}] + 5) ) + ...
}}
where

\eqn\sodef{\eqalign{ \chi[{\bf 11}] &=  \chi[{\bf 6}, {\bf 1}] + \chi[{\bf 1}, {\bf 5}], \cr \chi[{\bf 32}]  &
 =  \chi[{\bf 4}, {\bf 4}]+\chi[{\bf \overline{4}}, {\bf 4}],
}}
and the index is evaluated using the same R-symmetry as before.

One can note that the index forms characters of a larger global symmetry $SO(12)$, which cannot be realized from the $6d$ construction. However, the $6d$ picture suggests that besides $SO(11)$ one can also have $SU(2)\times SO(9)$ and $USp(4)\times SO(7)$ as global symmetries at special points on the conformal manifold. It is possible to show that the index is also consistent with these symmetries. These are not subgroups of one another but they are all subgroups of $SO(12)$ so the apparent $SO(12)$ structure can be understood as arising from the need to accommodate all these different global symmetries. 

We can study the conformal manifold from the $p q$ order terms. As the terms appearing are charged under $U(1)_m$, they are actually relevant or irrelevant deformations. Therefore the marginal operators must be in the adjoint of the global symmetry group. Assuming that there is a point with $SO(11)\times U(1)_m$ global symmetry, this leads to a $6$ dimensional conformal manifold. Note that there is no contradiction with also having points with $SU(2)\times SO(9) \times U(1)_m$ and $USp(4)\times SO(7) \times U(1)_m$ global symmetries as the rank of all of these groups is equal. However, $SO(12)\times U(1)_m$ has different rank and therefore is inconsistent with the other choices. Thus, if the $6d$ picture is correct, even though the index forms characters of $SO(12)$ it cannot have that symmetry on a point in the conformal manifold.

\

\noindent{\bf \it $E_6\times SU(2)\times U(1)$}

It follows from section 4 that the theory corresponding to a sphere with two puncture and half a unit of flux breaking $E_8$ to $E_6\times SU(2)\times U(1)$ is depicted in Fig. 10 (c). In terms of the flux basis we associate to it the flux $(-\frac{3}{4}; \frac{3}{8} , \frac{3}{8} , -\frac{1}{8} , -\frac{1}{8} , -\frac{1}{8} , -\frac{1}{8} , -\frac{1}{8} , -\frac{1}{8} )$ in the overcomplete basis and $( 0 , 0 , -1 , -1 , -1 , -1 , -1 , -1 )$ in the complete basis. The last six $U(1)$'s are associated to the $SU(6)$ flavor symmetry group. Gluing two such spheres together into a torus we obtain the theory shown in Fig. 14. This theory then corresponds to an $E_6\times SU(2)\times U(1)$ preserving torus compactification with unit flux. Next we shall analyze it in details.

First consider the case without the flipping fields. We inquire as to what is the superconformal R-symmetry, where the $6d$ R-symmetry can mix with the two $U(1)$'s. However we note that the charges under $U(1)_y$ are balanced so there is no mixing involving it. Thus the superconformal R-symmetry will be: $U(1)^{SC}_R = U(1)^{6d}_R + \alpha U(1)_m$. Performing a-maximization we find $\alpha = \frac{5}{27}$. We find that all gauge invariant operators have dimension above the unitary bound so it is plausible that this theory flows to an interacting IR SCFT. 

\

\centerline{\figscale{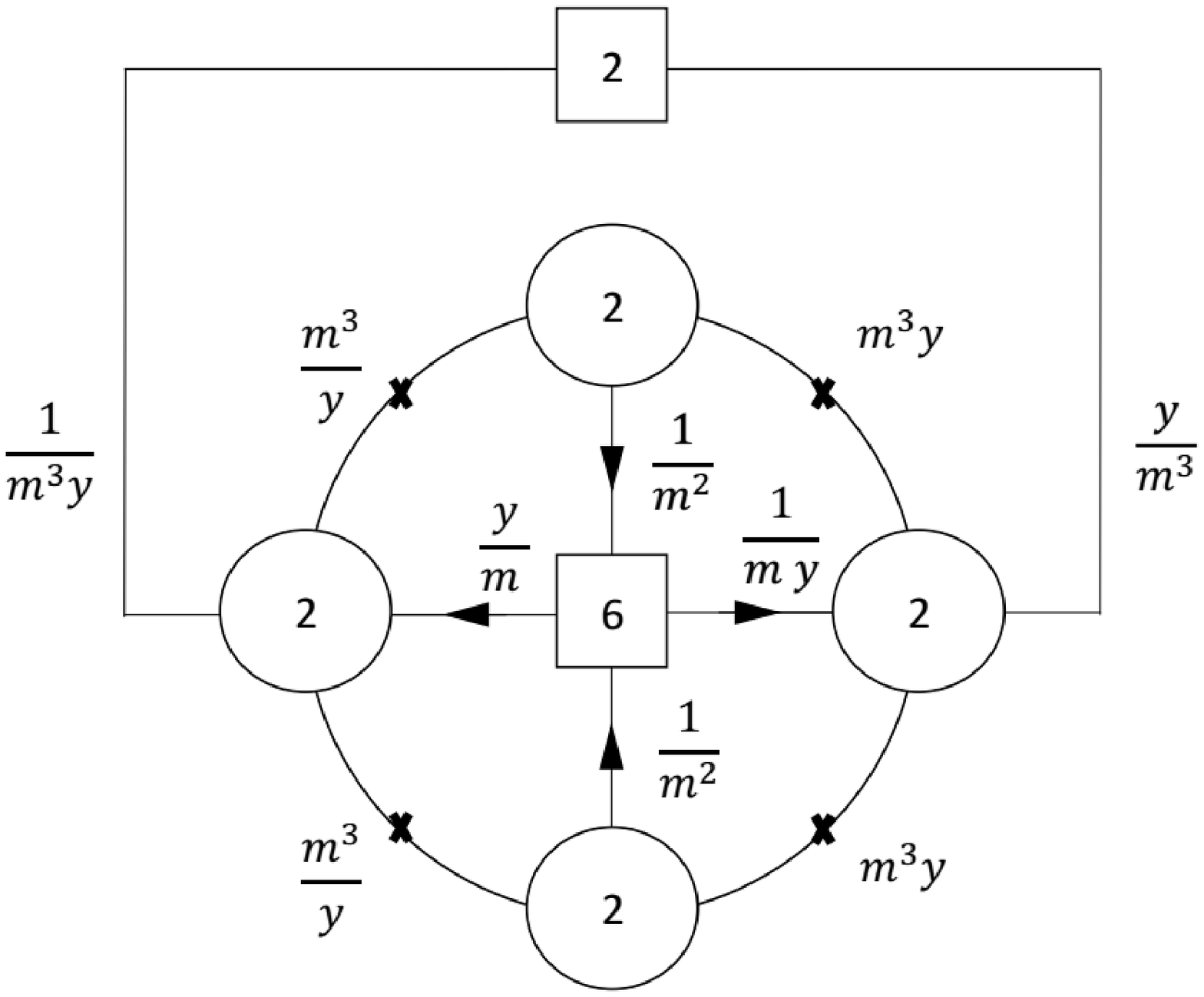}{2.1in}}
\medskip\centerline{\vbox{
\baselineskip12pt\advance\hsize by -1truein
\noindent\footnotefont{\bf Fig.~14:} The $E_6\times  U(1) \times SU(2)$ model from torus with one unit of flux. There is a cubic superpotential for each one of the four internal triangles, and two quartic ones for the bifundamentals of the external $SU(2)$ with the upper and lower half-circles. There is a natural R-symmetry, which is the one the theory inherits from $6d$, under which the gauge bifundamentals have R-charge $0$, the flippers R-charge $2$, and the rest R-charge $1$. Besides the $SU(6) \times SU(2)$ global symmetries there are also two non-anomalous $U(1)$'s which we denote as $U(1)_m$ and $U(1)_y$. The charges of all the fields under these $U(1)$'s are represented by fugacities.}
}

\

Now we add $4$ singlets and couple them through the flipping superpotential. We find that this superpotential is relevant compared to the SCFT point so the theory will flow to a new theory in the IR. We can repeat the a-maximization for this case, finding: $\alpha = \frac{1}{3 \sqrt{3}}$. Using this value we obtain for the conformal anomalies,

\eqn\sofyoe{
c=\frac52\sqrt{3}\,,\qquad \; a= 2\sqrt{3}\,.
} This agrees with the six dimensional computation noting that for $SU(2)\times E_6$ $\xi=3$. Also we find that all gauge invariant operators are above the unitary bound so it is again plausible that this theory flows to an interacting SCFT in the IR. Note that the singlet do not have free R-charge in the SCFT and are thus an inseparable part of it.

The $6d$ construction suggests that this theory has an $SU(2)\times E_6$ global symmetry somewhere on its conformal manifold. This is definitely not visible from the Lagrangian so to test this we wish to evaluate the superconformal index. For this it is convenient to work with the non-superconformal R-symmetry: $U(1)^{'}_R = U(1)^{6d}_R + \frac{2}{9} U(1)_m$. Note that $\frac{1}{3 \sqrt{3}} - \frac{2}{9} \approx -0.03$, so this R-symmetry is very close to the true SC R-symmetry. Using this R-symmetry we indeed find that the index can be written in characters of $SU(2)\times E_6$ at least to the order we evaluated. Particularly, the first terms in the supersymmetric index are,

\eqn\insuoe{\eqalign{ {\cal I} &= 1 + \frac{3}{m^6} \chi[{\bf 2}, {\bf 1}] (p q)^{\frac{1}{3}} + \frac{2}{m^4} \chi[{\bf 1},{\bf \overline{27}}] (p q)^{\frac{5}{9}} + \frac{3}{m^6} \chi[{\bf 2}, {\bf 1}] (p q)^{\frac{1}{3}}(p+q)\cr & + \frac{3}{m^{12}} (1+2\chi[{\bf 3}, {\bf 1}]) (p q)^{\frac{2}{3}} + \frac{1}{m^2} \chi[{\bf 2},{\bf 27}] (p q)^{\frac{7}{9}} + \frac{6}{m^{10}} \chi[{\bf 2},{\bf \overline{27}}] (p q)^{\frac{8}{9}} \cr & + p q \frac{2}{m^{18}}(5 \chi[{\bf 4}, {\bf 1}] + 4\chi[{\bf 2}, {\bf 1}]) + ...
}}
where

\eqn\sodef{\eqalign{ \chi[{\bf 2}, {\bf 1}] &= y^2 + \frac{1}{y^2} , \cr \chi[{\bf 1},{\bf \overline{27}}]  &
 = \chi[{\bf 2},{\bf 6}]_{SU(2)\times SU(6)} + \chi[{\bf 1},{\bf \overline{15}}]_{SU(2)\times SU(6)}.
}}
We next note several observations regarding the index. First it indeed forms characters of $SU(2)\times E_6$ where $SU(2) \times SU(6) \times U(1)_y$ is enhanced to this symmetry. All the anomalies are consistent with the enhancement and with the $6d$ result. We note that $U(1)_m$ is identified with $2 U(1)$ when we use the normalization convention of unit charge. From the order $p q$ terms we see that there are no marginal operators as all operators appearing at that order are charged under $U(1)_m$. In this case they are all charged negatively and so are irrelevant.

If we wish to interpret the $SU(2)\times E_6$ as the global symmetry somewhere on the conformal manifold, then we must view this term as $\chi[{\bf 3},{\bf 1}] + \chi[{\bf 1},{\bf 78}] + 1 - \chi[{\bf 3},{\bf 1}] - \chi[{\bf 1},{\bf 78}] -1$. This leads to an $8$ dimensional conformal manifold on a generic point of which the symmetry is broken to $U(1)^8$. This picture is consistent with the $6d$ conformal manifold generated via holonomies. We again do not observe the complex structure moduli of the torus. Like the previous case, this theory is strongly coupled and we do not have a weak coupling point to compare against. 

We can again identify some of the contributions as coming from the $6d$ conserved current multiplet of the $E_8$ global symmetry. Particularly, the contributions $\frac{3}{m^6} \chi[{\bf 2}, {\bf 1}] (p q)^{\frac{1}{3}}$, $\frac{2}{m^4} \chi[{\bf 1},{\bf \overline{27}}] (p q)^{\frac{5}{9}}$ and $\frac{1}{m^2} \chi[{\bf 2},{\bf 27}] (p q)^{\frac{7}{9}}$ have the same R-charge as marginal operators under the $6d$ R-symmetry. Furthermore the representations they carry exactly match those required to complete $SU(2)\times E_6$ to $E_8$ (once again we refer the reader to  Appendix A for the branching rule). Their number is again exactly as expected from the reasoning of \Brz. Marginal operators also behave as expected except for the lack of the marginal deformation expected from the complex structure moduli of the torus.

We can combine many copies of the sphere to construct theory with arbitrary flux preserving $E_6\times SU(2)$ symmetry. Again the model behaves differently depending on whether the flux is integer or half-integer, where in the half-integer case $U(1)_y$ is broken. From the 6d perspective this comes about as the non-integer flux must be accompanied by a center flux, here inside the $SU(2)$. This center flux in turn breaks it completely.

\

\centerline{\figscale{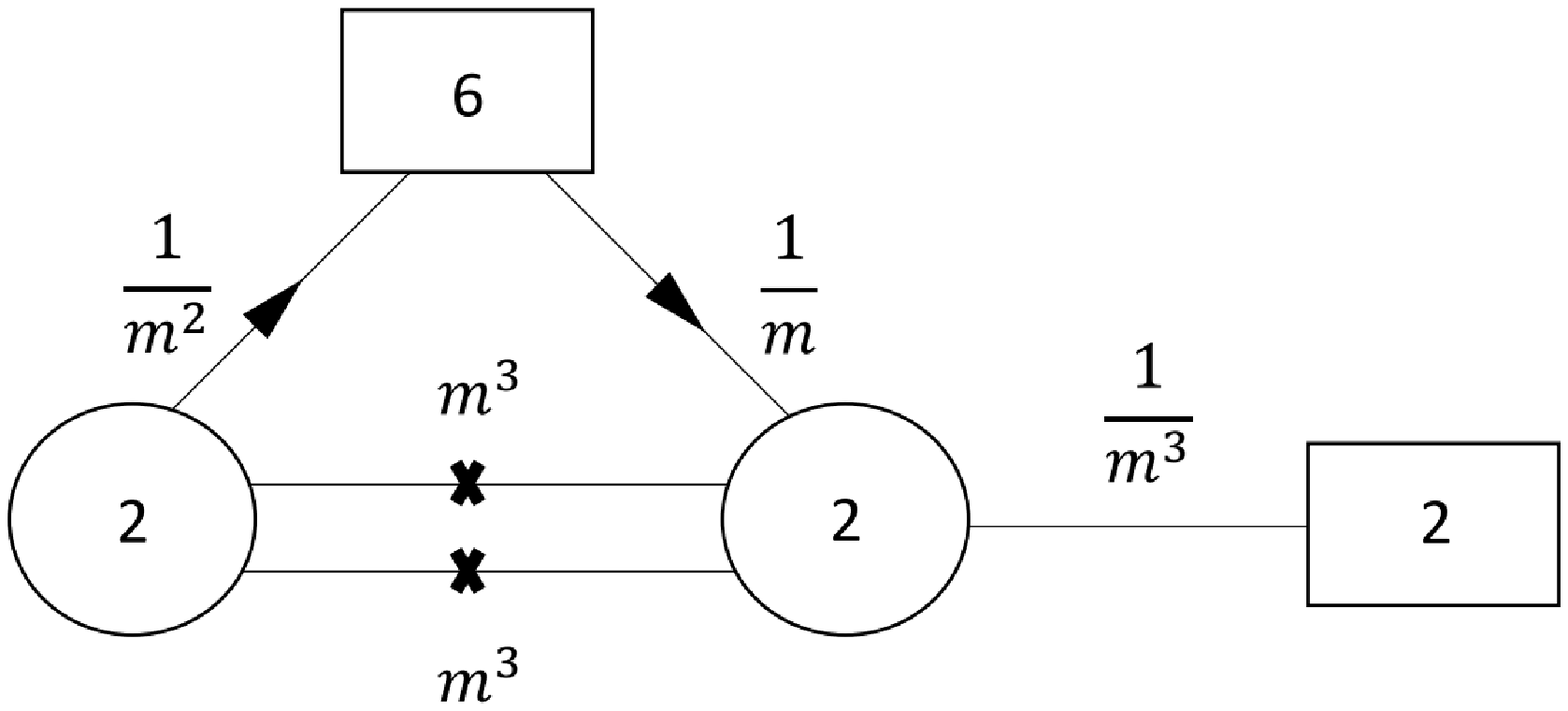}{2.1in}}
\medskip\centerline{\vbox{
\baselineskip12pt\advance\hsize by -1truein
\noindent\footnotefont{\bf Fig.~15:} The $E_6\times  U(1) \times SU(2)$ model from torus with half a unit of flux. There is a cubic superpotential for the two triangles, and a quartic one for the bifundamentals of global $SU(2)$ with the two gauge bifundamentals. There is a natural R-symmetry, which is the one the theory inherits from $6d$, under which the gauge bifundamentals have R-charge $0$, the flippers R-charge $2$, and the rest R-charge $1$. Besides the $SU(6)$ and $SU(2)$ global symmetries there is now only one non-anomalous $U(1)$ corresponding to $U(1)_m$. Its charges are represented in the figure by fugacities.}
}

\

As an example consider the case of flux half generated by connecting the two punctures of the basic tube. The quiver diagram of this model is shown in Fig. 15. Most of the dynamical properties are similar to the previous model save for the loss of $U(1)_y$. Particularly, the superconformal R-symmetry and the dimension of the operators are the same as there was no mixing with $U(1)_y$. Thus, it is possible that this theory also goes to an interacting fixed point.

\

From the $6d$ perspective we expect a $U(1)\times E_6$ global symmetry at some point on the conformal manifold. Again to test this we evaluate the superconformal index. We shall again employ the non-superconformal R-symmetry $U(1)^{'}_R$. We indeed find that the index forms characters of $E_6$, at least to the order evaluated where it reads: 

\eqn\insuoe{\eqalign{ {\cal I} &= 1 + \frac{3}{m^6} (p q)^{\frac{1}{3}} + \frac{1}{m^4} \chi[{\bf \overline{27}}] (p q)^{\frac{5}{9}} + \frac{3}{m^6} (p q)^{\frac{1}{3}}(p+q)\cr & + (\frac{6}{m^{12}}+m^6) (p q)^{\frac{2}{3}} + \frac{1}{m^2} \chi[{\bf 27}] (p q)^{\frac{7}{9}} + \frac{3}{m^{10}} \chi[{\bf \overline{27}}] (p q)^{\frac{8}{9}} \cr & + p q \frac{10}{m^{18}} + ...
}}
Here the characters of $E_6$ are given by the $SU(2)\times SU(6)$ subgroup as in \sodef.

\

\noindent{\bf \it $E_7'\times U(1)$}

We can also construct a theory which is obtained by compactification on sphere with two punctures and flux preserving $E_7$ but a different embedding than the one considered above. This is depicted in Fig. 10 (d). The two models are related by Seiberg duality \SeibergPQ\ once we build tori out of them. In terms of the flux basis this tube has the flux $(-\frac{1}{4}; \frac{1}{8} , \frac{1}{8} , \frac{1}{8} , \frac{1}{8} , \frac{1}{8} , 
\frac{1}{8} , -\frac{3}{8} , -\frac{3}{8} )$ in the overcomplete basis and $( 0 , 0 , 0 , 0 , 0 , 0 , -1 , -1 )$ in the complete basis. The last two flux values are for the $U(1)$'s associated with the $SU(2)$ global symmetry.

Two of the four gauge nodes have three flavors and thus performing a Seiberg duality on them we trade those with the fifteen gauge invariant mesonic and baryonic operators. The theory is then equivalent to torus with one unit of flux preserving $E_7\times U(1)$. 

\

\

\centerline{\figscale{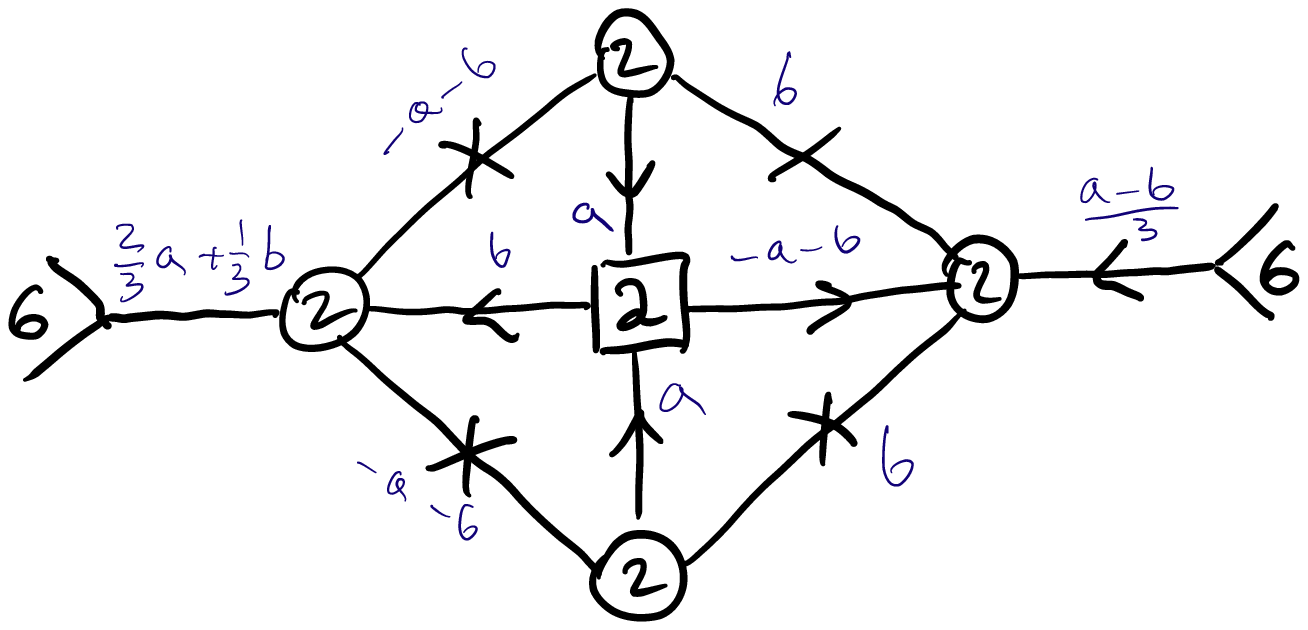}{2.1in}}
\medskip\centerline{\vbox{
\baselineskip12pt\advance\hsize by -1truein
\noindent\footnotefont{\bf Fig.~16:} The $E_7'\times  U(1)$ model from torus with one unit of flux.  A Seiberg duality of one of the nodes with three flavors brings us to the $E_7\times  U(1)$ model we considered in previous section.}
} 

This sphere with punctures is useful as it allows us to construct different models and perform many checks of the proposed correspondence between compactifications and four dimensional field theories.

\

\subsec{Rank one E-string on a torus: $E_8\to$ $G\times U(1)\times U(1)$}

We can construct theories with flux for more than one $U(1)$ by combining different spheres together. The resulting theories depend on the type of theories connected and on how these are connected. For example when connecting different tubes we have the freedom of choosing how the different global symmetries are embedded in one another. In fact we can even make non-trivial theories by connecting the same tube but with a non-trivial identification of the global symmetries, which we can describe by a permutation of the $U(1)$'s inside $SU(  8    )$. As there are fluxes associated with these $U(1)$'s, when connecting surfaces this way the total flux on the resulting surface will change. 

Clearly the possible theories one can build in this way is considerable, and we shall not examine all of them in detail. Instead we shall show some examples where we choose two tubes, and some way of connecting them, and study the anomalies of the resulting theories. The aim here is to show that these agree with the $6d$ predictions, which then serves as a consistency check on our proposal. Naturally one can study more complicated models, which can be used to realize other compactification types. Alternatively one can try to build equivalent surfaces in different ways, which are then expected to give dual theories. It may then be interesting to see if any new dualities arise in this way. We reserve these issues for future work.    

There is one subtlety in this construction regarding the central fluxes, which exist in all of these tubes. If present these lead to a breakdown of part of the global symmetry once the tube is closed. As a result, if one wishes to preserve the global symmetry, one must connect tubes with integer flux. Note that this is also true for tubes connected with a permutation of the $SU(  8    )$, as this may change the central flux element. As a result two identical half-flux tubes connected in this more general way may still carry non-trivial center flux.

\

\noindent{\bf $E_6\times U(1)\times U(1)$}

Let us consider gluing $2m$ copies of the $E_7$ sphere to $2n$ ones of the $E_6\times SU(2)$ theory.  We depict an example in Figure 17.
To glue the two types of spheres we split the $SU(  8    )$ of the $E_7$ trinion  to $SU(6)\times SU(2)\times U(1)$. The theory has manifestly $SU(2)\times SU(6)\times U(1)_b\times U(1)_a$. For all values of flux the symmetry group enhances to $E_6\times U(1)\times U(1)$. 

\

\centerline{\figscale{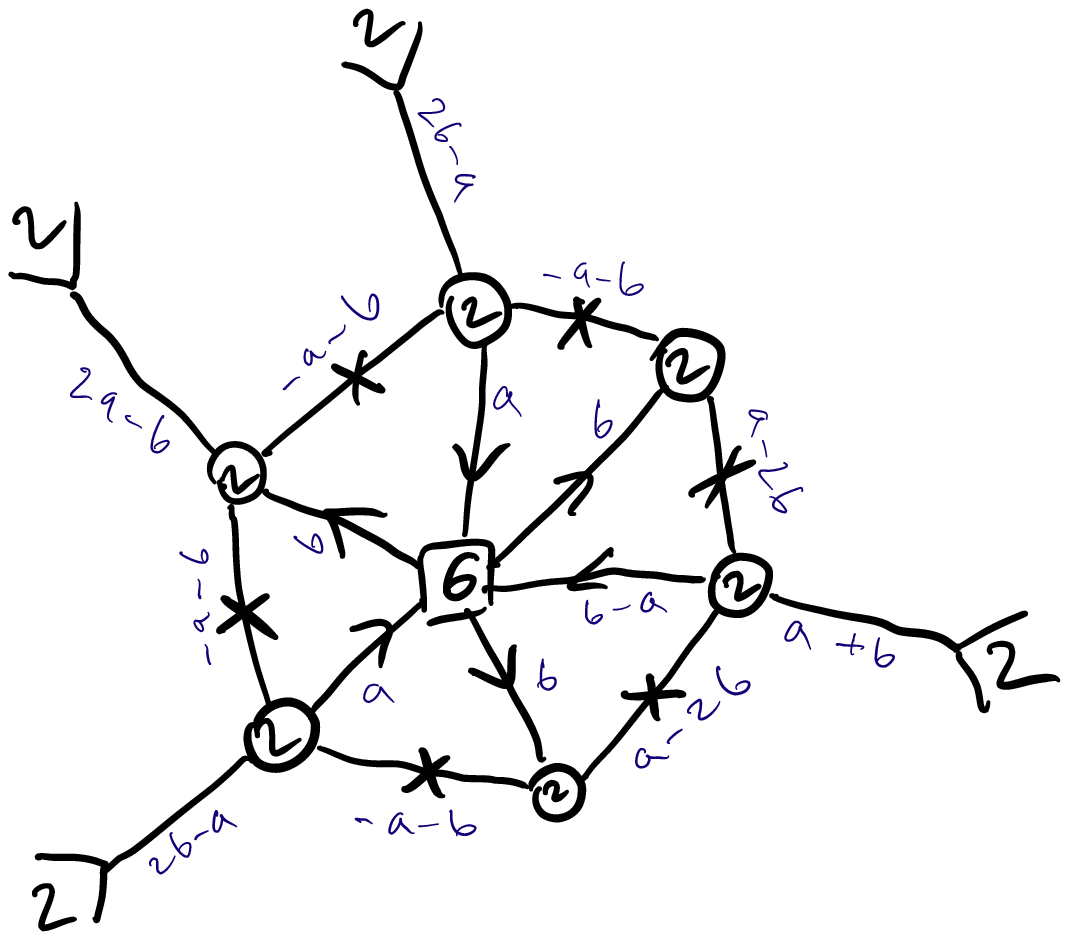}{2.1in}}
\medskip\centerline{\vbox{
\baselineskip12pt\advance\hsize by -1truein
\noindent\footnotefont{\bf Fig.~17:} Two spheres of $E_7\times  U(1)$ combined with two $E_6\times SU(2)$ spheres. The chiral fields are weighed by $q_a a+q_b b$ with $q_a$ and $q_b$ charges under $U(1)_a$ and $U(1)_b$ symmetries. The quiver is to be imagined as drawn on a sphere with the $SU(2)$ flavor node depicted by incomplete square located at infinity.}
}
The anomalies can be computed to give,

\eqn\exeess{
c=\frac52\sqrt{m^2+3m n+3n^2}\,,\qquad\, a=2\sqrt{m^2+3mn+3 n^2}\,.
} The mixing is given as,

\eqn\mieessu{
R=R'+\frac{-m-n}{3 \sqrt{m^2+3 m n+3 n^2}}q_a-\frac{m+2 n}{3 \sqrt{m^2+3 n m+3 n^2}}q_b\,.
} There are several cases with enhanced symmetry. Obviously, $n=0$ is $E_7$ and $m=0$ is $E_6\times SU(2)$. Take $m=-n$ or $m=-2n$ and we get $E_7$. Note that the square root multiplying the anomalies can be rewritten in a diagonal basis of fluxes $\sqrt{(\frac{m}2)^2+3(n+\frac{m}2)^2}$. 

From the $6d$ view point we compactify the E-string on a torus with flux $(-m - \frac{3n}{2}; \frac{3n}{4} , \frac{3n}{4} , -\frac{n}{4} , -\frac{n}{4} , -\frac{n}{4} , -\frac{n}{4} , -\frac{n}{4} , -\frac{n}{4} )$ in the overcomplete basis and $( -m , -m , -m-2n , -m-2n , -m-2n , -m-2n , -m-2n , -m-2n )$ in the complete basis. We now see that,

\eqn\retto{\eqalign{
\sum_i \xi_i z^2_i = & (m + \frac{3n}{2})^2 + (\frac{3n}{4})^2 + 3 (\frac{n}{4})^2 = \cr  & \frac{1}{8} (2 m^2 + 6 (m+2n)^2) = m^2+3mn+3 n^2 .}
} 
Using this together with \acgeni\ we recover \exeess. 

\

\noindent{\bf $SO(12)\times U(1)\times U(1)$}

Let us consider gluing $2m$ copies of the $E_7$ sphere to $2n$ ones of the $SO(14)$ theory.  We depict an example in Fig. 18.
To glue the two types of spheres we split the $SU(  8    )$ of the $E_7$ trinion  to $SU(4)\times SU(4)\times U(1)$. The theory has manifestly $SU(4)\times SU(4)\times U(1)_b\times U(1)_a$. For all values of flux the symmetry group enhances to $SO(12)\times U(1)\times U(1)$. 

\

\centerline{\figscale{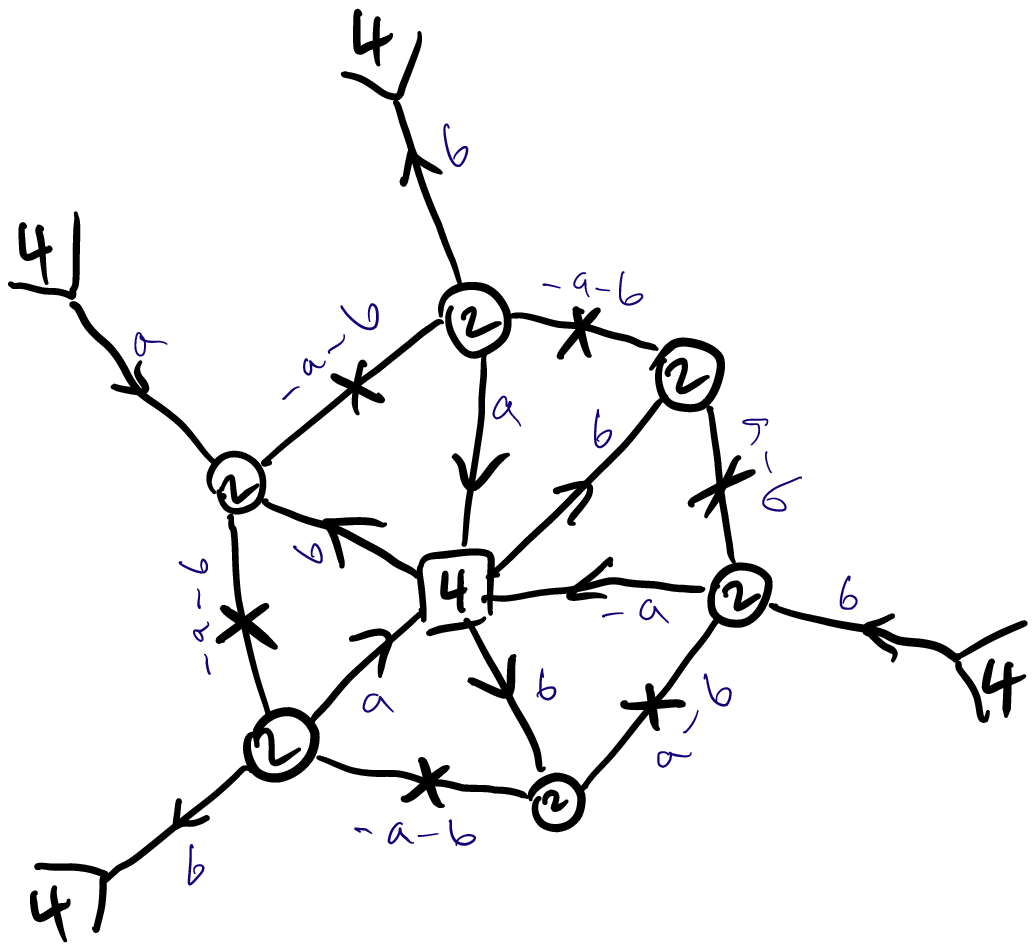}{2.1in}}
\medskip\centerline{\vbox{
\baselineskip12pt\advance\hsize by -1truein
\noindent\footnotefont{\bf Fig.~18:} Two spheres of $E_7\times  U(1)$ combined with two $SO(14)\times  U(1)$ spheres. The chiral fields are weighed by $q_a a+q_b b$ with $q_a$ and $q_b$ charges under $U(1)_a$ and $U(1)_b$ symmetries.}
}

\

 The anomalies can be computed to give,

\eqn\exeeddwss{
c=\frac52\sqrt{m^2+2m n+2n^2}\,,\qquad\, a=2\sqrt{m^2+2mn+2 n^2}\,.
} The mixing is given as,

\eqn\mieddwwessu{
R=R'-\frac{m}{3 \sqrt{m^2+2 n m+2 n^2}}a_a+ \frac{-m-2 n}{3 \sqrt{m^2+2 n m+2 n^2}}q_b\,.
} There are several cases with enhanced symmetry. Obviously, $n=0$ is $E_7$ and $m=0$ is $SO(14)$. Take $m=-n$ we get $E_7$, or $m=-2n$ and we get $SO(14)$. Note that the square root multiplying the anomalies can be rewritten in a diagonal basis of fluxes $\sqrt{(n+m)^2+n^2}$. 

From the $6d$ view point we compactify the E-string on a torus with flux $(-m - n ; \frac{n}{2} , \frac{n}{2} , \frac{n}{2} , \frac{n}{2} , -\frac{n}{2} , -\frac{n}{2} , -\frac{n}{2} , -\frac{n}{2} )$ in the overcomplete basis and $( -m , -m , -m , -m , -m-2n , -m-2n , -m-2n , -m-2n )$ in the complete basis. We now see that,

\eqn\rettt{\eqalign{
\sum_i \xi_i z^2_i& = (m + n)^2 + 2 (\frac{n}{2})^2 + 2 (\frac{n}{2})^2 = \cr&\frac{1}{8} (4 m^2 + 4 (m+2n)^2) = m^2+2 n m+2 n^2 .}
} 
Using this together with \acgeni\ we recover \exeeddwss. 

\

\noindent{\bf \it $SO(10)\times SU(2)\times U(1)\times U(1)$ and $SO(10)\times SU(3)\times U(1)$}

We can consider combining $2m$ $SO(14)$ theories with $2n$ $E_6\times SU(2)$ ones. To glue the two types of spheres we split the $SU(6)$ of the $E_6\times SU(2)$ trinion  to $SU(4)\times SU(2)\times U(1)$, and the $SU(4)$ with the least amount of flavors to $SU(2)\times SU(2)\times U(1)$. For general choices of the flux, that is of $m$ and $n$, the symmetry expected from this model is $SO(10)\times SU(2) \times U(1)^2$. Explicitly in the Lagrangian we see $SU(4)\times SU(2)\times SU(2)\times U(1)^3$. See Fig. 19 for an example.

Tuning the fluxes, it is possible to reach values for which the $SO(10)\times SU(2) \times U(1)^2$ symmetry is enhanced to $SO(10)\times SU(3) \times U(1)$. When we combine $2m$ $SO(14)$ tubes with $2n$ $E_6$ ones the anomalies we get are,

\eqn\anegbut{
c=\frac52\sqrt{2m^2+4m n+3 n^2}\,, a=2\sqrt{3 n^2+4 n m+2 m^2}\,.
} In particular when $n=-2m$ the anomalies are of $SO(10)\times SU(3)\times U(1)$ compactification. The negative sign just indicates that we need to combine the $SO(14)$ and $E_6$ tubes with S gluing. Diagonal basis here is $\sqrt{2(n+m)^2+n^2}$.

\

\

\centerline{\figscale{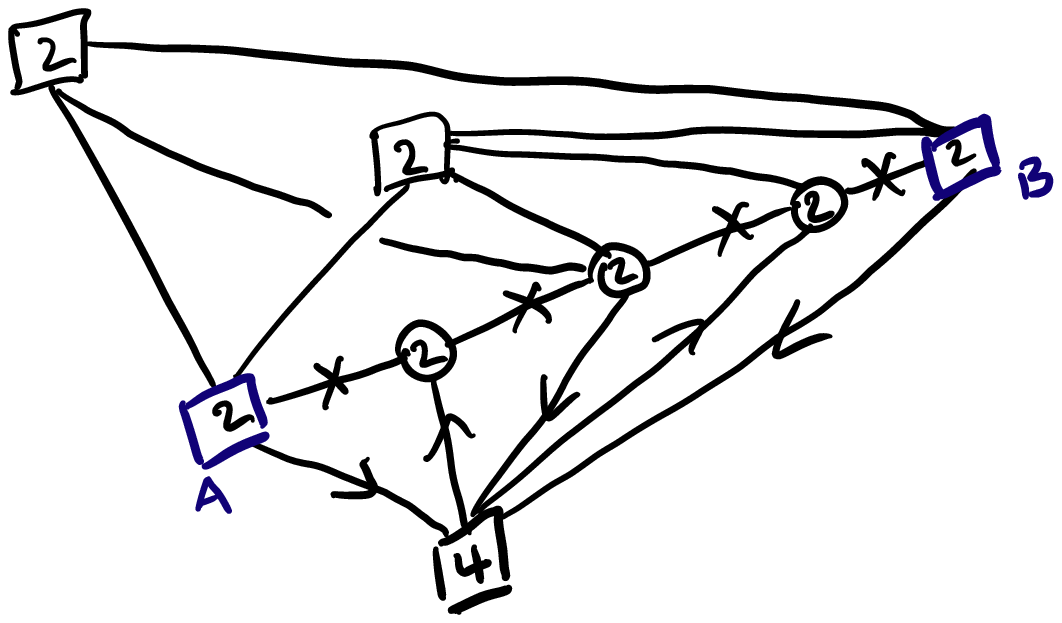}{2.1in}}
\medskip\centerline{\vbox{
\baselineskip12pt\advance\hsize by -1truein
\noindent\footnotefont{\bf Fig.~19:} Sphere with two punctures from gluing one $E_6\times  U(1)\times SU(2)$ sphere and one $SO(14)\times  U(1)$ sphere. The $A$ and $B$ $SU(2)$ symmetries correspond to punctures.}
}

From the $6d$ view point we compactify the E-string on a torus with flux $( - m - \frac{3n}{2} ; \frac{m}{2} + \frac{3n}{4} , \frac{m}{2} + \frac{3n}{4} , \frac{m}{2} - \frac{n}{4} , \frac{m}{2} - \frac{n}{4} , -\frac{m}{2} - \frac{n}{4} , -\frac{m}{2} - \frac{n}{4} , -\frac{m}{2} - \frac{n}{4} , -\frac{m}{2} - \frac{n}{4} )$ in the overcomplete basis and $( 0 , 0 , -2n , -2n , -2n-2m , -2n-2m , -2n-2m , -2m-2m )$ in the complete basis. Note that the relative orientation of the fluxes is dictated by the manner in which the global symmetry is identified between the tubes. We now see that,

\eqn\rettth{\eqalign{
\sum_i \xi_i z^2_i =& (m + \frac{3n}{2})^2 + (\frac{m}{2} + \frac{3n}{4})^2 + (\frac{m}{2} - \frac{n}{4})^2 + 2(\frac{m}{2} + \frac{n}{4})^2 = \cr   &\frac{1}{8} (2 (2n)^2 + 4 (2m+2n)^2) = 3 n^2+4 n m+2 m^2 .}
} 
Using this together with \acgeni\ we recover \anegbut.

Let us now discuss a simple model corresponding to an $SO(10)\times SU(3) \times U(1)$ compactification. We can have $n=-2$ and $m=1$. Naively we have twelve gauge groups, but because some of the gluings are S gluings, six of the gauge groups have two flavors in some duality frame and thus we get in the end only six gauge groups. 
The theory we consider is shown in Fig. 20. As usual we first consider the case without the flipping fields. Performing a-maximization we find that only the diagonal $U(1)_d = U(1)_x + U(1)_y + U(1)_z$ mixes with the R-symmetry, where the superconformal R-symmetry is: $U(1)^{SC}_R = U(1)^{6d}_R + (2-\frac{\sqrt{46}}{3}) U(1)_d$. We find that all gauge invariant operators have dimension above the unitary bound so it is plausible that this theory flows to an interacting IR SCFT.

Now we add $6$ singlets and couple them through the flipping superpotential. We find that this superpotential is relevant compared to the SCFT point so the theory will flow to a new theory in the IR. We can repeat the a-maximization for this case, where we again find that only $U(1)_d$ mixes, but now: $U(1)^{SC}_R = U(1)^{6d}_R - \frac{\sqrt{2}}{3\sqrt{3}} U(1)_d$. Using this value we obtain for the conformal anomalies,

\eqn\sosfyoe{
c=\frac52\sqrt{6}\,,\qquad \; a= 2\sqrt{6}\,.
} This agrees with the six dimensional computation noting that for $SO(10)\times SU(3)$ $\xi=6$. Also we find that all gauge invariant operators are above the unitary bound so it is again plausible that this theory flows to an interacting SCFT in the IR. Note that the singlet do not have free R-charge in the SCFT and are thus an inseparable part of it.

\

\centerline{\figscale{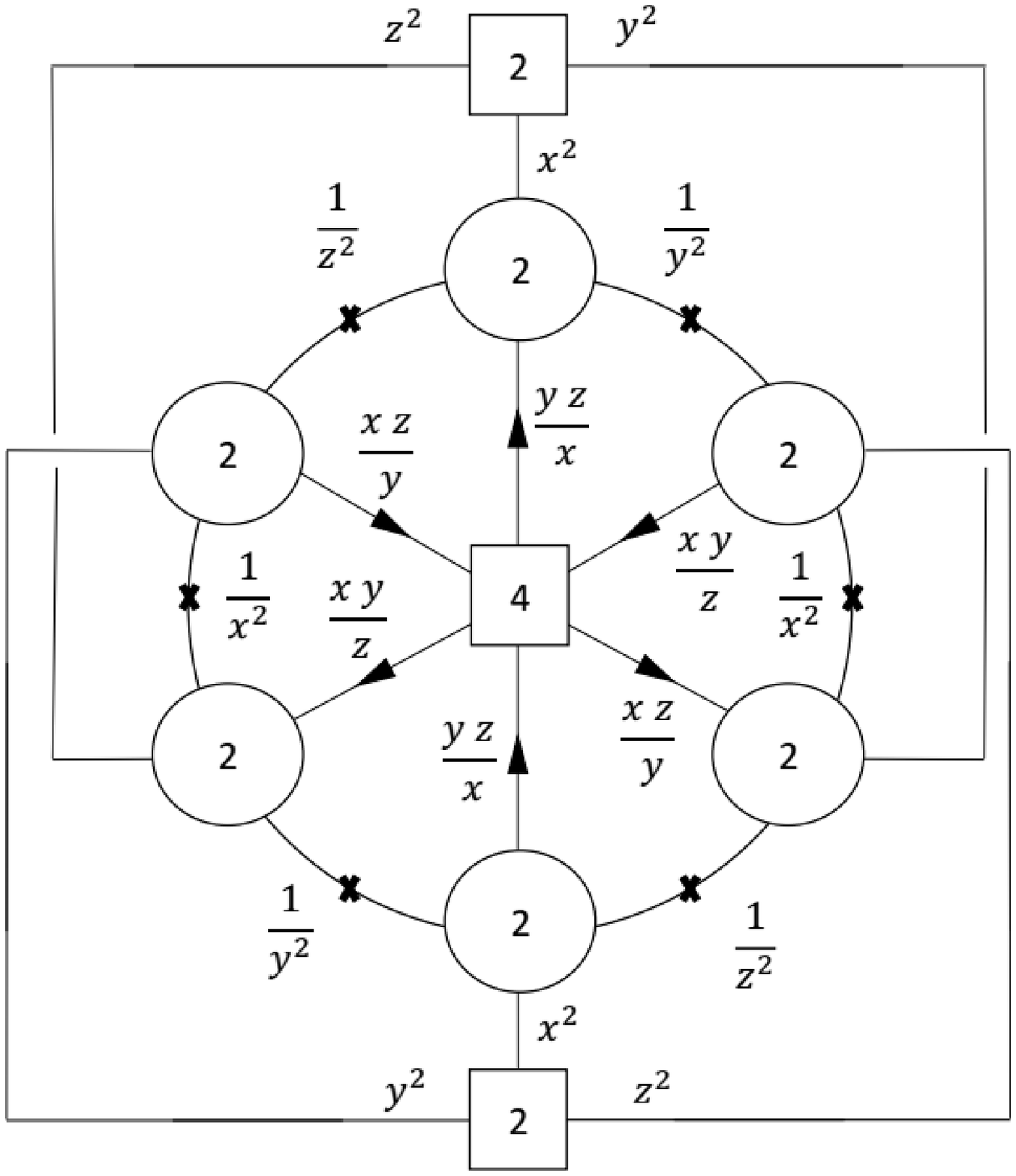}{2.1in}}
\medskip\centerline{\vbox{
\baselineskip12pt\advance\hsize by -1truein
\noindent\footnotefont{\bf Fig.~20:} The $SO(10)\times  U(1)\times SU(3)$ model from torus with one unit of flux. There is a cubic superpotential for each one of the six internal triangles, and six quartic ones for the bifundamentals involving the external $SU(2)$'s with $120^0$ semi-circles. There is a natural R-symmetry, which is the one the theory inherits from $6d$, under which the gauge bifundamentals have R-charge $0$, the flippers R-charge $2$, and the rest R-charge $1$. Besides the $SU(4)\times SU(2)\times SU(2)$ global symmetries there are also three non-anomalous $U(1)$'s which we denote as $U(1)_x, U(1)_y$ and $U(1)_z$. The charges of all the fields under these $U(1)$'s are represented by fugacities.}
}

\

The $6d$ construction suggests that this theory has an $SO(10)\times SU(3)$ global symmetry somewhere on its conformal manifold. This is definitely not visible from the Lagrangian so to test this we wish to evaluate the superconformal index. For this it is convenient to work with the non-superconformal R-symmetry: $U(1)^{'}_R = U(1)^{6d}_R - \frac{1}{3} U(1)_d$. Note that $-\frac{\sqrt{2}}{3\sqrt{3}} + \frac{1}{3} \approx 0.06$, so this R-symmetry is very close to the true SC R-symmetry. Using this R-symmetry we indeed find that the index can be written in characters of $SO(10)\times SU(3)$ at least to the order we evaluated. Particularly, the first terms in the supersymmetric index are,

\eqn\insuso{\eqalign{ {\cal I} &= 1 + 4 (x y z)^{\frac{4}{3}} \chi[{\bf 3}, {\bf 1}] (p q)^{\frac{1}{3}} + 3 x y z \chi[{\bf 1}, {\bf \overline{16}}] (p q)^{\frac{1}{2}} \cr & (2(x y z)^{\frac{2}{3}}\chi[{\bf \overline{3}}, {\bf 10}] + 10(x y z)^{\frac{8}{3}} \chi[{\bf 6}, {\bf 1}] + 6(x y z)^{\frac{8}{3}} \chi[{\bf \overline{3}}, {\bf 1}])(p q)^{\frac{2}{3}} + ...
}}
where

\eqn\sodef{\eqalign{ \chi[{\bf 3}, {\bf 1}] &= \frac{1}{(x y z)^{\frac{4}{3}}} (x^4 + y^4 + z^4) , \cr \chi[{\bf 1},{\bf 10}]  &
 = \chi[{\bf 2},{\bf 2},{\bf 1}]_{SU(2)\times SU(2)\times SU(4)} + \chi[{\bf 1},{\bf 1},{\bf 6}]_{SU(2)\times SU(2)\times SU(4)}, \cr \chi[{\bf 1},{\bf \overline{16}}]  & = \chi[{\bf 2},{\bf 1},{\bf 4}]_{SU(2)\times SU(2)\times SU(4)} + \chi[{\bf 1},{\bf 2},{\bf \overline{4}}]_{SU(2)\times SU(2)\times SU(4)}.
}}

We next note several observations regarding the index. First it indeed forms characters of $SO(10)\times SU(3)$ where $SU(2) \times SU(2) \times SU(4)$ is enhanced to $SO(10)$ while the non-diagonal combinations of $U(1)_x$, $U(1)_y$ and $U(1)_z$ combine to form the $SU(3)$. All the anomalies are consistent with the enhancement and with the $6d$ result. We note that $U(1)_d$ is identified with $- U(1)$ when we use the normalization convention of unit charge.

We can again identify some of the contributions as coming from the $6d$ conserved current multiplet of the $E_8$ global symmetry. Particularly, the contributions $4 (x y z)^{\frac{4}{3}} \chi[{\bf 3}, {\bf 1}] (p q)^{\frac{1}{3}}$, $3 x y z \chi[{\bf 1}, {\bf \overline{16}}] (p q)^{\frac{1}{2}}$ and $2(x y z)^{\frac{2}{3}}\chi[{\bf \overline{3}}, {\bf 10}]$ have the same R-charge as marginal operators under the $6d$ R-symmetry. Furthermore the representations they carry exactly match those required to complete $SU(3)\times SO(10)$ to $E_8$, at least up to the order we evaluated the index (once again we refer the reader to Appendix A for the branching rule). Their number is again exactly as expected from the reasoning of \Brz. There is one more relevant operator that should contribute at higher orders. As we did not get to order $p q$, we cannot comment on the marginal operators.

\

\noindent{\bf \it Combining $E_7\times U(1)$ and $E_7'\times U(1)$}

Consider combining $2n$ models of $E_7$ type and $2m$ of $E_7'$ kind. The symmetry for general flux is $SO(12)\times U(1)\times U(1)$.  The anomalies can be computed to give,

\eqn\exeddess{
c=\frac52\sqrt{m^2+m n+n^2}\,,\qquad\, a=2\sqrt{m^2+mn+n^2}\,.
} The mixing is given by,

\eqn\mieeddssu{
R=R'-\frac{n + m}{
 3 \sqrt{n^2 + m n + m^2}}q_a -\frac{n}{3 \sqrt{n^2 + n m + m^2}}q_b\,.
} There are several cases with enhanced symmetry. Obviously, $n=0$ is $E_7$ and $m=0$ is $E_7$. Take $m=-n$ and we get $E_7$, $m=n$  get $E_6\times SU(2)$. Note that the square root multiplying the anomalies can be rewritten in a diagonal basis of fluxes $\sqrt{3(\frac{m+n}2)^2+(\frac{m-n}2)^2}$. 

From the 6d view point we compactify the E-string on a torus with flux $(-m - \frac{n}{2} ; \frac{n}{4} , \frac{n}{4} , \frac{n}{4} , \frac{n}{4} , \frac{n}{4} , \frac{n}{4} , -\frac{3n}{4} , -\frac{3n}{4} )$ in the overcomplete basis and $( -m , -m , -m , -m , -m , -m , -m-2n , -m-2n )$ in the complete basis. We now see that,

\eqn\rettf{\eqalign{
\sum_i \xi_i z^2_i =& (m + \frac{n}{2})^2 + 3 (\frac{n}{4})^2 + (\frac{3n}{4})^2 =\cr & \frac{1}{8} (6 m^2 + 2 (m+2n)^2) = n^2 + n m + m^2.}
} 
Using this together with \acgeni\ we recover \exeddess.

\

\centerline{\figscale{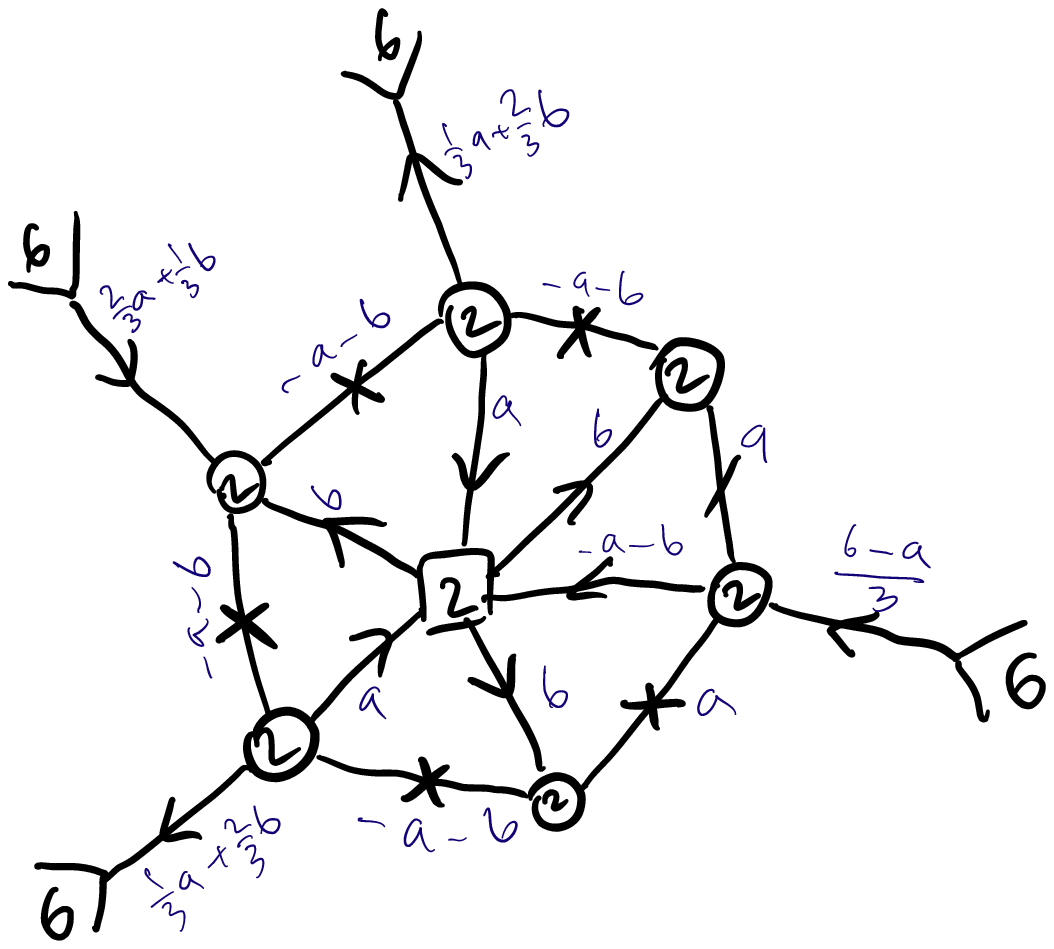}{2.1in}}
\medskip\centerline{\vbox{
\baselineskip12pt\advance\hsize by -1truein
\noindent\footnotefont{\bf Fig.~21:} Two spheres of $E_7\times  U(1)$ combined with two $E_7'\times  U(1)$ spheres. The chiral fields are weighed by $q_a a+q_b b$ with $q_a$ and $q_b$ charges under $U(1)_a$ and $U(1)_b$ symmetries.}
}

\

\noindent{\bf \it Combining $E_6\times SU(2)\times U(1)$ and $E_7\times U(1)$ to a tube for $SU(  8    )\times U(1)$}

Using the domain wall picture in section 4, combining the tube for $E_6\times SU(2)$ and the one for $E_7$ we can obtain $SU(  8    )$ tube. The resulting theory is drawn in Fig. 22.

When we glue two copies of the theory with two punctures to obtain a torus the superconformal R charges are,

\eqn\dtsecgy{
e, b \to 1\,,\qquad  c, d, o,i\to \frac13\,,\qquad r\to \,\frac43\, ,\qquad h, a, f,l\, \to\frac23\,   ,  \qquad n, k, g\to \frac12\,,  \qquad j\to \frac56\,.}  The anomaly of gluing $2z$ copies of the model in Fig. 22 are,

\eqn\andhyuieey{
c=5 z\,,\qquad a=4 z\,.}

\

\

\centerline{\figscale{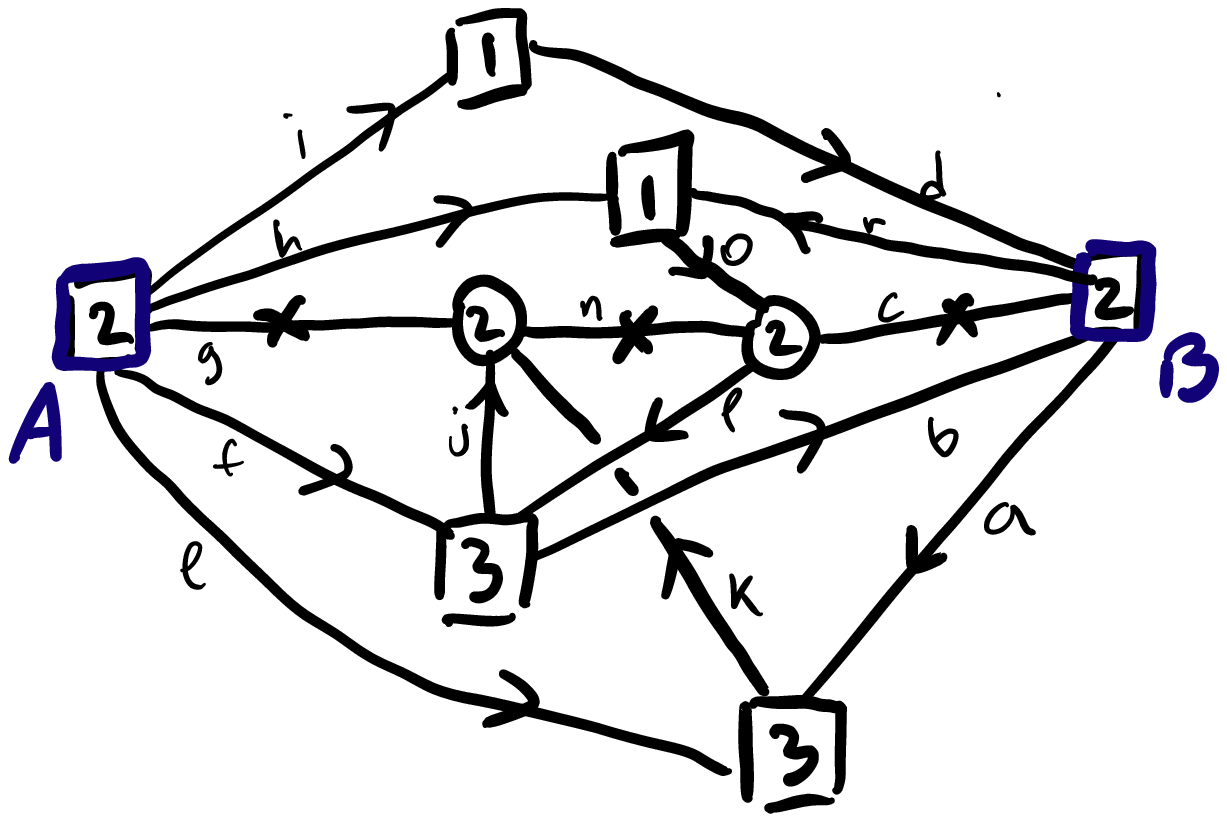}{2.3in}}
\medskip\centerline{\vbox{
\baselineskip12pt\advance\hsize by -1truein
\noindent\footnotefont{\bf Fig.~22:} The $SU(  8    )\times  U(1)$ model from tube and half unit of flux. The  superpotentials correspond to faces with one of the vertices being $1,3$. One has here fifteen fields, eight superpotentials, and two gauge nodes. Together with the rank four symmetry of the two $SU(3)$ symmetries, this gives rank nine symmetry. When  we glue even number of copies to form a torus one symmetry is broken by anomalies and we are left with rank eight symmetry. 
}} 

\

We can next evaluate the index. The model has two $SU(3)$ global symmetry groups as well as four non-anomalous $U(1)$'s. The model including the charges under the various symmetries is shown in Fig.  23. The index is given by:

\eqn\insueight{\eqalign{ {\cal I} &= 1 + 3 N^{3} \chi[{\bf \overline{8}}] (p q)^{\frac{1}{2}} + 2 N^{2} \chi[{\bf 28}] (p q)^{\frac{2}{3}} \cr & + N\chi[{\bf \overline{56}}](p q)^{\frac{5}{6}} + ...
}}
where

\eqn\sodeff{\eqalign{ \chi[{\bf \overline{8}}] &= \frac{m^{\frac{3}{2}} w^6}{y^{\frac{9}{2}} z^{\frac{9}{4}}} +  \frac{m^{\frac{3}{2}} y^{\frac{3}{2}}}{w^6 z^{\frac{9}{4}}} + \frac{w^2}{m^{\frac{5}{2}} y^{\frac{1}{2}} z^{\frac{1}{4}}} \chi[{\bf \overline{3}},{\bf 1}] + \frac{m^{\frac{3}{2}} y^{\frac{3}{2}} z^{\frac{7}{4}}}{w^2} \chi[{\bf 1},{\bf \overline{3}}], \cr N  &
 = m^{\frac{3}{2}} y^{\frac{3}{2}} z^{\frac{3}{4}}.
}}

Several things are apparent from \insueight. First it indeed forms characters of $SU(  8    )$ as expected. Second the operators appearing in the index are exactly the ones completing $SU(  8    )$ to $E_8$ (see the branching rule in  Appendix A). Also they all have the same R-charge as the $SU(  8    )$ adjoint marginal operators under the 6d R-symmetry. Finally we note that the number of such operators is as expected from the formula of \Brz. 

We can write the flux vector associated with this theory. It is generated from the $E_6\times SU(2)$ tube and an $E_7$ tube in a complicated manner. Particularly we consider splitting the $8$ flavors into two pairs of fours and complex conjugating one of the pairs. This is an inner automorphism from the $E_7$ point of view and results in a tube that still describe an $E_7$ embedding though a slightly different one. In the flux basis we have chosen it will be given by $(-\frac{1}{2}, -\frac{1}{2}, -\frac{1}{2}, -\frac{1}{2}, \frac{1}{2}, \frac{1}{2}, \frac{1}{2}, \frac{1}{2})$ in the complete basis and $(0; -\frac{1}{4}, -\frac{1}{4}, -\frac{1}{4}, -\frac{1}{4}, \frac{1}{4} , \frac{1}{4}, \frac{1}{4}, \frac{1}{4})$ in the overcomplete one. Note that this also reversed the $U(1)_t$ charge of four fields.

\

\centerline{\figscale{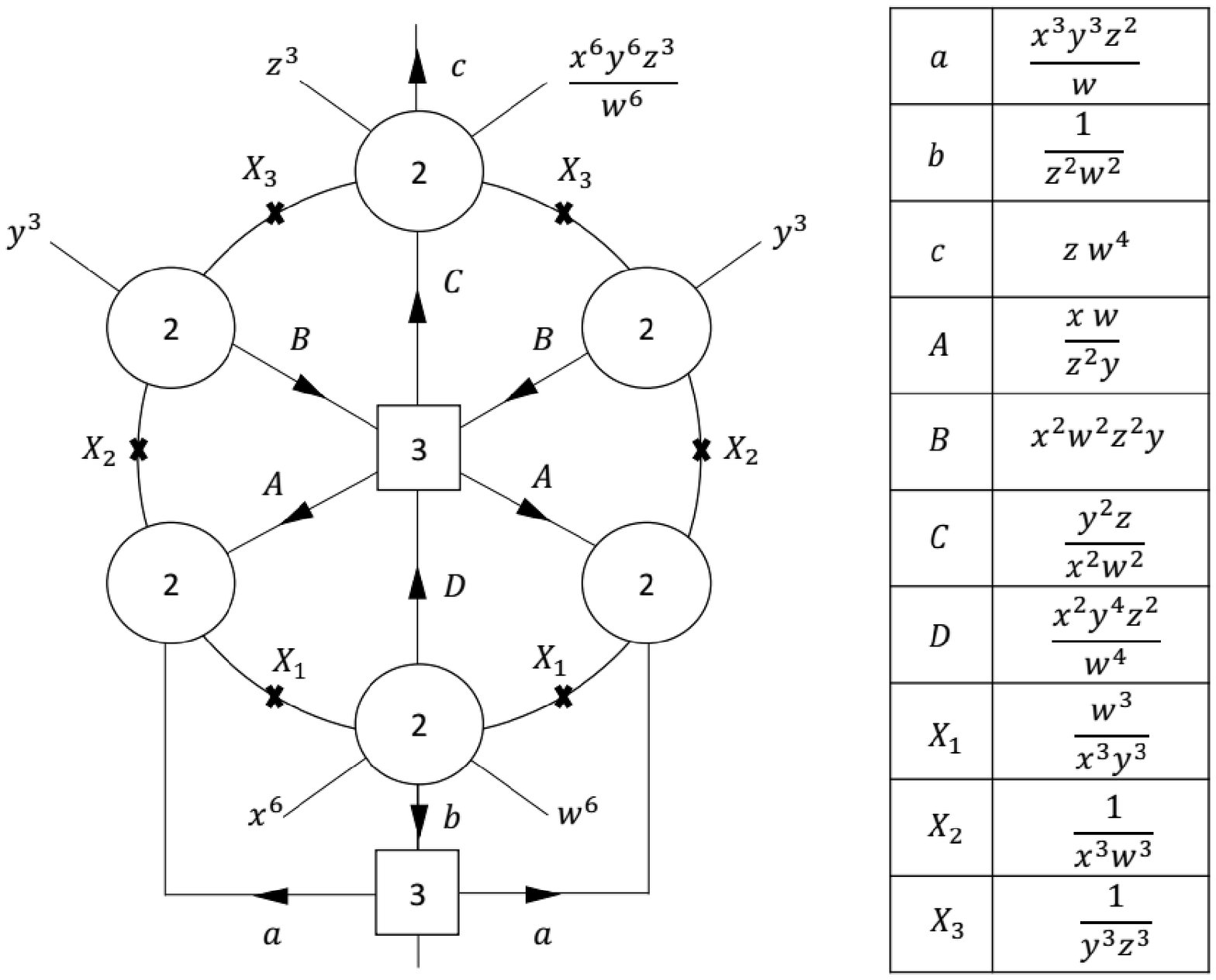}{3.5in}}
\medskip\centerline{\vbox{
\baselineskip12pt\advance\hsize by -1truein
\noindent\footnotefont{\bf Fig.~23:} The $SU(  8    )\times  U(1)$ model we get by connecting the two tubes in Fig. 22. Also shown are the charges of the fields
 summarized using fugacities.  The lines which do not end on boxes or 
 circles correspond to fundemantal fields of one gauge group.
}}

Now when we combine the two tubes four flavors have the same $U(1)_t$ charge and are glued using $\Phi$ gluing while the other four have opposite $U(1)_t$ charge and are glued using S gluing. This results in the tube in figure 22. It is now straightforward to write the flux by combining the fluxes of the two tubes paying special attention to which flavor is connected to which:

\eqn\sueflux{\eqalign{ &F_c = (-1, -1, -1, 0, 0, -1, -1, -1) + (-\frac{1}{2}, -\frac{1}{2}, -\frac{1}{2}, -\frac{1}{2}, \frac{1}{2}, \frac{1}{2}, \frac{1}{2}, \frac{1}{2}) =\cr &\qquad\qquad\;\, (-\frac{3}{2}, -\frac{3}{2}, -\frac{3}{2}, -\frac{1}{2}, \frac{1}{2}, -\frac{1}{2}, -\frac{1}{2},-\frac{1}{2}), \cr  &
 F_{oc}  
 =   (-\frac{3}{4}; -\frac{1}{8}, -\frac{1}{8}, -\frac{1}{8}, \frac{3}{8}, \frac{3}{8} , -\frac{1}{8}, -\frac{1}{8}, -\frac{1}{8}) + (0; -\frac{1}{4}, -\frac{1}{4}, -\frac{1}{4}, -\frac{1}{4}, \frac{1}{4} , \frac{1}{4}, \frac{1}{4}, \frac{1}{4}) =
  \cr &\qquad\;\qquad\, (-\frac{3}{4}; -\frac{3}{8}, -\frac{3}{8}, -\frac{3}{8}, \frac{1}{8}, \frac{5}{8} , \frac{1}{8}, \frac{1}{8}, \frac{1}{8}).
}}  

The flux of the first three is associate with the $SU(3)$ with the more flavors while the fifth one is associated with the $U(1)$ with the least number of flavors. 

Alternatively, the flux in the complete basis can be read from the two domain walls we connected. One is the $E_6\times SU(2)$ preserving domain wall and the other is the $E_7$ preserving domain wall with the flux chosen to be $(-\frac{1}{2}, -\frac{1}{2}, -\frac{1}{2}, -\frac{1}{2}, \frac{1}{2}, \frac{1}{2}, \frac{1}{2}, \frac{1}{2})$. This is the required assignment to get the tube in figure 22. Summing the fluxes, while taking due of care of how the symmetries are identified between the two tubes, reproduces \sueflux. 

\

\

\newsec{Closing punctures}

\  

We have discussed theories we can attribute to a sphere with two punctures. We will now turn to deriving the procedure of closing a puncture, starting from a theory corresponding to a surface with at least one  puncture and some value of flux to obtain a theory with one  puncture  less and, possibly, a different value of flux. In general we can achieve the removal of a puncture by turning on a vacuum expectation value for an operator charged under the symmetry associated to the puncture. The natural operators which are charged under puncture symmetries are the eight operators $M_j$ we have defined in the previous section. From dealing with compactifications of $M5$ branes on orbifolds we know that the procedure of closing a puncture might also involve adding chiral fields to the model flipping certain operators. We will proceed to derive the exact map between removal of punctures and vacuum expectation values and addition of fields by figuring out what one has to do in the case of a sphere with two punctures to obtain a sphere with one puncture. In this case we know the field theories for which we can trigger the RG flow and know the anomalies from six dimensional arguments.

We start from the $E_7$ tube with two punctures and half a unit of flux. Let us discuss this theory at the level of the index as it encodes very compactly all the fields and charges. The index of the theory is,

\eqn\tutye{
{\cal I}(z,u)=\Gamma_e(p q t^4)\left(\prod_{j=1}^8 \Gamma_e((q p)^{\frac12} t a_j z^{\pm1})\Gamma_e((q p)^{\frac12} t \frac1{a_j} u^{\pm1})\right) \Gamma_e(\frac{1}{t^2} u^{\pm1}z^{\pm1})\,.} We have the contribution of the operators $M_j$ for the two punctures in the brackets on the right hand side. Consider closing one of the punctures by giving a vacuum expectation value say to operator $M_i$. We need to choose which component of the $SU(2)$ fundamental representation obtains a vacuum expectation value and with no loss of generality we take,

\eqn\serui{
 u =\frac{a_i}{( q p)^{\frac12} t}\,.} Doing so, and getting rid of the goldstone modes, the index of the theory in the IR is as follows,

\eqn\indut{\eqalign{&
{\cal I} = \Gamma_e(p q t^4)\prod_{j\neq i}\Gamma_e(a_i/a_j)\prod_{j=1}^8 \Gamma_e((q p)^{\frac12} t a_j z^{\pm1}) \Gamma_e(q p t^2\frac1{a_i a_j})\cr
&\qquad \Gamma_e((q p)^{\frac12} \frac{z^{\pm1}}{t a_i}) \Gamma_e(\frac{a_i z^{\pm1}}{(q p)^{\frac12} t^3})\,.}
}  

We need to add chiral flips to flip some of the operators.  Otherwise the anomalies will not match the prediction from six dimensions. We find that  flipping the fields with contribution $\Gamma_e(a_i/a_j)$ and $\Gamma_e(q p t^2\frac1{a_i^2})$, that is giving them mass but not breaking any symmetry does the needed adjustment. All the flipped operators are components of $M_j$. The general prescription is thus to give vacuum expectation value to $M_i^+$ ($M_i^-$) and flip $M_j^+$ ($M_j^-$) and $M_i^-$ ($M_i^+$). In the theory we consider, $M_j$ are fields and flipping them is the same as making them massive, however in more general theories $M_j$ might be composite operators and flipping is then performed by adding chiral fields with linear couplings  to the operators. 

We claim that the resulting theory is a sphere with a single puncture and some value of flux. We could have also chosen to close the other puncture, which differs from the first by its color, that is the representation under $SU(  8    )$. The resulting theory can be generated by acting with complex conjugation on the $SU(  8    )$ fugacities but leaving $U(1)_t$ unchanged. In the overcomplete basis we will associate it with the same $U(1)_t$ flux but opposite $U(1)_{a_i}$ fluxes. 

 To figure out the value of the flux we consider taking two spheres and a collection of free tubes and combine them to form a sphere with flux. Then by matching anomalies we can infer the value of the flux for the cups. We find that the anomalies match, for an arbitrary number of $E_7$ tubes, if we associate the flux $( -\frac{3}{4} ; \frac{7}{8} , -\frac{1}{8} , -\frac{1}{8} , -\frac{1}{8} , -\frac{1}{8} , -\frac{1}{8} , -\frac{1}{8} , -\frac{1}{8} )$, where we have conveniently set $i=1$. Alternatively using the complete basis, the flux is $( 1 , -1 , -1 , -1 , -1 , -1 , -1 , -1 )$.

From this we conclude the following. Closing a puncture with sign $s$ and color $c$ by giving a vev to a meson charged under $U(1)_{a_i}$ shifts the flux by: $F_t \rightarrow F_t - s \frac{1}{4}, F_{a_i} \rightarrow F_{a_i} - c \frac{7}{8},  F_{a_j} \rightarrow F_{a_j} + c \frac{1}{8}$, where $j \neq i$ and we are using the overcomplete basis. We have also defined $c$ to be positive when $M$ is in the fundamental. We  note that the two colors that appear in the $E_7$ tube are not the only ones possible. In the other tubes, corresponding to fluxes in different $U(1)$'s, the colors of the two punctures again differ and not just by complex conjugation. We can always redefine the $SU(8)$ so that one color be identical to one of the colors in the $E_7$ tube, but then the other one will differ from both the colors presented here. When closing one of these types of punctures we expect the fluxes to be shifted differently. It may be interesting to understand  in more detail what are the possible colors and how the flux shift is determined by the color. We shall not pursue this here.

\

\newsec{Interacting trinions} 

\

We have matched some of the compactifications on genus one with no punctures and genus zero with less than or equal to two punctures with four dimensional theories. We will now proceed to figure out what is the model corresponding to a sphere with three punctures. Such a model with the two and one punctured spheres of the previous section will allow us to find field theoretical constructions for compactifications on any surface with any flux. 

We start by considering the anomaly computed in six dimensions for a surface with genus ${ g}$  and some value of flux. The anomalies for the R symmetry do not depend on the flux so we consider those. We assume that the surface is to be built from trinions with $\Phi$ gluing. This allows us to compute anomalies involving R symmetry by decomposing the surface into tubes and three punctured spheres. The contribution of the $\Phi$ gluing to the anomaly is given by,

\eqn\cgyu{
Tr R = 0 \times 2\times 8 + 3=   3\,,\,\qquad Tr R^3 =0^3\times 8\times 2+3=  3\,.}  The first terms come from the fields $\Phi$ and the last term from the $SU(2)$ gluinos.
The anomaly of genus ${ g}$ surface is obtained from six dimensions to  give us,

\eqn\sdyuoi{\eqalign{ & Tr R = (1-{ g}) 11 = (2{ g}-2) (Tr R)_{trinion}+(3{ g}-3)(3)\,,\cr & TrR^3 = ({ g}-1) 13 = (2{ g}-2) (Tr R^3)_{trinion} +(3{ g}-3)(3)\,.}
} From here we have that $(TrR^3)_{trinion}= 2$ and $(Tr R)_{trinion} = -10 $.  The question is then whether we can identify field theories in four dimensions with such an anomaly.

Serendipitously, we know such models. When one considers the trinions of compactifications of two $M5$ branes probing a $\Z_2$  singularity one obtains exactly such anomalies. There are an infinite number of trinions of that type differing by flux and types of punctures. However, all of them have the above  anomalies involving R symmetries, and these are good candidates to be related to a trinion of E-string compactification.  One can engineer these models as follows \refs{\RazamatDPL,\GaiottoUSA}. We start with a compactification of two M5 branes probing $\Z_2$ singularity and put these on a sphere with two maximal, having $SU(2)^2$ symmetry, and two minimal, having $U(1)$ symmetry, punctures. These four punctured sphere has a Lagrangian description as an $SU(2)^2$ gauge theory which happens to be  identical to Fig. 3 but  without the flip fields.  Then one can look at duality frames, geometrically pairs of pants decompositions,  where two $U(1)$ punctures sit together. We have a choice of which fluxes we associate to the two pairs of pants and which color of puncture runs in the tube connecting them (note that these fluxes and colors are for the compactification of two M5 branes on $\Z_2$ singularity). The theory corresponding to the pair of pants with the two maximal punctures is a trinion of that setup with the choices in the decomposition giving rise to three punctured spheres with different fluxes and colors.
A description in terms of a Lagrangian can be obtained for the trinions by using the duality and exploiting symmetries appearing at strong coupling of the conformal manifold of the $SU(2)^2$ model.
 This procedure of deriving the trinion is analogous to the one deriving the trinion, the MN $E_6$ theory \MinahanFG, for compactifications of three M5 branes~~\refs{\ArgyresCN}. The derivation of the Lagrangian is similar to the derivation of Lagrangian for the $E_6$ MN model \RazamatL. 

 There is one theory in this list of theories which stands out, and it was denoted $T_A$ in \RazamatDPL (see also \GaiottoUSA).  The construction of  the $T_A$ model is described in  great detail in \RazamatDPL\ and we refer the reader there for all the properties of this model. Here in Appendix D we detail the supersymmetric index of (a deformation to be relevant in what follows)  the model from which a Lagrangian description can be read off.   Such a Lagrangian description gives us, for example, all the information of the model which does not depend on the coupling constants. This includes anomalies and supersymmetric partition functions. 
The model $T_A$ has global symmetry $SU(2)^3 \times SO(  8    )\times U(1)^2$.  The index of this was computed to have the form \RazamatDPL\ ,

\eqn\interu{\eqalign{&
{\cal I}=1 +( \frac2{a^4 t^2}+\frac{a^4}{t^2} +       t({\bf 2}_A{\bf 8}_v+{\bf 2}_C{\bf 8}_c+ {\bf 2}_B {\bf 8}_s)+t a^4 {\bf 2}_A{\bf 2}_C{\bf 2}_B+\frac1{a^4}{\bf 28}) p q \cr &\;\;\,\qquad +
 (-{\bf 28}-{\bf 3}_A-{\bf 3}_C-{\bf 3}_B-1-1) p q+\cdots\,.}} We have only written terms which are relevant using the superconformal R-symmetry.  The reason this model is special is that we can clearly see that there are eight operators in the fundamental representation for the three $SU(2)$ symmetries which are in the various ${\bf 8}$ reps of $SO(  8    )$. We will soon see that a trinion of the E-string is to be identified with the relevant deformation of this model switching on the operator with weight $\frac{a^4}{t^2}$ and that the trinion has flux $\frac34$ for the $U(1)_t$ symmetry.  The apparent symmetry of the theory in addition to symmetries coming from punctures is $SO(8)\times U(1)$. 
  The symmetry preserved by each puncture  is $SU(  8    )\times U(1)$, however we have different colors of the puncture (different embeddings of $SU(8)\times U(1)$ in $E_8$), as is apparent from the different representations of the $M_i$ operators,  and also have fractional value of flux. All in all such effects might lead for the symmetry to  be broken to the $SO(8)\times U(1)$  we explicitly see in the index.  The precise mechanism of the breaking of he symmetry should be related to the precise classification and definition of the notion of color and the related center fluxes which we do not pursue here.\foot{See \RazamatDPL\ for similar issues when considering compactifications of two M5 branes on $\Z_2$ singularity.}
  When the trinions are glued to form closed Riemann surfaces with integer value of flux we expect the symmetry to be enhanced to have rank eight.
 We will denote the trinion after the relevant deformation as $T_e$. We refer the reader to Appendix D for more essential details about $T_e$ in particular the expression of the index from which the field theoretic construction can be read of and the anomalies computed.

We can compare the anomalies of $T_e$ theory with the anomalies from the 6d E-string theory compactified on a three punctured sphere. As discussed in section 2.6, the 4d anomaly from the 6d theory is obtained by the sum of geometric and inflow contributions. Regarding the flux $\frac{3}{4}$ for the $U(1)_t$ symmetry, we compute the anomalies of three punctured sphere as
\eqn\teano{\eqalign{
	&Tr(U(1)_R^3) = (\frac{13}{2})_{geo} + 3\!\times\! (-\frac{3}{2})_{inf} = 2 \ , \quad Tr(U(1)_R) = (-\frac{11}{2})_{geo} +3\!\times\! (-\frac{3}{2})_{inf}= -10 \,, \cr
	&Tr(U(1)_F^3) = (9)_{geo} + 3\!\times\!(-1)_{inf} = 6 \ , \quad Tr(U(1)_F) = (9)_{geo} + 3\!\times\!(-4)_{inf} =-3\,, \cr
	&Tr(U(1)_F^2U(1)_R) = (-2)_{geo} = -2 \ , \quad Tr(U(1)_F U(1)_R^2) = (-3)_{geo} = -3 \ .
}}
These results are in perfect agreement with the anomalies of $T_e$ theory, where $U(1)_F = -\frac{1}{2} U(1)_t$.

Let us consider combining $2{ g}-2$ theories $T_e$ with the $S$ -- gluing to form closed Riemann surfaces. As we are using the $S$ -- gluing with even number of trinions the flux of this model is vanishing. Computing the index we obtain that it is given by the following for general genus ${ g}$,

\eqn\dfru{
{\cal I} = 1+ ({\bf 248}({ g}-1)+3{ g}-3) q p +\cdots\,.} Such an index is precisely what we expect for the conformal manifold of a theory with vanishing flux. The first term at order $q p$  is given by the flat connections of $E_8$ and the second by the complex structure moduli. As we see only a sub-group of $E_8$ explicitly, $U(1)\times SO(  8    )$, let us write down the decomposition of $E_8$,

\eqn\decuo{
{\bf 248}=1+\frac{1}{t^4} + t^4 + (2t^2 + 1 + \frac{2}{t^2}){\bf 28} + {\bf 35_V} + {\bf 35_S} + {\bf 35_C} .}
This identifies the $SO(  8    )$ embedding as the $SO(  8    )\subset SU(  8    ) \subset E_8$.  Let us note that the index computation of a theory corresponding to genus $g$ is independent of the pairs-of-pants decomposition of the surface, at least to the order we have computed it. This is consistent with the different decompositions corresponding to different duality frames.  

We can combine the trinions with $\Phi$ gluing to form a genus ${ g}$ Riemann surface. The anomaly polynomial can be easily obtained for this model. The symmetry that we see is $SO(  8    )\times U(1)$ and we parametrize the trial R symmetry as $R=R'+s q$ with $R'$ the six dimensional R symmetry and $q$ the charge under flavor $U(1)$. Then we obtain that the trial anomalies are,

\eqn\tryuio{\eqalign{& a
(s)= \frac{3}{16}  (1-{ g}) \left(1728 s^3+288 s^2-144 s-25\right)\,,\cr & c
(s)=
\frac{1}{8}  (1-{ g}) \left(2592 s^3+432 s^2-252 s-43\right)\,.}}
This matches the computation in six dimensions if we associate the theory $T_e$ with sphere with three punctures and flux $\frac34$ for the $U(1)$ symmetry. We also then expect that the genus ${ g}$ theory will have $E_7\times U(1)$ symmetry.

We can check the flux assignment and the symmetry by computing the index. We expect the full symmetry to be there for odd genus as the flux then will be integer. For example computing the index of genus three surface we glue four $T_e$ models together with $\Phi$ gluing, which adds to flux $3$, we obtain,

\eqn\indtho{
{\cal I}_{{ g}=3}=1+(\frac8{t^4} +\frac{5{\bf 56}}{t^2}) p q+ 2({\bf 133}+1+3) q p +(-4 t^4-{\bf 56} t^2) p q+\cdots\,.} We have written terms up to order $q p$ in six dimensional R symmetry. The first terms are relevant operators, second marginal, and third irrelevant, when the mixing with the $U(1)$ is taken into account. The value of $s$ which extremizes the $a$ anomaly here is $\frac{\sqrt{10}-1}{9}$ and hence operators which are marginal in six dimensional R symmetry and have positive $U(1)$ charge become irrelevant, and with negative charge become relevant. The number of marginal, relevant, and irrelevant  operators are given by a geometric formula \Brz (see Appendix E for some more details). The marginal (minus the conserved currents) are,

\eqn\nhu{
(dim G+3)({ g}-1) \qquad\to \quad ({\bf 133}+1+3)2\,.} The relevants are given by (see \Brz) the split of ${\bf 248}$ adjoint representation to $E_7\times SU(2)$ representations keeping only the negative charge components and weighing them properly with flux and genus. The decomposition is \eqn\decompoi{{\bf 248}=(t^2+\frac1{t^2}){\bf 56}+1+\frac1{t^4}+t^4+{\bf 133}\,.} 
Then the number of relevant deformations is taking ${\frak F}$ as flux,

\eqn\regeo{
\frac1{t^2}{\bf 56}({ g}-1+{\frak F})+\frac1{t^4}({ g}-1+2{\frak F}) \qquad \to \quad \frac{5{\bf 56}}{t^2}+\frac8{t^4}\,.}
The index contribution of the irrelevant deformations is obtained in the same way (see \Brz) keeping only the operators with positive charges,

\eqn\regui{
t^2{\bf 56}({ g}-1-{\frak F})+t^4({ g}-1-2{\frak F}) \qquad \to \quad -{\bf 56}t^2-4 t^4\,.}
The negative sign is to be interpreted as fermionic operators.
This indeed matches the numbers appearing in the computation of the index.

We can study in detail the different genera theories and always we find consistent results. Another check that one can perform is to combine the tubes that we have found with the trinions to obtain theories of various flux. If we combine $2{ g}-2$ $T_e$ models and $n$  $E_7$ tubes we obtain a theory of flux $\frac{3{ g}-3-n}2$ (flux of tube is $-\frac12$). Note that $n$ can be also negative as we can flip the sign of the tube. The anomaly polynomials  can be computed to give,

\eqn\asdt{\eqalign{& a
(s)={ g} \left(-324  s ^3-54  s ^2+27  s +\frac{75}{16}\right)+108 (n+3)  s ^3-9 (n+3)  s +54  s ^2-\frac{75}{16}\,,\cr & c(
s)= \frac{1}{8} \left({ g} \left(-2592  s ^3-432  s ^2+252  s +43\right)+864 (n+3)  s ^3-84 (n+3)  s +432  s ^2-43\right)\,.}}
This matches the computation in six dimensions. As another test of this expressions let us take $n=3{ g}-3$. This gives flux to be vanishing, that means that the symmetry should be $E_8$ and the $U(1)$ should not be mixing with the R symmetry. Plugging this value into the above expression we obtain that the trial anomaly is,

\eqn\tryurtu{
a(s)= \frac{3}{16} (1-{ g}) \left(288 s^2-25\right)\,.}
This is extremized with vanishing $s$ as expected. The anomalies of theory with no flux are then $(c,\; a)= (\frac{43}8({ g}-1),\; \frac{75}{16}({ g}-1))$ as expected. Let us also write down the conformal anomalies for the generic case,

\eqn\anshu{\eqalign{ & a
= \frac{1}{432} \left(\frac{48 n (6 { g} -n-6) \left(\sqrt{-6 ({ g} -1) n+10 ({ g} -1)^2+n^2}-{ g} +1\right)}{(-3 { g} +n+3)^2}\right.\cr&\left.\qquad\qquad\qquad\qquad\,+480 \sqrt{-6 ({ g} -1) n+10 ({ g} -1)^2+n^2}+1329 { g} -1329\right)\,,\cr &c
= \frac{1}{216} \left(\frac{24 n (6 { g} -n-6) \left(\sqrt{-6 ({ g} -1) n+10 ({ g} -1)^2+n^2}-{ g} +1\right)}{(-3 { g} +n+3)^2}\right.\cr&\qquad\quad\qquad \left.+294 \sqrt{-6 ({ g} -1) n+10 ({ g} -1)^2+n^2}+759 { g} -759\right)\,.}}

\

\subsec{Theories with punctures}

Let us discuss theories with punctures. In particular we can discuss the conformal manifold and anomalies. 
The anomalies will work out almost automatically after we have verified these for surfaces with no punctures and verified the right procedure to gauge symmetries coming from punctures.
In presence of punctures the global symmetry $G_{max}$ preserved by the flux is typically broken to a subgroup. The punctures preserve symmetry $P_j=SU(  8    )\times U(1)$. The symmetry preserved by the theory is then 

\eqn\prehsio{
G_{max}\cap P_1\cap \dots \cap P_s\,,} 
where we have $s$ punctures. Although each puncture preserves $SU(  8    )\times U(1)$, it can be embedded differently inside $E_8$. The different choices as we mentioned before will be referred as colors of a puncture. The dimension of the conformal manifold is expected to be given by \RazamatDPL \ the general expression,

\eqn\dicyero{
dim {\cal M}= 3g-3+s +dim G_{max}(g-1+\frac{s}2) -\sum_{j=1}^s dim (G_{max}\cap P_j)+L\,.} Here $L$ is number of abelian factors in $G_{max}$. We can compute indices of theories with punctures and arbitrary Riemann surface.   Combining $2{ g}-2+s$ $T_e$ models to obtain genus $g$ model with $s$ punctures we find that at order $q p$ the index is,

\eqn\indgdhuy{
(g   - 1+\frac{s}2)(133+1)+3{ g}-3+s - \frac{s}2(63+1)\,.} Here $63+1$ is the dimension of $SU(  8    )\times U(1)$, symmetry preserved by the puncture. In particular this symmetry is a subgroup of $E_7\times U(1)$ symmetry of $T_e$. This is consistent with the general expression.

In similar way combining $T_e$ models using S-gluing to obtain theories corresponding to zero flux, we obtain that at order $q p$ the index is,

\eqn\nfgty{
{3 g} -3+s+248({ g}-1+\frac{s}2)-\frac{s}2(63+1)\,.
} Which is consistent with what we expect. One can perform more checks of the higher genus models with punctures, for example glue in tubes with different symmetries/fluxes. As far as we have checked one always lands on their feet regarding the expectations we have discussed here. We stress once again that punctures deserve a more thorough treatment than given here and we leave this for future work.

\

\newsec{Summary and comments}

In this paper we have charted a map between field theoretic constructions in four dimensions and compactifications of rank one E string theory. This allowed us to conjecture, and provide evidence for, simple quiver gauge theories with the IR symmetry group being larger,  often much larger, than the symmetry observed in the UV description. For genus higher than one the field theoretic construction requires introduction of the $T_e$ model which is non-Lagrangian in the usual sense. It can be constructed as a deformation of a model one obtains by gauging a symmetry only appearing at strong coupling on the conformal manifold of some Lagrangian construction. The map between the geometric compactification and the field theoretic constructions has passed numerous checks.  These include the systematic treatment of enhanced symmetry and also the dualities between different field theoretic constructions.

There are several observations worth mentioning. First, we have found that there are numerous relations between models one finds studying E-string compactifications and models obtained compactifying M5 branes probing A type singularity. For example, E string on a torus with $z$ units of  flux breaking the symmetry to $E_7\times U(1)$  is related to four punctured sphere compactifications of two M5 branes probing  $\Z_{2z}$ singularity.  The theory we found to come from three punctured sphere in E-string comapctification is a deformation of a theory coming from three punctured sphere for two M5 branes on $\Z_2$ singularity. The relations between the two compactifications require either deformations by relevant operators or by introduction of gauge singlet fields.  It is not surprising that different compactifications lead to similar SCFTs, however the relations in our case are ubiquitous and it would be interesting to understand them. 

One can discuss various generalizations of the construction discussed here. For example, we can study higher rank versions of the E-string theory. Such models can be engineered by studying more than one M5 brane on $D_4 $ singularity. We discuss some of the more simple generalizations in this direction in Appendix B. For example, we argue that  deformations of compactifications on torus with flux are obtained by changing the $SU(2)$ gauge groups appearing in rank one with $USp(2Q)$ gauge groups and additional fields in the antisymmetric representation. However, we lack important parts of the story here, for example the field theoretic construction of the  three punctured spheres and the simple arguments in five dimensions. It will be interesting to study this generalization in detail.

We can consider generalizations to various $(1,0)$ starting points.  This paper has given evidence that the predictions coming from $6d$, even for an exotic theory such as E-string theory, are indeed satisfied and thus we have a vast class of ${\cal N}=1 $ SCFT's in front of us.   We emphasize that the six dimensional predictions are rather straightforward and robust once the symmetry and anomalies of the six dimensional starting point are known. The challenge is to find a corresponding construction in four dimensions.   At least when the five dimensional version of the six dimensional model has Lagrangian domain wall constructions we believe that our methods can provide for a systematic way of building four dimensional models corresponding to torus compactifications with flux and spheres with less than three punctures. We will report on  this in an upcoming work \HeeKiZVS.

Finally, let us mention that the relations between six dimensions and four coming from compactifications often lead to deep interplay between four and two dimensional physics. It will be very interesting to understand this better here. For example, the supersymmetric index in four dimensions of a given model can be viewed as  as a TFT correlator on the Riemann surface leading to that particular model \refs{\GaddeUV,  \GaddeIK,\GaddeKB}. It will be interesting to understand the details of such a TFT in our case and its relation to surface defects \GaiottoXA\ in the four dimensional models and to integrable models (for example the eight parameter elliptic relativistic generalization of the Heun equation \refs{\vandrue,\dieskla,\diesklb})\foot{We thank S. Ruijsenaars for discussions of these matters.}

\

\

\noindent {{\bf Acknowledgmenets}}  

We would like to thank Chris Beem, Patrick Jefferson, Zohar Komargodski, and Yuji Tachikawa for useful discussions.  We also like to thank SCGP summer workshop 2017 for hospitality during part of this work.
 The research of HK and CV is supported in part by NSF grant PHY-1067976. GZ is supported in part by  World Premier International Research Center Initiative (WPI), MEXT, Japan.  SSR is  a Jacques Lewiner Career Advancement Chair fellow. The research of SSR was also supported by Israel Science Foundation under grant no. 1696/15 and by I-CORE  Program of the Planning and Budgeting Committee.

\appendix{A}{Branching Rules}

In this appendix we summarize some branching rules that are useful in the study of compatification of E-string with flux. 

\subsec{$E_8 \rightarrow U(1)\times G$}

\noindent {\bf $G=E_7$}

\eqn\fy{\bf{248} \rightarrow \bf{1}^{\pm 2} \oplus \bf{1}^{0} \oplus \bf{133}^{0} \oplus \bf{56}^{\pm 1}}

\

\noindent{\bf $G=SO(14)$}

\eqn\tyupo{\bf{248} \rightarrow \bf{1}^{0} \oplus \bf{91}^{0} \oplus \bf{14}^{\pm 2} \oplus \bf{64}^{-1} \oplus \bf{\overline{64}}^{1}}

\

\noindent{\bf $G=SU(2) \times E_6$}

\eqn\drty{\bf{248} \rightarrow (\bf{1},\bf{1})^{0} \oplus (\bf{3},\bf{1})^{0} \oplus (\bf{1},\bf{78})^{0} \oplus (\bf{1},\bf{27})^{2} \oplus  (\bf{1},\bf{\overline{27}})^{-2} \oplus (\bf{2},\bf{27})^{-1} \oplus (\bf{2},\bf{\overline{27}})^{1} \oplus (\bf{2},\bf{1})^{\pm 3}}

\

\noindent{\bf $G=SU(  8    )$}

\eqn\eryw{
\bf{248} \rightarrow \bf{1}^{0} \oplus \bf{63}^{0} \oplus \bf{8}^{3} \oplus \bf{\overline{8}}^{-3} \oplus \bf{28}^{-2} \oplus \bf{\overline{28}}^{2} \oplus \bf{56}^{1} \oplus \bf{\overline{56}}^{-1}}

\

\noindent{\bf $G=SU(3) \times SO(10)$}

\eqn\fyu{\eqalign{\bf{248} & \rightarrow  (\bf{1},\bf{1})^{0} \oplus (\bf{8},\bf{1})^{0} \oplus (\bf{1},\bf{45})^{0} \oplus (\bf{3},\bf{1})^{-4} \oplus  (\bf{\overline{3}},\bf{1})^{4} \oplus (\bf{3},\bf{10})^{2} \oplus (\bf{\overline{3}},\bf{10})^{-2} \cr & \oplus  (\bf{1},\bf{16})^{3} \oplus (\bf{3},\bf{16})^{-1} \oplus (\bf{1},\bf{\overline{16}})^{-3} \oplus (\bf{\overline{3}},\bf{\overline{16}})^{1}}}

\

\noindent{\bf $G=SU(2) \times SU(7)$}

\eqn\sdyu{\eqalign{\bf{248} & \rightarrow  (\bf{1},\bf{1})^{0} \oplus (\bf{3},\bf{1})^{0} \oplus (\bf{1},\bf{48})^{0} \oplus (\bf{2},\bf{7})^{3} \oplus  (\bf{2},\bf{\overline{7}})^{-3} \oplus (\bf{1},\bf{7})^{-4} \oplus (\bf{1},\bf{\overline{7}})^{4}\cr & \oplus  (\bf{2},\bf{21})^{-1} \oplus (\bf{1},\bf{35})^{2} \oplus (\bf{2},\bf{\overline{21}})^{1} \oplus (\bf{1},\bf{\overline{35}})^{-2}}}

\

\

\

\noindent{\bf $G=SU(4) \times SU(5)$}

\eqn\dtyui{\eqalign{
\bf{248} & \rightarrow  (\bf{1},\bf{1})^{0} \oplus (\bf{15},\bf{1})^{0} \oplus (\bf{1},\bf{24})^{0} \oplus (\bf{4},\bf{1})^{5} \oplus  (\bf{\overline{4}},\bf{1})^{-5} \oplus (\bf{4},\bf{5})^{-3} \oplus (\bf{\overline{4}},\bf{\overline{5}})^{3}\cr & \oplus  (\bf{6},\bf{5})^{2} \oplus (\bf{6},\bf{\overline{5}})^{-2} \oplus (\bf{4},\bf{\overline{10}})^{1} \oplus (\bf{1},\bf{\overline{10}})^{-4} \oplus  (\bf{\overline{4}},\bf{10})^{-1} \oplus  (\bf{1},\bf{10})^{4}}
}

\

\noindent{\bf $G=SU(2) \times SU(3) \times SU(5)$}

\eqn\dgyu{\eqalign{
\bf{248} & \rightarrow  (\bf{1},\bf{1},\bf{1})^{0} \oplus (\bf{3},\bf{1},\bf{1})^{0} \oplus (\bf{1},\bf{8},\bf{1})^{0} \oplus (\bf{1},\bf{1},\bf{24})^{0} \oplus  (\bf{2},\bf{3},\bf{1})^{-5} \oplus (\bf{2},\bf{\overline{3}},\bf{1})^{5} \oplus (\bf{2},\bf{3},\bf{5})^{1} \cr  & \oplus  (\bf{2},\bf{\overline{3}},\bf{\overline{5}})^{-1} \oplus (\bf{2},\bf{1},\bf{10})^{-3}  \oplus (\bf{2},\bf{1},\bf{\overline{10}})^{3} \oplus  (\bf{1},\bf{1},\bf{5})^{6} \oplus  (\bf{1},\bf{1},\bf{\overline{5}})^{-6} \oplus  (\bf{1},\bf{\overline{3}},\bf{5})^{-4}\cr  & \oplus   (\bf{1},\bf{3},\bf{\overline{5}})^{4} \oplus (\bf{1},\bf{\overline{3}},\bf{10})^{2} \oplus (\bf{1},\bf{3},\bf{\overline{10}})^{-2}
}}

\subsec{$E_8 \rightarrow U(1)^2\times G$}

\noindent{\bf $G=E_6$}

\eqn\ftyuo{
\bf{248} \rightarrow 2 \bf{1}^{(0,0)} \oplus \bf{1}^{(\pm 2,0)} \oplus \bf{78}^{(0,0)} \oplus \bf{27}^{(0,2)} \oplus  \bf{\overline{27}}^{(0,-2)} \oplus \bf{27}^{(\pm 1,-1)} \oplus \bf{\overline{27}}^{(\pm 1,1)} \oplus \bf{1}^{(\pm 1,\pm 3)}}

The two $U(1)'s$ are spanned by $(a,b)$ where flux only in $a$ breaks $E_8$ to $U(1) \times E_7$ and flux only in $b$ breaks $E_8$ to $U(1) \times SU(2) \times E_6$. Furthermore, flux where $a = \pm b$ preserves a different $E_7 \subset E_8$ and $a= \pm 3b$ preserves a different $SU(2) \times E_6 \subset E_8$.

\

\noindent{\bf $G=SO(12)$}

\eqn\ertu{
\bf{248} \rightarrow 2 \bf{1}^{(0,0)} \oplus \bf{1}^{(\pm 2,0)} \oplus \bf{1}^{(0,\pm 2)} \oplus \bf{66}^{(0,0)} \oplus \bf{32}^{(0,\pm 1)} \oplus  \bf{32'}^{(\pm 1,0)} \oplus \bf{12}^{(\pm 1,\pm 1)}}

The two $U(1)'s$ are spanned by $(a,b)$ where flux only in either of them breaks $E_8$ to $U(1) \times E_7$. Furthermore, flux where $a = \pm b$ preserves an $SO(14) \subset E_8$.

\

\noindent{\bf $G=SU(2)\times SO(10)$}

\eqn\tewu{\eqalign{
\bf{248} & \rightarrow  2(\bf{1},\bf{1})^{(0,0)} \oplus (\bf{1},\bf{1})^{(0,\pm 2)} \oplus (\bf{3},\bf{1})^{(0,0)} \oplus (\bf{1},\bf{45})^{(0,0)} \oplus (\bf{2},\bf{10})^{(0,\pm 1)} \oplus \cr &  (\bf{2},\bf{1})^{(\pm 2,\pm 1)} \oplus (\bf{1},\bf{10})^{(\pm 2,0)}  \oplus  (\bf{1},\bf{16})^{(-1, \pm 1)} \oplus (\bf{2},\bf{\overline{16}})^{(-1, 0)} \oplus (\bf{1},\bf{\overline{16}})^{(1, \pm 1)} \oplus (\bf{2},\bf{16})^{(1, 0)}}
}
The two $U(1)'s$ are spanned by $(a,b)$ where flux only in $a$ breaks $E_8$ to $U(1) \times SO(14)$ and flux only in $b$ breaks $E_8$ to $U(1) \times E_7$. Furthermore, flux where $a = \pm b$ preserves an $SU(2)\times E_6 \subset E_8$ while flux where $2 a = \pm b$ preserves an $SU(3)\times SO(10) \subset E_8$.

\appendix{B}{Higher rank}

We can consider the generalization of the discussion  to higher rank E string models. In five dimensions, that is taking the E string on a circle, the higher rank model becomes $USp(2Q)$ gauge theory with an antisymmetric hypermultiplet and eight hypermultiplets in the fundamental representation. The symmetry is $SU(8)\times U(1) \times SU(2)$ with the  $SU(2)$ rotating the half- hypermultiplets of the antisymmetric field. A simple conjecture following from five dimensions is that we need just to change the $SU(2)$ groups we have obtained for rank one with  $USp(2Q )$. Moreover it is natural to add a field in the antisymmetric representation.

Consider the case of the compactification with flux breaking the $E_8$ to $E_7\times U(1)$.  We just have the same quiver diagram as in Fig. 3 just change the gauge groups to $USp( 2 Q)$ and add the antisymmetric fields. The gauge groups here have zero one loop beta function.  We couple the antisymmetric field to bilinears of bifundamental chirals  which also are coupled to singlet flippers. The model has manifest $SU(  8    )\times U(1)$  which at the level of te index again can be seen to enhance to $U(1)\times E_7$. However, there is no $SU(2)$ symmetry which we expect to have for higher rank. 

\

\centerline{\figscale{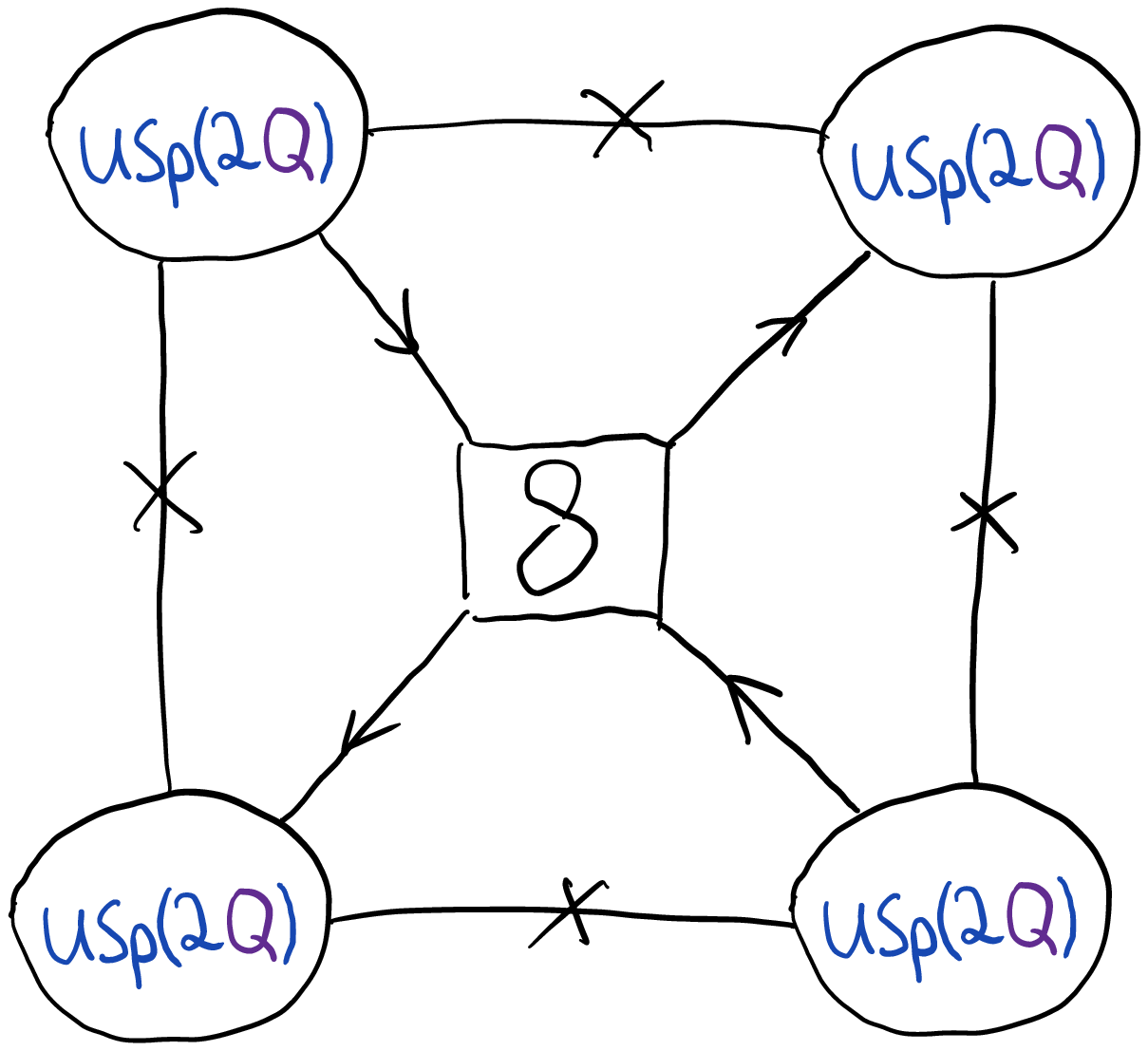}{2.1in}}
\medskip\centerline{\vbox{
\baselineskip12pt\advance\hsize by -1truein
\noindent\footnotefont{\bf Fig.~24:} Example of $USp(2Q)$ quiver theory corresponding to deformation of compactification of rank $Q$ E string on a torus with two units of flux. Each node has an antisymmetric tensor which couples to bilinears of bifundamentals with a cubic coupling.}} 

\

There is a way to connect this model to six dimensions. For simpicity we consider the case with two $USp(2Q)$ gauge groups. The other cases work in a similar manner. Let us write the trial anomaly polynomial of this theory with the R symmetry natural  from six dimensions. We assign R charge $0$ to the bi-fundamentals, R charge $+1$ to the fundamentals charged under $SU(  8    )$, and R charge $+2$ to the antisymmetrics.  The $U(1)$ charge of the antisymmetrics is  $+1$, of the bifundamentals $-\frac12$, of the fundamentals $+\frac14$. The anomalies are then defining $R=R'+ s F$

\eqn\anf{\eqalign{&a=\frac{9}{64} s Q\left(s^2 (6 Q-3)+12 s(Q-1)-16\right)\,,\qquad c=\frac{3}{64}s Q \left(9 s^2 (2 Q-1)+36 s(Q-1)-56\right)\,.}} This matches the six dimensional result \ans\ when we set $h=2s+2$. Such a specialization means that we turn on a deformation of the four dimensional theory corresponding to the compactification breaking $SU(2)$ symmetry and locking the R symmetry with the Cartan of the $SU(2)$ and the $U(1)$ in a certain way. We conclude that the theory we obtain here is a deformation of the theory one obtains in a compactification, breaking the $SU(2)$ but having $E_7\times U(1)$. The conformal R-charge here is the free one and the anomaly is given  after extremization by

\eqn\csf{
a=Q(Q+1)\,,\qquad c= Q( Q +\frac32 ) \,.}

\

\subsec{Theories with $G\times U(1)$}

We can construct, in a similar manner to the above, theories preserving other groups by replacing the $SU(2)$ gauge groups with $USp(2Q)$.  In each case the four dimensional theory can be argued to be related to a deformation of a compactification of the six dimensional theory.

We here consider an example of a theory with $G=E_6\times SU(2)$. See Fig. 25 for the quiver theory. This model has manifestly $SU(6)\times SU(2)\times U(1)_a\times U(1)_b$ symmetry. The symmetry $SU(6)\times SU(2)\times U(1)_{\frac{a}4-\frac{b}2}$  enhances to $E_6 \times SU(2)$. Let us denote $U(1)_s$ such that $q_s=\frac12 q_a+\frac14 q_b$. Then the anomaly polynomials are given here,

\eqn\ansiopu{\eqalign{&a=
\frac{9}{64} s Q \left(s^2 (6 Q-3)+12 s(Q-1)-16\right)\,,\cr       &
 c=\frac3{64} s Q (-56 + 36 s (Q-1) + 9 s^2 (-2Q-1))\,.
 }} This matches the six dimensional result \ans\ when we set $h=3s+2$ and $\xi=3$.
 
 \
 \centerline{\figscale{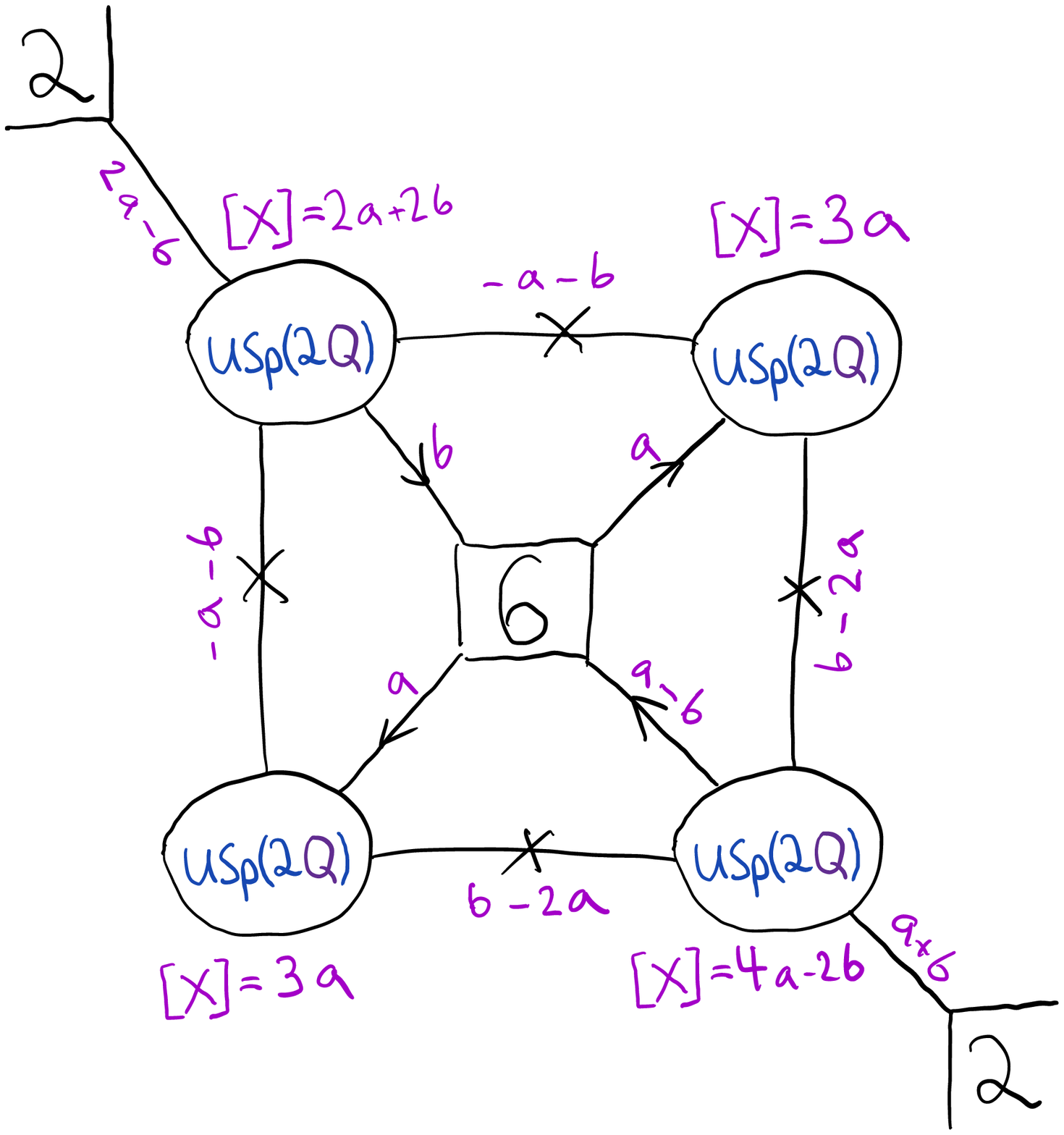}{2.9in}}
\medskip\centerline{\vbox{
\baselineskip12pt\advance\hsize by -1truein
\noindent\footnotefont{\bf Fig.~25:} Theory corresponding to unit of flux and $E_6\times SU(2)$ symmetry for rank $Q$ E-string on a torus. The fields $X$ are the antisymmetric fields for each gauge group. These fields couple to the bifundamental fields according to their $U(1)_a$ and $U(1)_b$ charges.}
}

It will be interesting to discover field theory constructions having the full rank nine symmetry of the six dimensional model and also to generalize field theory constructions to higher genus. We hope to return to this questions in future studies.

\

\

\appendix{C}{Flux quantization}

In this appendix we consider the possible choices for fluxes in the compactification of a $6d$ theory on a Riemann surface $\Sigma$. When compactifying the 6d theory with a flux in some $U(1)$ subgroup of the full global symmetry group, the flux must be quantize $\int_{\Sigma} C_1 (U(1)) = c n$ where $n$ is an integer, and $c$ is some normalization dependent constant. This quantization condition is analogous to the Dirac quantization condition for magnetic monopoles and it is sometimes convenient to think about it in this way to get a physical picture. 

Specifically this means that the flux must be quantize such that for every state the phase factor $e^{2\pi i q \int_{\Sigma} C_1 (U(1))}$, where $q$ is the charge of the state under the $U(1)$, is equal to $1$. This means that we must take $c=\frac{1}{q_{min}}$ where $q_{min}$ is the minimal charge in the system. This is indeed just the well known Dirac quantization condition. This means that flux quantization is dependent on the $6d$ spectrum. Particularly if the $U(1)$ is part of a non-abelian group $G$ then the minimal charges present depend on what representations of $G$ appear in the $6d$ theory. For instance for the $U(1)$ Cartan of $SU(2)$ the minimal charge is in the doublet, and normalizing its charge to one the fluxes will be integer. However if that is not present in the $6d$ spectrum then half-integer fluxes are also consistent as all states will have even charge in this normalization. In other words, flux quantization depends on the global structure of the group and not just the local one, for instance $SU(2)$ versus $SO(3) = SU(2)/\Z_2$.

In this paper we dealt mostly with $E_8$ which does not have a center, so we will concentrate on the case where the $6d$ global symmetry group is simply connected. Later, for completeness, we shall mention some additional choices that exist when the group is not simply connected. Even in this case flux quantization can become involved. For instance consider the case when there is flux in several $U(1)$'s. The condition now becomes:

\eqn\fluxquan{
e^{2\pi i \sum_i q_i \int_{\Sigma} C_1 (U(1)_i)} = 1\,  ,}
and one can envision situations when each term individually will not obey the condition, but their sum will. Note however, that this must occur for every state in the $6d$ theory. Thus there must be some combined transformation of the $U(1)$'s that acts trivially on all states. In other words the full symmetry group is not just the direct product of all the $U(1)$ but $U(1) \times U(1) ... U(1)/ Z$ where $Z$ is some discrete group. This modding out expresses the fact that there is some combined transformation that acts trivially and so needs to be modded out.

This can also occur when the fluxes are not just in a $U(1)$ but also in the center of a non-abelian group. For instance consider the case when the global symmetry is locally $U(1) \times SU(2)$ and the spectrum is generated by the two states ${\bf 3}^{\pm 2}$ and ${\bf 2}^{\pm 1}$. The minimal charge under the $U(1)$ here is $1$ and so we expect the flux to be integer in this normalization. However we can also have half-integer flux if we also turn on a flux in the center of the $SU(2)$. Recall that the center of $SU(2)$ is $\Z_2$ whose non-trivial element acts as: ${\bf 2} \rightarrow - {\bf 2}, {\bf 3} \rightarrow {\bf 3}$. Thus the ${\bf 3}^{\pm 2}$ is consistent since it has even charge while ${\bf 2}^{\pm 1}$ will get a $-1$ both from the $U(1)$ flux and from the center which will cancel exactly. This is again since while locally the group is $U(1) \times SU(2)$, globally it is actually $\frac{U(1) \times SU(2)}{\Z_2}$ where the $\Z_2$ is the combined $U(1)$ $\pi i$ transformation and the center of $SU(2)$.

In fact such structures are ubiquitous when one starts with a simply connected group $G$ and break it to a subgroup $U(1) \times G'$ via flux. Generically the commutant in $G$ is not $U(1) \times G'$ but $\frac{U(1) \times G'}{Z}$ for some discrete group $Z$. In these cases one can accommodate non-integer fluxes (where we have chosen a basis so that the minimal charge is $1$) if combined with a flux in the center of $G'$. The flux can be generated by two holonomies that do not commute up to an element of the center. Mathematically, it is referred to as a nonzero Stiefel-Whitney class for the global symmetry bundle $\frac{G'}{Z}$. These center fluxes, specifically the holonomies needed to generate them, break some part of $G'$ as we shall now discuss.  

\subsec{Center fluxes}

As we mentioned we can also incorporate a flux in the center of a non-abelian group. This can be realized by turning on two almost commuting holonomies, that is two holonomies that commute up to an element of the center. For example consider the group $SU(2)$ and the two following holonomies:

\eqn\holo{
A=
\pmatrix{
  i & 0 \cr
  0 & -i
}
,  B=
\pmatrix{
  0 & i \cr
  i & 0
}\,. }
These holonomies obey: $AB=-AB$, so they commute up to the center $-I$ element of $SU(2)$. Putting these two elements on the two cycles of a torus for instance generate a flux in the $\Z_2$ center of $SU(2)$. 

Now several points are worth noticing. Generically the holonomies on a Riemann surface are not independent as they must obey the fundamental group condition. For the torus this condition is that the two holonomies must commute\foot{For higher genus Riemann surfaces these generalize to the condition: $\prod^g_{i=0} [A_i, B_i] =1$, where $A_i,　B_i$ are the holonomies under the $2g$ cycles.}. Thus we cannot turn on flux in a non-abelian group just by turning on two constant but non-cummuting holonomies. Here again it is important that the group is not $U(1) \times G'$ but $\frac{U(1) \times G'}{Z}$. For instance the two holonomies \holo\ do not commute in $SU(2)$ but they do commute in $SO(3)$, and thus are valid holonomies in that case. In all of these cases the holonomies are valid ones for the actual group even though they are not valid in the universal cover.

A second observation is that the holonomies generically break part of the symmetry. To illustrate this we again refer to the two holonomies \holo. In their presence we preserve only the part of $SU(2)$ that commute with them\foot{This is usually refereed to as the centralizer of the elements in $G$.}. In this case one can show that this breaks $SU(2)$ completely.    

\subsec{Non-simply connected groups}

We now want to say a few words about what happens when the $6d$ global symmetry group $G$ is not simply connected. First, more charges will be consistent compared to the case involving the universal cover. We do note that in some cases the difference may be quite subtle. For instance consider the case of $USp(4)$ versus $\frac{USp(4)}{\Z_2} = SO( 5  )$, and flux in the $U(1)$ whose commutant is $SU(2)$ breaking ${\bf 4} \rightarrow {\bf 2}^{\pm 1}$, ${\bf 5} \rightarrow 1^{\pm 2} + {\bf 3}^{\pm 0}$. In both cases a half-integer flux is possible, but in the $USp(4)$ case this must be accommodated by a center flux in the $SU(2)$. Therefore, the difference between the two cases appears not in the possible choices of flux but in the global symmetry preserved by the flux.

Besides the difference in quantization and global symmetry, we also have the possibility of turning on flux in the universal cover group via almost commuting holonomies. As mentioned in the previous subsection we cannot turn on non-commuting holonomies on the torus, but we can turn on two holonomies that commute in $G$ yet do not commute in the universal cover. This means that we apply the same procedure as in the previous subsection but now to the full group $G$. As previously mentioned, this is known in the mathematical literature as turning on a non-trivial Stiefel-Whitney class. These are discrete elements whose values are given by $\pi_1(G)$, which are just the elements up to which the two holonomies commute. Turning on such elements has appeared in the context of the compactification of a $6d$ SCFT with non-simply connected global symmetry in \BahGPH.

As we previously discussed, a non-trivial Stiefel-Whitney class generically breaks $G$. An interesting problem then is to determine what is the centralizer for each possible choice. This problem is rather involved yet was studied by a variety of people from both the physics and mathematics viewpoint \refs{\WittenNV,\WittenNVT,\WittenNVS,\Kac,\Keure,\Keura,\almost}. We shall now describe some aspects of this issue.    

\subsec{Global symmetry preserved by center fluxes}

We start with several general observations. The most important of these is that the global symmetry preserved depends on the choice of holonomies. However, there is a particular choice preserving the maximal global symmetry $G'$, given by say holonomies $A$ and $B$, where all other choices can be generated by holonomies $A t$ and $B t'$ where $t, t'$ belong to the maximal torus of $G'$. Note that while the various possible centralizer groups are subgroups of $G$ they may not be subgroups of one another.

The point in this structure is that when we turn on central fluxes we are forced to turn on holonomies, but have some freedom in their exact form. From the form of the holonomies it is clear that for generic choices of $t, t'$ we break $G'$ to $U(1)^r$ where $r$ is the rank of $G'$. This follows as the holonomies must commute up to the center. For special choices of $t, t'$ the symmetry $U(1)^r$ enhances to various non-abelian groups. Since the maximal torus is connected, and as any holonomies that commute up to a specific center element can be written in that form, we can continuously move from any chosen pair to any other one.

The implications of this on the 4d theories resulting from such compactifications are as follows. The parameters associated with tuning holonomies are generically mapped to marginal deformations in 4d, and the space of holonomies then is mapped to part of the conformal manifold of the theory. Thus, we expect the theory to contain an $r$ dimensional conformal manifold where at a generic point of which the symmetry is broken to $U(1)^r$, but is enhanced to various non-abelian symmetries, particularly $G'$, for special points on the conformal manifold.

A list with the possible values of $G'$ appears in \Kac. 
We shall next discuss these possibilities for various choices of $G$. We shall not classify all possible non-abelian groups one can get, but instead discuss ones that are of interest here.

\

\

\noindent{\it $SU(N)$}

The case of $SU(N)$ is probably the most well known. The center of $SU(N)$ is $\Z_N$ and we can choose a pair that commute up to the element $\omega^l$ where $\omega$ is the generator of $\Z_N$. This can be realized by the matrices:

\eqn\holoSU{
A=
\pmatrix{
  I_{k \times k} & 0 & 0 & ... & 0 \cr
  0 & \omega^l I_{k \times k} & 0 & ... & 0 \cr
	0 & 0 & \omega^{2l} I_{k \times k} & ... & 0 \cr
	. & & & & \cr
	. & & & & \cr
	. & & & & \cr
	0 & 0 & 0 & ... & \omega^{\frac{(N-k) l}{k}} I_{k \times k}
}
,  B=
\pmatrix{
  0 & I_{k \times k} & 0 & ... & 0 \cr
	0 & 0 & I_{k \times k} & ... & 0 \cr
	. & & & & \cr
	. & & & & \cr
	. & & & & \cr
	0 & 0 & 0 & ... & I_{k \times k} \cr
  I_{k \times k} & 0 & 0 & ... & 0
} \, ,}
where $k=gcd(N,l)$ and we have used $I_{k \times k}$ for a $k \times k$ identity matrix. These matrices also appeared in \tHooft.

These preserve an $SU(gcd(N,l))$ subgroup of $SU(N)$. In this case this is the maximal group we can get.

\

\noindent{\it $USp(2N)$}

The center of $USp(2N)$ is $\Z_2$ which acts non-trivially on the fundamental representation. Recall that $USp(2N)$ is defined as the matrices $M$ obeying:

\eqn\USpdef{
M^{\dagger} M = I_{2N \times 2N} , M^T J_{2N \times 2N} M = J_{2N \times 2N} ,}
where 

\eqn\jey{
J_{2N \times 2N}=
\pmatrix{
  0 & I_{N \times N} \cr
  -I_{N \times N} & 0
} .}
This choice of presentation highlights the $SU(2) \times SO(N)$ subgroup of $USp(2N)$, and we can now choose the following pair:

\eqn\holoUSpgen{
A=
\pmatrix{
  i I_{N \times N} & 0 \cr
  0 & -i I_{N \times N}
}
,  B=
\pmatrix{
  0 & i I_{N \times N} \cr
  i I_{N \times N} & 0
} .}
These break the $SU(2)$, but preserve the $SO(N)$. This choice is available for any $N$, but when $N$ is even there is another choice that preserves a larger group. Consider the presentation highlighting the $SO(2) \times USp(N)$ subgroup of $USp(2N)$, where we represent $J_{2N \times 2N}$ by: 

\eqn\jeyru{
J_{2N \times 2N}=
\pmatrix{
  J_{N \times N} & 0 \cr
  0 & J_{N \times N}
}  .}
In this presentation we can choose the holonomies:

\eqn\holoUSpgen{
A=
\pmatrix{
   I_{N \times N} & 0 \cr
  0 & - I_{N \times N}
}
,  B=
\pmatrix{
  0 & I_{N \times N} \cr
   I_{N \times N} & 0
} .   }
These preserve $USp(N)$ which is the largest group one can preserve when $N$ is even.

Besides these there are various other groups one can preserve.

\

\noindent{\it $Spin(2N+1)$}

The center of $Spin(2N+1)$ is $Z_2$ where $\frac{Spin(2N+1)}{Z_2} = SO(2N+1)$ so this element acts non-trivially on the spinors. For the case of $Spin(3)=SU(2)$ we have already presented a pair commuting up to its center. In the general case we can now use the breaking $Spin(2N+1)\rightarrow Spin(3) \times Spin(2N-2)$ and embed the same holonomies inside $Spin(3)$. Since the spinor decomposes to bispinors this will have the desired effect. This breaks $Spin(3)$ but preserves $Spin(2N-2)$.
It turns out that in this case one can actually preserve a larger group $Spin(2N-1)$ \Kac. Note that the ranks of the two groups are the same so there is no contradiction with the structure of the holonomy space. We can understand how this comes about as follows. Consider the breaking of $Spin(2N+1)\rightarrow U(1) \times Spin(2N-1)$. Under it the spinor of $Spin(2N+1)$ decomposes to two spinors of $Spin(2N-1)$. So we can represent the $Spin(2N+1)$ spinor as a two component vector of $Spin(2N-1)$ spinors. Now consider the following holonomies acting on this vector: 

\eqn\holoSpinodd{
A= a
\pmatrix{
  I_{2^{N-1} \times 2^{N-1}} & 0 \cr
  0 & -I_{2^{N-1} \times 2^{N-1}}
}
,  B= b
\pmatrix{
  0 & I_{2^{N-1} \times 2^{N-1}} \cr
  I_{2^{N-1} \times 2^{N-1}} & 0
}\,. }

These commute up to the center of $Spin(2N+1)$, where $a,b$ are some constants chosen so that the matrices $A,B$ sit in the appropriate group. These holonomies break the $U(1)$, but preserve $Spin(2N-1)$.   
 
 \
 
 \

\noindent{\it $Spin(2N)$}

The center of $Spin(2N)$ differs depending on whether $N$ is even or odd. In the $N$ even case it is $\Z_2 \times \Z_2$ while in the $N$ odd case it is $\Z_4$. The generator of $\Z_4$ in the $N$ odd case, $\tilde{\omega}$, acts as $i$ on the spinors and $-1$ on the vectors. Thus we have two distinct center choices $\tilde{\omega}$, and $\tilde{\omega}^2$ where the latter is the element that projects $Spin$ to $SO$. In the $N$ even case the generators of $\Z_2 \times \Z_2$, $\omega_1$ and $\omega_2$, act as $-1$ on the vector and one of the spinors. Thus again we have two distinct center choices, $\omega_1$ and $\omega_1 \omega_2$ where the latter is the element that projects $Spin$ to $SO$.

Let us consider the element projecting $Spin$ to $SO$, which can be discussed uniformly in the same manner as the $Spin(2N+1)$ case. Particularly, we consider the breaking $Spin(2N)\rightarrow Spin(3)\times Spin(2N-3)$ and embed the holonomies in $Spin(3)=SU(2)$. This achieves the desired result while breaking the $SU(2)$ but preserving $Spin(2N-3)$. This is the largest global symmetry one can preserve in this case. Particularly we cannot preserve $Spin(2N-2)$.

To see this we again consider the breaking of now $Spin(2N)\rightarrow U(1) \times Spin(2N-2)$. Under it the spinor of $Spin(2N)$ decomposes to two spinors of $Spin(2N-2)$, but now these spinors are in different spinor representations of $Spin(2N-2)$. We can again represent the $Spin(2N)$ spinor as a two component vector of $Spin(2N-2)$ spinors, but now the two components are in different spinor representations. These requires us to modify the holonomies to: 

\eqn\holoSpineven{
A= a
\pmatrix{
  I_{2^{N-2} \times 2^{N-2}} & 0 \cr
  0 & -I_{2^{N-2} \times 2^{N-2}}
}
,  B= 
\pmatrix{
  0 & {\cal V} \cr
 {\cal V} & 0
}\,. }

Here we introduce the operator $\cal V$ which maps one spinor representation to the other, and retains the constant $a$ chosen so that the matrices $A$ sit in the appropriate group. The operator $\cal V$ can be naturally associated with a vector of $Spin(2N-2)$, as these are the non-diagonal element that appear in the adjoint decomposition of $Spin(2N)$ and can indeed introduce the required mapping. 

The holonomies \holoSpineven\ commute up to the center of $Spin(2N)$, and break the $U(1)$ part. However we cannot preserve $Spin(2N-2)$ as we need to also choose a specific $\cal V$. As it can be identified with a vector of $Spin(2N-2)$, it will generically break it to $Spin(2N-3)$. More generally we can just think of the matrix $B$ as implementing the outer automorphism of $Spin(2N-2)$, as it exchanges the two spinor representations. This preserves the subgroup of $Spin(2N-2)$ that is invariant under this outer automorphism. The largest group that can be preserved is indeed $Spin(2N-3)$, however, by choosing a different representation of this element we can preserve different groups. Particularly it follows from the work of Kac (see section 3.3 in \TachikawaYMS\ for a discussion on this aimed for physicists) that one can preserve in this way the group $Spin(2k+1) \times Spin(2N-2k-3)$ for $k=0,1,2 ... , N-2$.

So to conclude we see that the largest group we can preserve here is $Spin(2N-3)$, but other choices exist for instance there are choices preserving $Spin(2k+1) \times Spin(2N-2k-3)$ for $k=1,2 ... , N-3$. Note that these in general are not subgroups of one another. 

We can next consider the case of $N$ even and center choice $\omega_1$. This case was studied extensively in \WittenNVS, which analyzed the various choices. The maximal symmetry one can preserve here is $USp(N)$. This can be seen by using the $SU(2)\times USp(N)$ subgroup of $Spin(2N)$ and again embed the holonomies in the $SU(2)$. There are other choices, involving other subgroups that can be used, preserving different symmetries. Using these choices one can preserve a $USp(2k) \times Spin(N-2k)$ for any $k=0,1,... ,\frac{N}{2}$.

In the $N$ odd case we can consider the center choice $\tilde{\omega}$. In this case the maximal symmetry we can preserve is $Spin(N-2)$. There are however other choices. For instances there is one preserving $USp(N-3)$ \WittenNV. 

\

\noindent{\it $E_6$}

The center of $E_6$ is $\Z_3$, and we can choose a pair of holonomies commuting up to the generator of $\Z_3$ or its inverse. Either way the maximal subgroup one can preserve is $G_2$.

\

\

\noindent{\it $E_7$}

The center of $E_7$ is $\Z_2$ which acts non-trivially on the fundamental ${\bf 56}$ dimensional representation of $E_7$. It is known that the maximal subgroup that can be preserved is $F_4$.

There are other possible choices, and we shall analyze another case which appears in our discussion. In it we utilize the $SU(  8    )$ subgroup of $E_7$. Under the embedded $SU(  8    )$ the fundamental of $E_7$ decomposes as ${\bf 56} \rightarrow {\bf 28} + {\bf \overline{28}}$ and the adjoint as ${\bf 133} \rightarrow {\bf 63} + {\bf 70}$, where we note that the ${\bf 28}, {\bf 70}$ and ${\bf 63}$ are the rank $2$ antisymmetric, rank $4$ antisymmetric and adjoint of $SU(  8    )$ respectively. Thus we can introduce the vector $V$ and matrix $M$:

\eqn\VaM{
V=
\pmatrix{
   F \cr
  \overline{F} 
}
,  M=
\pmatrix{
  A & \Lambda \cr
   \overline{\Lambda} & \overline{A}
},}where $F$ is in the ${\bf 28}$ of $SU(  8    )$, $A$ in the ${\bf 63}$, and $\Lambda$ is in the ${\bf 70}$. The matrix $M$ can act on $V$  where here $A$ maps $F\rightarrow F$ or $\overline{F}\rightarrow \overline{F}$ while $\Lambda$ map $\overline{F}\rightarrow F$ or $F\rightarrow \overline{F}$. The matrix $M$ then represents the ${\bf 56}$ of $E_7$. 

We can now consider turning on two holonomies, one with $A=i I, \Lambda=0$ and another with $A=0, \Lambda\neq 0$. These commute up to the center of $E_7$. The first holonomy breaks $E_7$ to $SU(  8    )$. To determine the centralizer of the second inside $SU(  8    )$ we need to choose a specific $\Lambda$. We note that these induce the transformation ${\bf 28}\rightarrow {\bf \overline{28}}$, which is just the complex conjugation outer automorphism of $SU(  8    )$. So the problem reduces to finding the possible subgroups that are invariant under this outer automorphism. It is again known that there are different choices depending on how one realized the outer automorphism. We can again employ the Kac prescription to determine the possible groups, but for our purposes only two suffice. These two are just the natural choices $USp(8)$ and $SO(  8    )$, which are the two real simple subgroups of $SU(  8    )$. The latter in particular appears prominently in this article.   

\

\subsec{Central fluxes and Riemann surfaces with punctures}

So far we only discussed the effect of the central fluxes when the Riemann surface is closed. However we encounter also situations with surfaces with punctures. Particularly consider the tubes we introduced. Many of these have fractional fluxes that require central fluxes for consistency, and we can ask how are these manifested in the tube.

The central fluxes require two almost commuting holonomies. One of these holonomies must surround the puncture while the other must stretch between the punctures. Let's start with the one around the puncture. It is known that punctures require an holonomy around them \BeniniMZ. It is this holonomy that breaks part of the internal symmetry and effect the structure of the conformal manifold. So the presence of the holonomy is not special but rather generic. It should be noted that there are different choices for this holonomy and two punctures may have holonomies, preserving the same symmetry, but embedded differently. These are said to have different color, and connecting them leads to a breakdown of some of the global symmetry.

Now we turn to the second holonomy stretching between the two punctures. Due to homotopy relations, the holonomy around one of the puncture must be equal the holonomy around the other conjugated by the holonomy stretching between them, but this holonomy must not commute with the holonomies around the punctures. Therefore, we see that the holonomies around the two punctures must differ by the action of the second holonomy and so must have different colors. Thus, to conclude, there is a relation between central fluxes in tubes and the difference in colors between the two punctures of the tube.

For example, consider the $E_7$ tube. We have presented a pair of almost commuting elements in the previous section. One of these, the one with $\Lambda=0$, preserves only an $SU(  8    )$ subgroup of $E_7$. This is the same as the punctures and so is natural to associate it with the holonomy around the puncture. The second element acts on the $SU(  8    )$ by complex conjugation. Thus, we see that the presence of the central fluxes is manifested in the tube by whether the two punctures are the same or differing by complex conjugation of the $SU(  8    )$.
Indeed in the basic tube the punctures differ exactly in this way and we indeed have central flux. Gluing an even number of these eliminate the central fluxes and indeed when these are not present the colors of the two punctures are the same.

A similar  discussion can be entertained also  in the case of  the other tubes, where the presence of the central fluxes necessitates a difference in color between the two punctures. This should also have generalizations to more punctures and higher genus. We shall not analyze these cases here.       

\

\

\subsec{Summary}

\noindent Finally we wish to summarize the discussion here:

- The quantization of flux depends on the global structure of the group $G$.

- Adopting a normalization where the minimal charge is $1$, the fluxes in $U(1)$'s are integers. However fractional fluxes may be possible if the subgroup inside $G$ is not a direct product of the $U(1)$'s and the preserved non-abelian groups. 

- In many cases consistency of such fractional fluxes necessitates the introduction of fluxes in the center of a non-abelian symmetry. This can be accommodated by a pair of almost commuting holonomies. Such a pair however will break the global symmetry. The preserved global symmetry depends on the choice of holonomies, where, depending on the choice, different subgroups can be preserved.

- If $G$ is not simply connected one can also incorporate a non-trivial Stiefel-Whitney class. This again can be accommodated by a pair of almost commuting holonomies. Again this will result in breaking of $G$ to a smaller group.   

- The presence of almost commuting holonomies on a tube is manifested through a difference in the colors of the two punctures.

\appendix{D}{The  $T_e$ model}

Let us here give the index of the $T_e$ model. We will encode the information in the supersymmetric index written as an integral over elliptic Gamma functions. From this expression one can deduce the Lagrangian of the model and the computation of anomalies.  The index can be written as,

\eqn\indTbag{\eqalign{
&
{\cal I}_e = \Gamma_e((q  p)^{\frac12}t (\frac{1}{\beta^2}  v_2)^{\pm1}v_1^{\pm1})
\Gamma_e(\frac{q   \,  p}{t^2})
\cr
&(p;p)(q;q)\oint\frac{dz}{4\pi i z}\frac{\Gamma_e(\frac{(q  p)^{\frac12}}{t^2}(\beta^2 v_2^{-1})^{\pm1} z^{\pm1})}{\Gamma_e(z^{\pm2})}\Gamma_e(t z^{\pm1}v_1^{\pm1}){\cal I}_0({\bf c},{\bf w}, \sqrt{z v_2},\sqrt{v_2/z})\,.
 }} We have defined,

\eqn\ofrbi{
\eqalign{
&{\cal I}_0({\bf z},{\bf v}, a,b) =(p;p)^2(q;q)^2\oint\frac{dw_1}{4\pi i w_1}\oint\frac{dw_2}{4\pi i w_2}
\frac{\Gamma_e(\frac{(pq)^{\frac12}}{t^2} w_1^{\pm1}w_2^{\pm1})}{\Gamma_e(w_1^{\pm2})\Gamma_e(w_2^{\pm2})}\cr
&\qquad \Gamma_e((q  p)^{\frac14}t\beta b^{-1}w_1^{\pm1}z_1^{\pm1})\Gamma_e((q  p)^{\frac14}\beta b w_1^{\pm1}z_2^{\pm1})\Gamma_e((q  p)^{\frac14}t\beta^{-1} b w_2^{\pm1}z_1^{\pm1})\Gamma_e((q  p)^{\frac14}\beta^{-1} b^{-1}w_2^{\pm1}z_2^{\pm1})\cr
&\qquad \Gamma_e((q  p)^{\frac14}t\beta^{-1} a w_1^{\pm1}v_1^{\pm1})\Gamma_e((q  p)^{\frac14}\beta^{-1} a^{-1}	 w_1^{\pm1}v_2^{\pm1})\Gamma_e((q  p)^{\frac14}t\beta a^{-1}w_2^{\pm1}v_1^{\pm1})\Gamma_e((q  p)^{\frac14}\beta a w_2^{\pm1}v_2^{\pm1})\,.
}
}  The fugacities $c_1$, $w_1$, and $v_1$ encode the three $SU(2)$ symmetries associated with the punctures. The fugacities $w_2$, $v_2$, $c_2$, and $\beta^2$ parametrize the $SO(8)$ and $t$ the additional $U(1)$.
The index ${\cal I}_0$ is the index of a Lagrangian theory, $SU(2)^2$ gauge theory with five flavors for each gauge node. The charges of fields can be read from the expression of the supersymmetric index, and from the charges one can deduce  the superpotentials. We then tune the coupling of the IR fixed point to a locus where $U(1)_{\frac{a}b} $ enhances to $SU(2)$ and gauge it with additional matter which can be read from \indTbag, with the charges and the superpotentials again deduced from the index. 

\

\appendix{E}{Formula for relevant and marginal deformations}

Compactifying six dimensional theories on a Riemann surface with flux there is a certain very general class of relevant and marginal deformations of the resulting four dimensional theories which can be predicted to exist following simple geometric considerations. We refer to \Brz\ for details and here we just give the formulas we use in the bulk of the paper.  Consider a six dimensional theory with symmetry group $G$ and  discuss compactification on a genus $g$ Riemann surface. For simplicity we choose to turn on flux for one $U(1)$ (which we denote by $U(1)_a$) in $G$ but the results can be easily generalized for any value of flux. We denote the value of the flux by ${\cal F}_a$ (and we assume it is positive) and by $G'\times U(1)_a$ the group preserved by the flux. Next, we consider  decomposition of the character of the adjoint representation of $G$ to $G'\times U(1)_a$ representations,

\eqn\dgyui{
\chi_{adj}(G) =\sum_{i} a^{q_i}\chi_{R_i}(G')\,.
} Here $q_i$ is the $U(1)_a$ charge of the representation $R_i$ appearing in the decomposition of the adjoint representation of $G$ to $G'$ representations.
By definition of $G'$ there are two representations appearing in the above sum with charge zero, adjoint of $G'$ and a singlet of $G'$.
The claim of \Brz\ is that for general choice of genus $g$ and flux ${\cal F}_a$ the index of the four dimensional theory written with six dimensional R charge is,

\eqn\isjny{\eqalign { &
{\cal I}=1+\biggl(\sum_{i|q_i<0} \chi_{R_i}(G') a^{q_i}(g-1-q_i {\cal F}_a)\biggr) q p +\biggl(3g-3 +(1+\chi_{adj}(G')) (g-1)\biggr) q p +\cr&\qquad \biggl( \sum_{i|q_i>0} \chi_{R_i}(G') a^{q_i}(g-1-q_i {\cal F}_a)\biggr) q p+\cdots\,.}
} The first term will give relevant deformations with superconformal R symmetry of four dimensions, the second term marginal deformations, and last term irrelevant deformations. For low values of genus and flux this formula might get adjustments, but in general we expect it to be correct, see \Brz\ 
for details.

\listrefs

\end